\documentclass{aa}  

\usepackage{graphicx}
\usepackage{txfonts}
\usepackage{hyperref}
\usepackage{enumitem}% http://ctan.org/pkg/enumitem
\usepackage{natbib}
\usepackage{siunitx}
\usepackage{adjustbox}
\usepackage{textcomp}
\usepackage{tablefootnote}
\usepackage{float}
\usepackage{placeins}
\usepackage[flushleft]{threeparttable}
\usepackage{lipsum}
\usepackage{pifont}% http://ctan.org/pkg/pifont
\usepackage{dblfloatfix}
\usepackage{epstopdf}
\usepackage[super,negative]{nth}
\usepackage{subcaption}
\usepackage{pdfpages}
\usepackage{makecell}
\usepackage{booktabs}
\usepackage{multirow}
\usepackage{cellspace}
\usepackage{rotating}
\usepackage{siunitx}
\usepackage{relsize}

\DeclareGraphicsExtensions{.pdf,.eps,.png,.jpg}
\bibpunct{(}{)}{;}{a}{}{,} % to follow the A&A style
\providecommand{\e}[1]{\ensuremath{\times 10^{#1}}}
\setlist{parsep=0pt,listparindent=\parindent}

\newcommand\omicron{o}
\usepackage[normalem]{ulem}
\usepackage{color}
\hypersetup{
    colorlinks=true,
    linkcolor=black,
    citecolor=blue,
    filecolor=blue,
    urlcolor=blue,
}

\begin{document} 

\title{Visible and near-infrared spectro-interferometric analysis of the edge-on Be star $\omicron$ Aquarii%%\thanks{Based on observations performed at CHARA, USA and ESO, Chile under program IDs 087-D.0311 and 094.D-0140.}
}
\titlerunning{spectro-interferometric view of $\omicron$ Aquarii}

\author{
     E. S. G. de Almeida\inst{\ref{inst1}} 
\and A. Meilland\inst{\ref{inst1}} 
\and A. Domiciano de Souza\inst{\ref{inst1}}
\and P. Stee\inst{\ref{inst1}}
\and D. Mourard\inst{\ref{inst1}}
\and N. Nardetto\inst{\ref{inst1}}
\and R. Ligi\inst{\ref{inst2}}
\and I. Tallon-Bosc\inst{\ref{inst3}}
\and D. M. Faes\inst{\ref{inst4}}
\and A. C. Carciofi\inst{\ref{inst4}}
\and D. Bednarski\inst{\ref{inst4}}
\and B. C. Mota\inst{\ref{inst4}}
\and N. Turner\inst{\ref{inst5}}
\and T. A. ten Brummelaar\inst{\ref{inst5}}
}

\institute{
Université Côte d'Azur, Observatoire de la Côte d'Azur, CNRS, Laboratoire Lagrange, France\\ \email {Elisson.Saldanha@oca.eu}\label{inst1} 
\and
INAF-Osservatorio Astronomico di Brera, Via E. Bianchi 46, I-23807 Merate, Italy \label{inst2}
\and 
Univ Lyon, Univ Lyon1, Ens de Lyon, CNRS, Centre de Recherche Astrophysique de Lyon UMR5574, F-69230 Saint-Genis-Laval, France \label{inst3}
\and
Instituto  de  Astronomia, Geofísica e Ciências Atmosféricas, Universidade de São Paulo, São Paulo, Brazil \label{inst4}
\and
CHARA Array - Georgia State University, Mount Wilson, CA, USA \label{inst5}
}

\date{}

%%%%%%%%%%%%%%%%%%%%%%%%%%%%%%%%%%%%%%%%%%%%%%%%%%%%%%%%%%%%%%%%%%%%%abstract
  \abstract
  % context heading (optional)
  {}
  % aims heading (mandatory)
   {We present a detailed visible and near-infrared spectro-interferometric analysis of the Be-shell star $\omicron$ Aquarii from quasi-contemporaneous CHARA/VEGA and VLTI/AMBER observations.}
   % methods heading (mandatory)
   {We analyzed spectro-interferometric data in the H$\alpha$ (VEGA) and Br$\gamma$ (AMBER) lines using models of increasing complexity: simple geometric models, kinematic models, and radiative transfer models computed with the 3-D non-LTE code HDUST.}
  % results heading (mandatory)
   {
   We measured the stellar radius of $\omicron$ Aquarii in the visible with a precision of 8\%: 4.0 $\pm$ 0.3 $\mathrm{R_{\sun}}$. We constrained the circumstellar disk geometry and kinematics using a kinematic model and a MCMC fitting procedure. The emitting disk sizes in the H$\alpha$ and Br$\gamma$ lines were found to be similar, at $\sim$10-12 stellar diameters, which is uncommon since most results for Be stars indicate a larger extension in H$\alpha$ than in Br$\gamma$. We found that the inclination angle $i$ derived from H$\alpha$ is significantly lower ($\sim$\ang{15}) than the one derived from Br$\gamma$: $i$ $\sim$ \ang{61.2} and \ang{75.9}, respectively. While the two lines originate from a similar region of the disk, the disk kinematics were found to be near to the Keplerian rotation (i.e., $\beta$ = -0.5) in Br$\gamma$ ($\beta$ $\sim$ -0.43), but not in H$\alpha$ ($\beta$ $\sim$ -0.30). After analyzing all our data using a grid of HDUST models (BeAtlas), we found a common physical description for the circumstellar disk in both lines: a base disk surface density $\Sigma_{0}$ = 0.12 g cm\textsuperscript{-2} and a radial density law exponent $m$ = 3.0. The same kind of discrepancy, as with the kinematic model, is found in the determination of $i$ using the BeAtlas grid. The stellar rotational rate was found to be very close ($\sim$96\%) to the critical value. Despite being derived purely from the fit to interferometric data, our best-fit HDUST model provides a very reasonable match to non-interferometric observables of $\omicron$ Aquarii: the observed spectral energy distribution, H$\alpha$ and Br$\gamma$ line profiles, and polarimetric quantities. Finally, our analysis of multi-epoch H$\alpha$ profiles and imaging polarimetry indicates that the disk structure has been (globally) stable for at least 20 years. 
   }
  %conclusions
  {Looking at the visible continuum and Br$\gamma$ emission line only, $\omicron$ Aquarii fits in the global scheme of Be stars and their circumstellar disk: a (nearly) Keplerian rotating disk well described by the viscous decretion disk (VDD) model. However, the data in the H$\alpha$ line shows a substantially different picture that cannot fully be understood using the current generation of physical models of Be star disks. The Be star $\omicron$ Aquarii presents a stable disk (close to the steady-state), but, as in previous analyses, the measured $m$ is lower than the standard value in the VDD model for the steady-state regime ($m$ = 3.5). This suggests that some assumptions of this model should be reconsidered. Also, such long-term disk stability could be understood in terms of the high rotational rate that we measured for this star, the rate being a main source for the mass injection in the disk. Our results on the stellar rotation and disk stability are consistent with results in the literature showing that late-type Be stars are more likely to be fast rotators and have stable disks.
  }

%%%%%%%%%%%%%%%%%%%%%%%%%%%%%%%%%%%%%%%%%%%%%%%%%%%%%%%%%%%%%%%%%%%%%


\maketitle

%
%-------------------------------------------------------------------

%%%%%%%%%%%%%%%%%%%%%%%%%%%%%%%%%%%%%%%%%%%%%%%%%%%%%%%%%%%%%%%%%%%%%%%%%%%%%%%%%%%%%section: introduction
\section{Introduction}\label{sec_introduction}

Classical Be stars are main-sequence B-type stars that show (or showed at some time) Balmer lines in emission and infrared excess in their spectral energy distribution. The Be phenomenon is found among the entire spectral range of B stars \citep[e.g.,][]{townsend04}: $M_{\star}$ from $\sim$3 $\mathrm{M_\odot}$ (B9, $T_{\mathrm{eff}}$ $\sim$ 12000 K), up to $\sim$18 $\mathrm{M_\odot}$ (B0, $T_{\mathrm{eff}}$ $\sim$ 30000 K). These observational characteristics are well explained as arising from a dust-free gaseous disk that is supported by rotation with a slow radial velocity \citep[see, e.g.,][]{rivinius13}. The most successful theory to explain the evolution of the disk structure is the so-called viscous decretion disk (VDD) model, where its dynamics are driven by viscosity \citep[e.g.,][]{lee91,okazaki01,bjorkman05}.\par

It is widely accepted that fast rotation plays an important role in the formation of the Be star disk. However, while interferometric analyses typically provide rotational rates $v_{\mathrm{rot}}/v_{\mathrm{crit}}$ $\gtrsim$ 0.7 \citep[e.g.,][]{meilland12, cochetti19}, some statistical studies show rates ranging from $\sim$0.3 up to 1.0 \citep[e.g.,][]{cranmer05, zorec16}. Moreover, it is still not clear whether the rotational rate is correlated to other stellar parameters such as the effective temperature \citep[e.g.,][]{cochetti19}. Hence, despite the success of the VDD model, the physical mechanism(s) driving the mass injection remains unclear and a detailed physical characterization for the central star and the disk structure is mandatory to better understand the Be phenomenon. By gaining access to geometry on the milliarcsecond scale and kinematics on a few tens of km\,s$^{-1}$ scale, spectro-interferometry offers a unique opportunity to probe the circumstellar environment and stellar surfaces of Be stars \citep[see, e.g.,][]{chesneau12, stee12}.\par

The bright, late-type Be star (type B7IVe) $\omicron$ Aquarii (HD 209409) is known to have a fairly stable disk \citep{sigut15}. The stability of the circumstellar disk is evidenced by the quasi-constant equivalent width in the H$\alpha$ line, double-peak separation, and the absence of long-term violet-to-red ($V/R$) peak variations \citep[e.g.,][]{rivinius06, sigut15}. This star shows a high value of $v \sin i$ $\sim$ 282 km s$^{-1}$ \citep{fremat05} and a shell absorption in H$\alpha$, thus indicating a high stellar inclination angle of about \ang{70}, as discussed below.\par

\citet{meilland12} presented the first spectro-interferometric analysis of $\omicron$ Aquarii with the VLTI/AMBER instrument as part of their AMBER survey of eight bright Be stars. Despite the low data quality and very limited number of observations (just one measurement), they were able  to significantly constrain the disk geometry and kinematics. They found that the disk emission in the Br$\gamma$ line, modeled as an elliptical Gaussian distribution, had a FWHM of 14 $\pm$ 1 D$_\star$ (with $R_\star$ = 4.4 R$_\odot$), where D$_\star$ and $R_\star$ are, respectively, the stellar diameter and radius. They estimated the inclination angle as $i$ = 70 $\pm$ \ang{20} and found a stellar rotational rate of $v_{\mathrm{rot}}/v_{\mathrm{crit}}$ = 0.77 $\pm$ 0.21 ($\Omega$/$\Omega_c$ = 0.93$^{+0.06}_{-0.17}$), where $v_\mathrm{crit}$ and $\Omega_\mathrm{crit}$ are, respectively, the linear and angular critical velocity. New VLTI/AMBER spectro-interferometric measurements of $\omicron$ Aquarii were presented in the Be star survey of \citet{cochetti19}. Here, they obtained seven good-quality measurements for $\omicron$ Aquarii (i.e., 21 baselines). Using a similar model as in \citet{meilland12}, they found a Br$\gamma$ emission FWHM significantly smaller than in \citet{meilland12}, 8 $\pm$ 0.5 D$_\star$ (with $R_\star$ = 4.4 R$_\odot$), and better constrained the object inclination angle (70 $\pm$ \ang{5}).\par

A detailed analysis of $\omicron$ Aquarii using H$\alpha$ spectroscopy and interferometry was performed by \citet{sigut15}. These authors combined large band (15 nm) interferometric data centered on H$\alpha$, obtained from the Navy Precision Optical Interferometer (NPOI), with H$\alpha$ spectroscopy from the Lowell Observatory Solar-Stellar Spectrograph. Using the radiative transfer code BEDISK \citep{sigut07}, they were able to reproduce simultaneously the visibility, H$\alpha$ line profile, and spectral energy distribution (SED), and showed that the disk is quite stable for up to about ten years. Interestingly, they found a disk extension in H$\alpha$ (Gaussian FWHM of 12.0 $\pm$ 0.5 D$_\star$) close to the one determined by \citet{meilland12} in Br$\gamma$ (FWHM of 14 $\pm$ 1 D$_\star$). They concluded that this is uncommon since most previously studied Be stars exhibit a larger (up to two times) disk emission region in H$\alpha$ than in Br$\gamma$.\par

In this paper, we present new CHARA/VEGA spectro-interferometric measurements of $\omicron$ Aquarii centered on the H$\alpha$ emission line ($\lambda$ = 0.656 $\mu$m). They are analyzed conjointly with the AMBER  Br$\gamma$ line ($\lambda$ = 2.166 $\mu$m) measurements from \citet{meilland12} and \citet{cochetti19}, using models of increasing complexity: simple geometric models, kinematic models, and radiative transfer models. This is the first time the code HDUST has been used to model simultaneously spectro-interferometric data from H$\alpha$ and Br$\gamma$. It is the second time for the kinematic model \citep[i.e., after the $\delta$ Scorpii data published in][]{meilland11}. This multi-wavelength and multi-line approach allows us to draw a more complete picture of the stellar surface and circumstellar environment of the Be star $\omicron$ Aquarii.\par

This paper is organized as follows. In Sect. \ref{sec_observations}, we present the observations and the data reduction process. Our analysis using geometric models of the VEGA calibrated (absolute) visibility is shown in Sect. \ref{sec_geometric_modeling}. In Sect. \ref{sec_kinematic_model_mcmc}, we fit the VEGA and AMBER differential visibility and phase with a kinematic model using a Markov Chain Monte Carlo (MCMC) model fitting method. In Sect. \ref{sec_hdust_modeling}, all the interferometric data are analyzed in terms of 3-D non-LTE radiative transfer models. Our kinematic and radiative transfer models are discussed in Sect. \ref{sec_comparison_hdust_kinematic}. In Sect. \ref{sec_spectrum_sed}, our best-fit models are compared to non-interferometric observables: the spectral energy distribution and line profiles (H$\alpha$ and Br$\gamma$). The comparison with polarimetric data is performed in Sect. \ref{sec_variability_polarimetry} in the context of the disk stability. In Sect. \ref{sec_discussion}, we discuss the morphological, kinematic, and physical descriptions for $\omicron$ Aquarii and its circumstellar disk. Our conclusions are summarized in Sect. \ref{sec_conclusions}.\par

%%%%%%%%%%%%%%%%%%%%%%%%%%%%%%%%%%%%%%%%%%%%%%%%%%%%%%%%%%%%%%%%%%%%%%%%%%%%%%%%%%%%%section: observations

\section{Observations}\label{sec_observations}

\subsection{CHARA/VEGA}

The VEGA instrument \citep{mourard09} is one of the two visible beam combiners on the CHARA Array \citep{brummelaar05}. It can simultaneously combine up to four beams, operating at different wavelengths from 450 to 850 nm. VEGA is equipped with two cameras (blue and red detectors) that can observe in two different spectral domains simultaneously (around the H$\beta$ and the H$\alpha$ lines). Currently, it is the only instrument at the CHARA Array with a spectral resolution high enough to resolve narrow spectral features such as atomic and molecular lines. It offers 3 spectral modes: $R$ = 1000 (LR), $R$ = 6000 (MR), and $R$ = 30000 (HR).\par

\begin{figure}[t]
\centerline{\resizebox{0.50\textwidth}{!}{\includegraphics{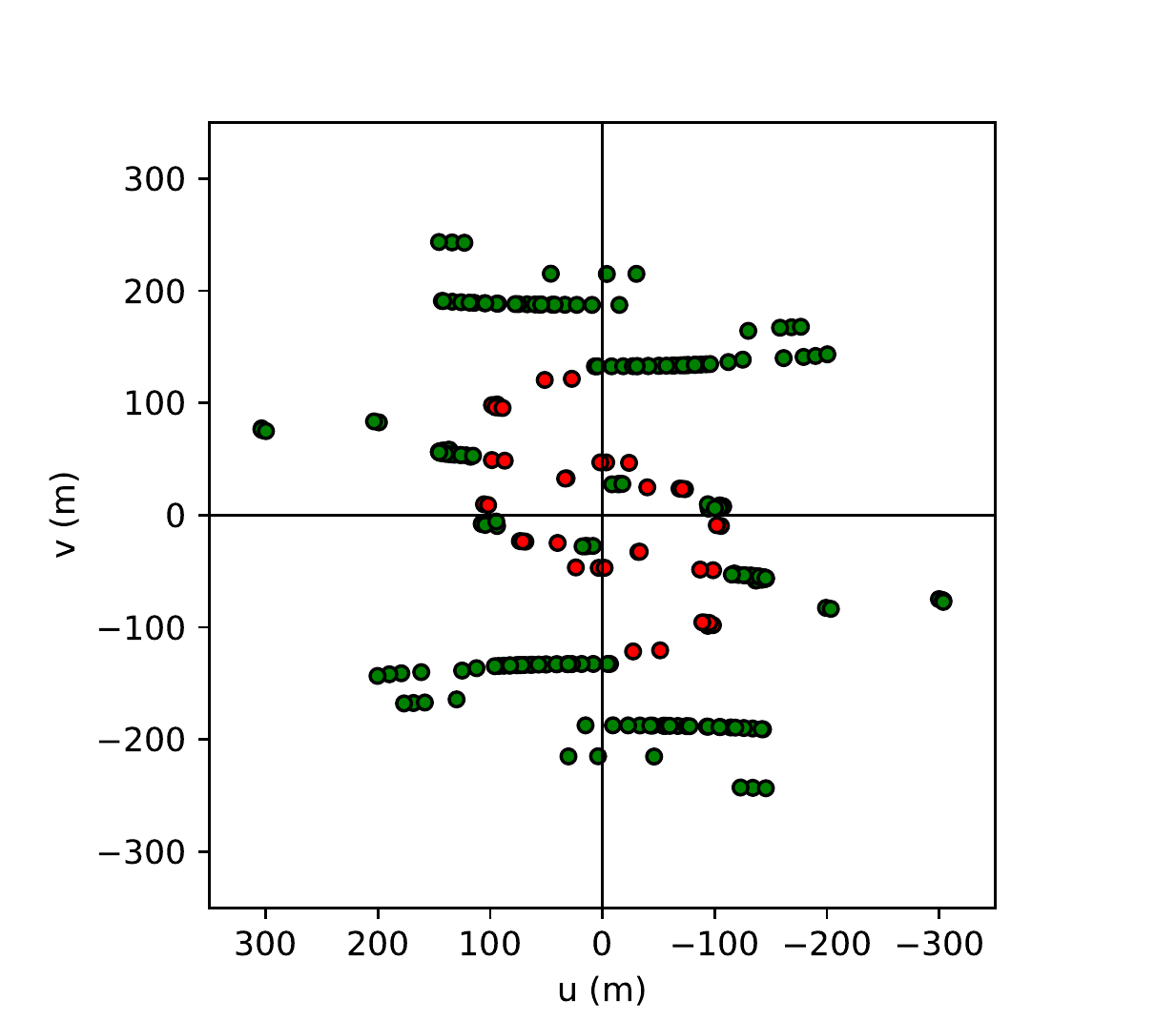}}}
\caption{$uv$ plan coverage obtained around H$\alpha$ (0.656 $\mu$m) with CHARA/VEGA (green) and Br$\gamma$ (2.166 $\mu$m) with VLTI/AMBER (red).}
\label{uv}
\end{figure}

$\omicron$ Aquarii was observed 50 times with VEGA between 2012 and 2016 in MR mode centered on the H$\alpha$ emission line at 0.656 $\mu$m. The 2012 and 2016 observations were focused on the disk geometry and kinematics and data were taken with small baselines (up to 105 m) and without stellar calibrators. On the other hand, the 2013 and 2014 campaigns were aimed at constraining not only the H$\alpha$ emission, but also the R-band continuum geometry. Consequently, observations were carried out with longer baselines (up to 330 m) with a standard calibration plan alternating observations of the science target and few calibrator stars chosen using the SearchCal \citep{bonneau06} tool developed by the Jean-Marie Mariotti Center (JMMC)\footnote{\url{https://www.jmmc.fr/english/tools/proposal-preparation/search-cal/}}. Table \ref{calib_vega_amber} (Appendix \ref{appendix_observational_logs}) shows useful information about the stars used as interferometric calibrators during these campaigns. The complete log of observations is presented in Table \ref{log_vega_amber} and the corresponding $uv$ plane coverage for the VEGA observations is plotted in Fig \ref{uv}.\par

%---------------------------------------%---------------------------------------
%---------------------------------------%---------------------------------------

Data were reduced using the standard VEGA data reduction software\footnote{See VEGA group page at \url{https://lagrange.oca.eu/fr/vega}.} described in \citet{mourard12}. For all programs, differential visibility and phases were computed from the intercorrelation between a fixed 15 nm window centered on H$\alpha$ and a sliding smaller window (i.e. 1, 2 , or 5 {\AA}, depending on the data quality). For the 2013 and 2014 data, the raw squared visibility was computed for $\omicron$ Aquarii, and its calibrators, using the auto-correlation method on a 15 nm band centered on the H$\alpha$ emission line (649-664 nm) and another band in the close-by continuum (635-650 nm). Then the transfer function was estimated assuming the diameter of the calibrators recorded before and after the science target observation, and its uncertainty using a weighted standard deviation. Finally, for each measurement, the calibrated squared visibility was derived by dividing $\omicron$ Aquarii's raw squared visibility by the estimated transfer function.\par

\subsection{VLTI/AMBER} 

The AMBER instrument \citep{petrov07} was a three-beam combiner (decommissioned in 2018) at the Very Large Telescope Interferometer (VLTI). It operated in the H- and K-bands with three spectral resolutions: $R$ = 35 (LR), $R$ = 1500 (MR), and $R$ = 12000 (HR). It offered the highest spectral resolution at the VLTI, being the most adapted for studying the gaseous environment in emission lines.\par

$\omicron$ Aquarii was observed with AMBER during two observing surveys of Be stars in 2011 (ESO program  087-D.0311) and in 2014 (ESO program 094.D-0140). The observations were performed in HR mode in K-band centered on the Br$\gamma$ emission line at 2.166 $\mu$m. The data from 2011 was published in \citet{meilland12} and the 2014 data in \citet{cochetti19}. During this second survey, seven measurements were acquired for $\omicron$ Aquarii with three different triplets. The log of AMBER observations is also presented in Table \ref{log_vega_amber} and the corresponding $uv$ plane coverage is plotted in Fig. \ref{uv}. \par

Calibration was performed using similar methods as the one described for VEGA. However, AMBER measurements were often affected by a highly variable transfer function mainly due to the variable quality of the fringe tracking performed by the FINITO fringe tracker, during the long exposure time needed to perform HR mode observations. As it was the case during our $\omicron$ Aquarii observations, we present in this paper only the analysis of differential measurements obtained using the standard AMBER data reduction software \texttt{amdlib} \citep{tatulli07, chelli09}. 

%%%%%%%%%%%%%%%%%%%%%%%%%%%%%%%%%%%%%%%%%%%%%%%%%%%%%%%%%%%%%%%%%%%%%%%%%%%%%%%%%%%%%section: geometric modeling
\section{Geometric modeling: VEGA calibrated visibility}\label{sec_geometric_modeling}

In this section, we fit the H$\alpha$ and continuum squared visibilities ($V^2$) from the VEGA observations where calibrators were observed. We note that as the AMBER data were not calibrated, such analysis cannot be performed on the K-band continuum and Br$\gamma$ line. \par

To determine if we can separate the circumstellar disk and the stellar photosphere emissions and constrain their geometry independently, we fitted our data with geometric models of increasing complexity: one-component models (uniform disk, UD, or a uniform ellipse) and two-component models (UD plus UD, Gaussian disk, or uniform or Gaussian ellipse).\par

%---------------------------------------%---------------------------------------
\begin{table*}[t]
\caption{\label{table_litpro_vega} Results from the geometric modeling of the VEGA $V^{2}$ data in the close-by continuum band (635-650 nm) and in the band centered on H$\alpha$ (649-664 nm).
For each band, many models were tested, but only these composed of one and two uniforms disks (UD) are presented here. The angular diameter of each UD component is denoted as $\theta_1$ and $\theta_2$. The normalized flux contribution of the first and second model components are, respectively, $F_1$ and F$_2$ ($F_1$ + $F_2$ = 1). All parameters were free in our modeling.} 
\centering
\begin{tabular}{c |c c c c | c c c c}
\toprule
\toprule
& \multicolumn{4}{c|}{ Continuum (635-650 nm)} & \multicolumn{4}{c}{H$\alpha$ (649-664 nm)}\\
\midrule
Model & $\theta_1$ (mas) & $\theta_2$ (mas) & $F_2$ & $\chi^2_\mathrm{r}$ & $\theta_1$ (mas) & $\theta_2$ (mas) & $F_2$ & $\chi^2_\mathrm{r}$ \\
\midrule
1 UD   & 0.28 $\pm$ 0.01 & ---           & ---            & 1.1  & 0.36 $\pm$ 0.01 & ---           & ---            & 2.8 \\
2 UDs  & 0.27 $\pm$ 0.02 & $23^{+82}_{-23}$ & 0.03 $\pm$ 0.03  & 1.1 & 0.26 $\pm$ 0.02 & 6.5 $\pm$ 2.1 & 0.15 $\pm$ 0.03 & 1.3 \\
\bottomrule
\end{tabular}
\end{table*}
%---------------------------------------%---------------------------------------

Here, the first component represents the stellar surface and the second one the circumstellar disk. To perform our fit, we used the LITpro model fitting software \citep{tallonbosc08} for optical and infrared interferometric observations developed by the Jean-Marie Mariotti Center (JMMC)\footnote{LITpro software is available at \url{https://www.jmmc.fr/english/tools/data-analysis/litpro/}.}.\par

\begin{figure}
\centerline{\resizebox{0.50\textwidth}{!}{\includegraphics{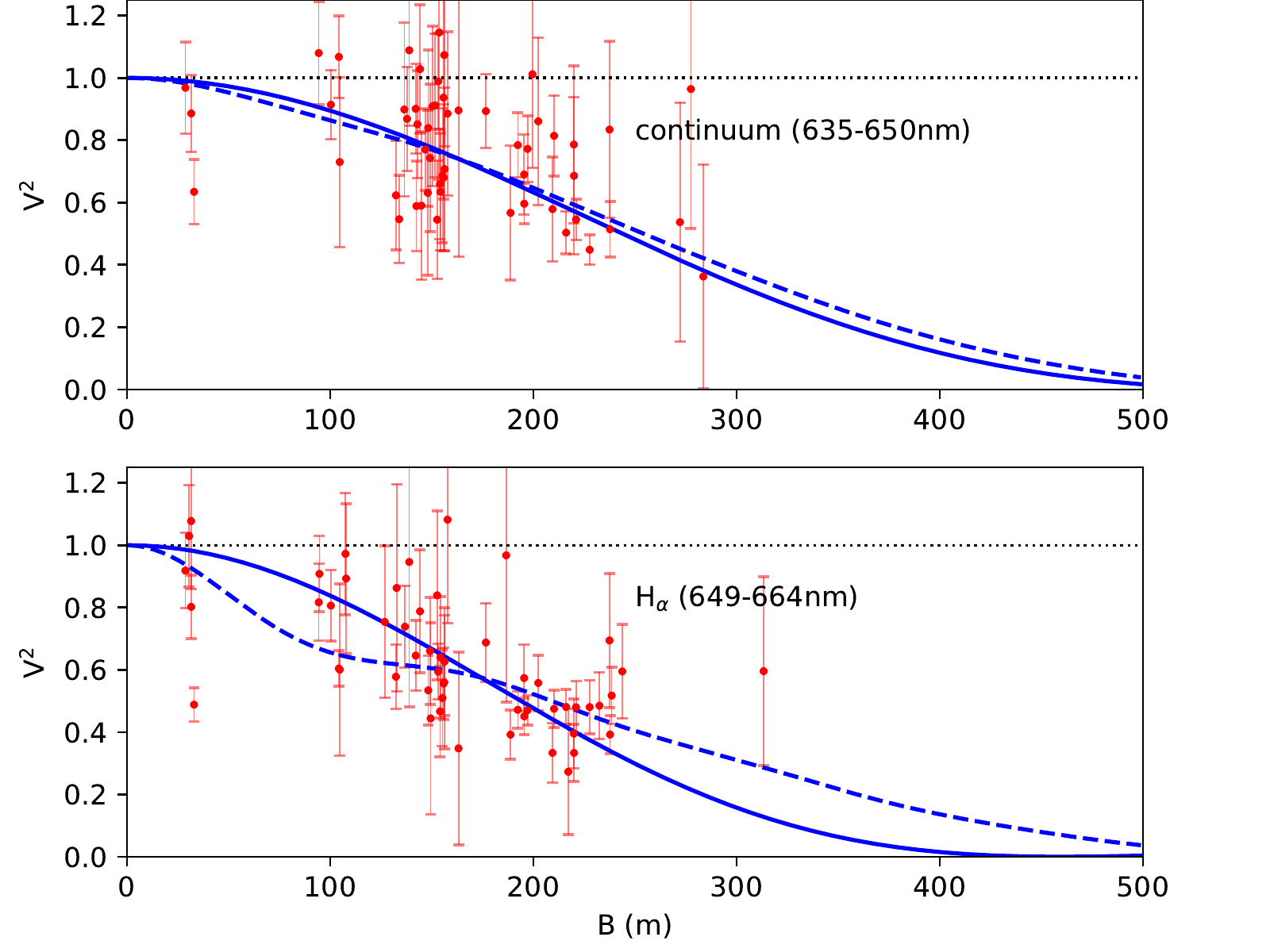}}}
\caption{VEGA $V^2$ measurements in the close-by continuum band (top) and in the H$\alpha$ band (bottom) are shown in red points. Our best-fit models consisting of one (solid line) and two (dashed line) uniform disks are overplotted in blue. See Table \ref{table_litpro_vega} and text for discussion.}
\label{vega_calibrated_vis_model_fit}
\end{figure}

In Fig. \ref{vega_calibrated_vis_model_fit}, we show the comparison between the visibility curves of our best-fit models to the VEGA data both in the continuum and H$\alpha$ bands. One sees that the object is partially resolved in the continuum and the H$\alpha$ line. The lower level of the visibility in the band centered on the  H$\alpha$ line clearly shows that the object is larger in H$\alpha$ than in the close-by continuum region. Assuming that the emission originates from both the stellar photosphere and a circumstellar disk, the lower visibility in H$\alpha$ is due to a larger fraction of the H$\alpha$ flux coming from the disk than from the star. In contrast, the flux contribution from the star is greater than that from the disk in the continuum R-band.\par

%---------------------------------------%---------------------------------------
%---------------------------------------%---------------------------------------

Our main results are summarized in Table \ref{table_litpro_vega}. We only show our results using UD models since there is no improvement in terms of reduced $\chi^2$ ($\chi^2_\mathrm{r}$) when considering more complex models, that is, with a higher number of free parameters. For the continuum band, there is no significant improvement in terms of reduced $\chi^2$ between a simple UD and a two-component UD model. The central star is clearly resolved by the longer baselines and its extension is significantly constrained with a UD diameter of $\theta$ = 0.28 $\pm$ 0.01 mas ($\chi^2_\mathrm{r}$ $\sim$ 1.1). This value corresponds to an upper limit to the stellar diameter measurements neglecting the putative contribution of the circumstellar disk in the R-band continuum. Adding a second component to the model only marginally reduces the extension of the first component. The contribution of the second component, representing the circumstellar disk, is small ($F_2$ = 0.03 $\pm$ 0.03), thus the extension of the disk cannot be constrained.\par

%---------------------------------------%---------------------------------------
\begin{figure*}
\centerline{\resizebox{0.90\textwidth}{!}{\includegraphics{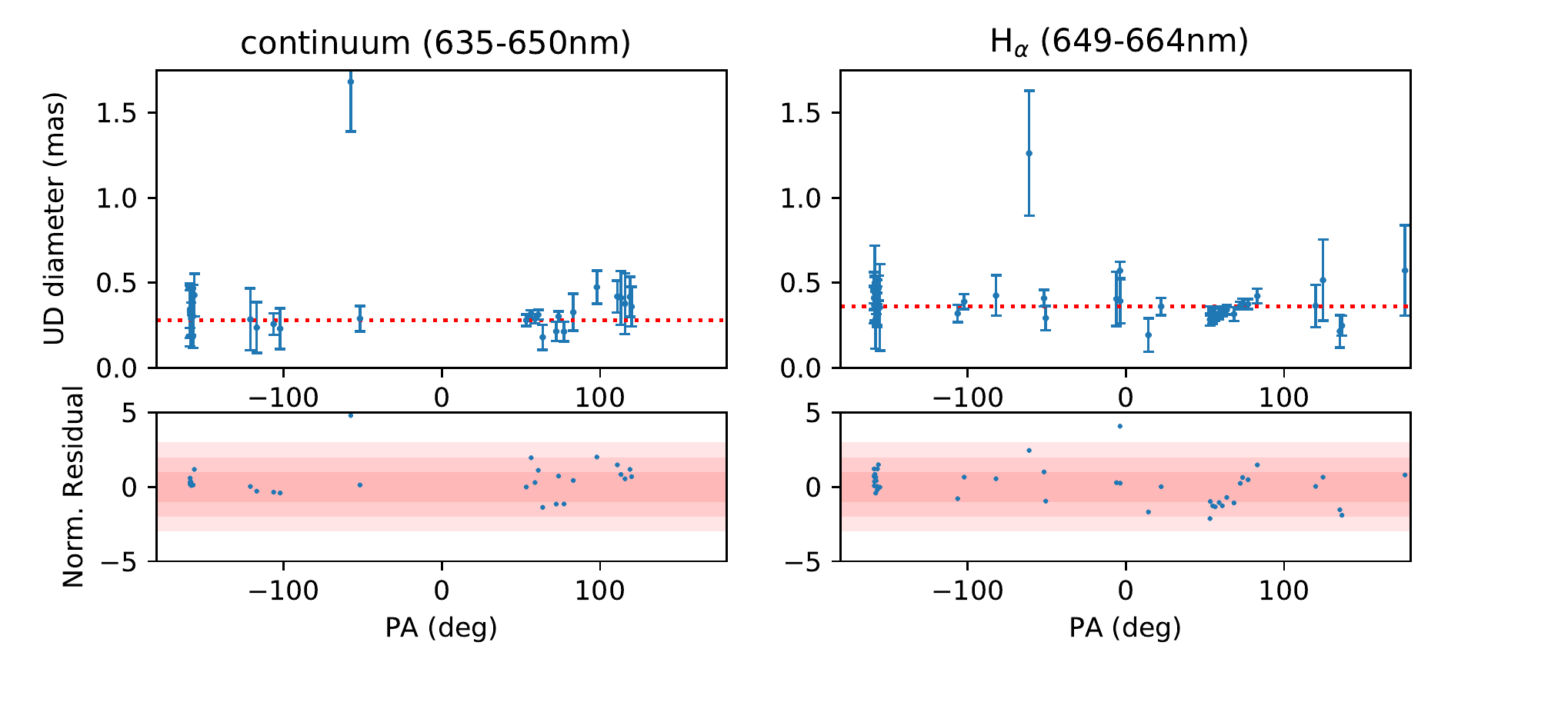}}}
\vspace{-0.5cm}
\caption{Top panels: uniform disk diameter derived from each individual VEGA $V^2$ measurements (continuum band in the left and H$\alpha$ band in the right) plotted as a function of the baseline position angle ($PA$). The red dotted line represents the best-fit diameter from modeling all the data in each band ($\theta$ = 0.28 mas in the continuum and $\theta$ = 0.36 mas in the H$\alpha$ band). Bottom panels: corresponding normalized residuals.}
\label{v2_PA}
\end{figure*}
%---------------------------------------%---------------------------------------

Unlike the continuum case, the situation is quite different in the band centered on the H$\alpha$ line. The single uniform disk gives a significantly higher $\chi^2_\mathrm{r}$ $\sim$ 2.8 for a best-fit model with $\theta$ = 0.36 mas. In this case, adding a second component reduces $\chi^2_\mathrm{r}$ by a factor of two, leading to $\chi^2_\mathrm{r}$ $\sim$ 1.3. Using a model with two uniform disks, we converge to a diameter of the first component similar to the one found from the continuum, that is, 0.26 $\pm$ 0.02 mas. The flux contribution of the second component and its extension are significantly constrained. However, the uncertainty remains quite large, that is, $F_{2}$ = 0.15 $\pm$ 0.03 and $\theta_2$ = 6.5 $\pm$ 2.1 mas (see Table \ref{table_litpro_vega}).\par

Considering that the first component of our model represents the stellar photosphere, our measurement is slightly higher than the value assumed in the work of \citet{sigut15} of 0.22 mas. However, their adoption for the stellar angular diameter is based on a spectral type-radius relation for B dwarf stars \citep{townsend04}. Moreover, this value of 0.22 mas represents the polar radius. $\omicron$ Aquarii is a fast rotator likely to be significantly flattened, and our measurements are spread over different orientations, so that we end up measuring a mean radius of the star projected on the sky. Assuming a distance of 144 pc \citep[derived from the Gaia DR2 parallaxes, ][]{gaia18}, $\theta_1$ = 0.26 $\pm$ 0.02 mas corresponds to a stellar radius $R_{\star}$ = 4.0 $\pm$ 0.3 R$_\odot$.\par

Finally, to try to detect any possible stellar or circumstellar disk flattening from the squared visibility measurements, we also computed individual uniform disk equivalent diameter for each $V^2$ measurement. This analysis of the uniform disk diameter for $\omicron$ Aquarii, as a function of the VEGA baseline orientation, is shown in Fig. \ref{v2_PA}. As expected from our analysis (considering uniform elliptical models), we do not find any evidences of flattening from modeling our $V^2$ dataset since no clear trends are found in the model residual as varying the baseline position angle.\par

%%%%%%%%%%%%%%%%%%%%%%%%%%%%%%%%%%%%%%%%%%%%%%%%%%%%%%%%%%%%%%%%%%%%%%%%%%%%%%%%%%%%%section: kinematic modeling
\section{Kinematic modeling: VEGA and AMBER differential data}\label{sec_kinematic_model_mcmc}

To constrain the geometry and kinematics of the circumstellar gas in the H$\alpha$ and Br$\gamma$ lines, we fit the VEGA and AMBER differential visibility and phase measurements using a simple bi-dimensional kinematic model for a rotating disk\footnote{Available at the JMMC service AMHRA: \url{https://amhra.oca.eu/AMHRA/}.}.\par

\subsection{The kinematic model}\label{sec_kinematic_code}

This kinematic model was already used in a series of papers about spectro-interferometric modeling of Be stars, including \citet{delaa11}, \citet{meilland12}, and \citet{cochetti19}, and is presented in detail in these references. \par

In short, the intensity map for the central star is modeled as a uniform disk, and the circumstellar disk as two elliptical Gaussian distributions, one for the flux in continuum, and the other one for the flux in line. The disk is geometrically thin so that the ellipse flattening ratio is set to $1/\cos i$, where  $i$ is the inclination angle. The disk intensity map in the line is computed taking into account the Doppler effect due to the disk rotational velocity in the considered spectral channels. The parameters of our kinematic model are the following : 

\begin{enumerate}[label=(\roman*)]
\setlength\itemsep{1em}

\item The simulation parameters: size in pixels ($n_{xy}$), field of view in stellar diameters ($fov$), number of wavelength points ($n_{\lambda}$), central wavelength of the emission line ($\lambda_{0}$), step size in wavelength ($\delta\lambda$), and spectral resolution ($\Delta\lambda$).

\item The global geometric parameters: stellar radius ($R_{\star}$), distance ($d$), inclination angle ($i$), and disk major-axis position angle ($PA$).

\item The disk continuum parameters: disk major-axis FWHM in the continuum ($a_{c}$), disk continuum flux normalized by the total continuum flux ($F_{c}$).

\item The disk emission line parameters: disk major-axis FWHM in the line ($a_{\mathrm{line}}$) and line equivalent width ($EW$).

\item The kinematic parameters: rotational velocity ($v_\mathrm{rot}$) at 1.5 $R_{\mathrm{p}}$ (polar radius) and exponent of the rotational velocity power-law ($\beta$).
\end{enumerate}

\subsection{Model fitting using the MCMC method}\label{sec_mcmc_method}

To perform our model fitting, we used the code emcee \citep{foreman13}. This is an implementation in Python of the Markov Chain Monte Carlo (MCMC) method from \citet{goodman10}. Some recent works on stellar interferometry used this code \citep[see., e.g.,][]{monnier12, domiciano14, sanchez17, domiciano18}.\par

The simulation parameters were set as follows: $n_{xy}$ = 256, $fov$ = 60 $\mathrm{D_\star}$, $n_{\lambda}$ = 60 (VEGA) and 110 (AMBER), $\lambda_{0}$ = 6563 {\AA} (VEGA) and 21661 {\AA} (AMBER), $\delta\lambda$ = 2.5 {\AA} (VEGA) and 1.0  {\AA} (AMBER), and $\Delta\lambda$ = 5.0 {\AA} (VEGA) and 1.8 {\AA} (AMBER). 
To reduce the number of free parameters, we set $R_{\star}$ = 4.0 $\mathrm{R_\odot}$ and $d$ = 144 pc. We also fixed the disk continuum extension $a_{c}$ and flux $F_{c}$ to 0 for VEGA (i.e., neglecting the disk contribution in the continuum, based on our analysis of the VEGA $V^2$ data). In the AMBER analysis, we adopted $a_{c}$ = 3 $\mathrm{D}_\star$ and  $F_{c}$ = 0.2 from \citet{cochetti19}. The line equivalent width was set to 19.9 {\AA} in H$\alpha$. \citep{sigut15}. For Br$\gamma$, we computed the $EW$ using the AMBER spectra from all observations and found a mean value of 13.6 $\pm$ 1.1 {\AA}, which is compatible with the value from \citet{meilland12}, 12.6 {\AA}, but not with the result from \citet{cochetti19} of 18.1 {\AA}. Finally, from the ten parameters of the kinematic model, the fitting of the VEGA and AMBER data were performed with at most five free parameters: $i$, $PA$, $a_\mathrm{line}$, $v_\mathrm{rot}$, and $\beta$.\par 

The likelihood function ($p_\mathrm{like}$) of the MCMC procedure was chosen as $\ln(p_\mathrm{like}) = -\chi^2_{\mathrm{total}}/2$, where $\chi^2_{\mathrm{total}}$ is the sum of the $\chi^2$ computed for the differential visibility and the differential phase. Thus, our attempt to converge to samples of parameters that maximizes the likelihood function means the minimization of the total $\chi^2$ between our interferometric data and the kinematic model. \par

We performed three different model fitting tests with different constraints on the value of $v_\mathrm{rot}$:  

\begin{enumerate}[label=(\roman*)]
\setlength\itemsep{1em}

\item Five free parameters: $i$, $PA$, $a_\mathrm{line}$, $v_\mathrm{rot}$, and $\beta$. Without the inclusion of any prior probability function in the analysis.\par

\item Four free parameters: $i$, $PA$, $a_\mathrm{line}$, and $\beta$. The stellar rotational velocity $v_\mathrm{rot}$ is fixed on the critical value of 391 km s\textsuperscript{-1} \citep{fremat05}.\par

\item Five free parameters: $i$, $PA$, $a_\mathrm{line}$, $v_\mathrm{rot}$, and $\beta$. We take into account a prior probability function $p_\mathrm{prior}$ on $v\sin i$. Adopting $\mu$ = 282 km s\textsuperscript{-1} and $\sigma$ = 20 km s\textsuperscript{-1}, from the measured $v\sin i$ = 282 $\pm$ 20 km s\textsuperscript{-1} \citep{fremat05}, we have the following expression for $p_\mathrm{prior}$:

\begin{equation}
ln(p_\mathrm{prior}) = \frac{ -{(v\sin i - \mu)}^{2} }{ 2{\sigma}^2 },
\end{equation}
where $v\sin i$ is calculated from the sampled MCMC values for the stellar rotational velocity and inclination angle.

Hence, considering a high weight on $p_\mathrm{prior}$, the following quantity for the posterior probability function $p_\mathrm{post}$ is maximized:

\begin{equation}
\ln(p_\mathrm{post}) = -100 \left( \frac{{(v\sin i - \mu)}^{2} }{ 2{\sigma}^2 } \right)  -\frac{\chi^{2}}{2}.
\label{eq:ppost}
\end{equation}

Note that this is equivalent to the case of equal weights for $p_\mathrm{prior}$ and $p_\mathrm{like}$, but considering a lower error bar on $v\sin i$, namely, $\sigma$ = 2 km s\textsuperscript{-1}.\par

\end{enumerate}

We typically used several hundreds of walkers ($\sim$300-900) for the MCMC run. Convergence was obtained for about 50 to 100 iteration steps in each walker, but we used a conservative value of 150 steps in the burn-in phase and 50 in the main phase to estimate the parameters values and uncertainties. Overall, we found a mean acceptance fraction of $\sim$0.5-0.6 in our MCMC tests. This is close to the optimal range for this parameter of $\sim$0.2-0.5 \citep[see, e.g.,][]{foreman13}.\par

\subsection{Best-fits in H$\alpha$ and Br$\gamma$}\label{sec_kinematic_models_results}

\begin{figure*}
  \begin{adjustbox}{minipage=\linewidth,scale=1.30}
  \centering
  \hspace{-4.3cm}%
  \includegraphics[width=0.45\linewidth]{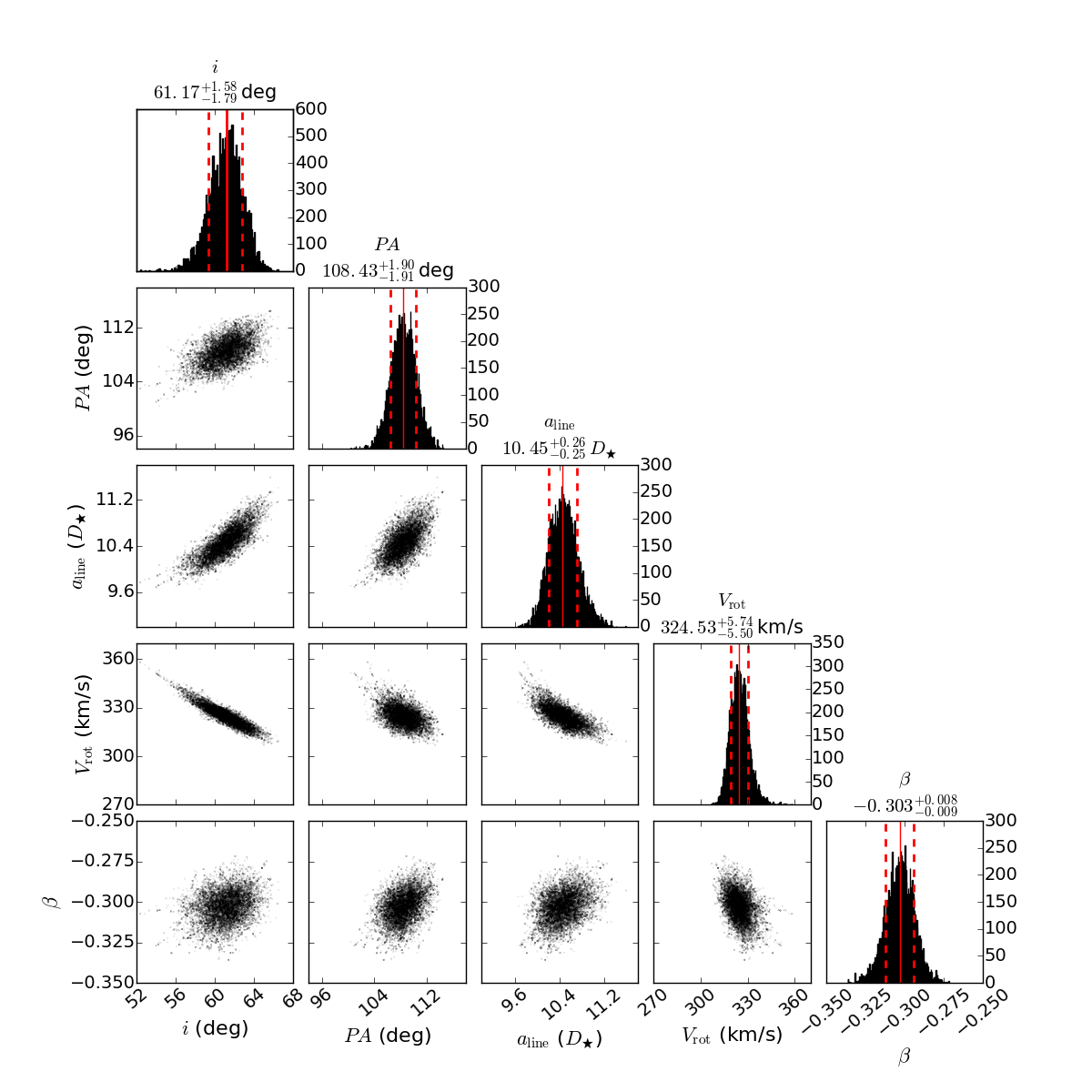}
  \medskip
  \hspace{-0.6cm}%
  \includegraphics[width=0.45\linewidth]{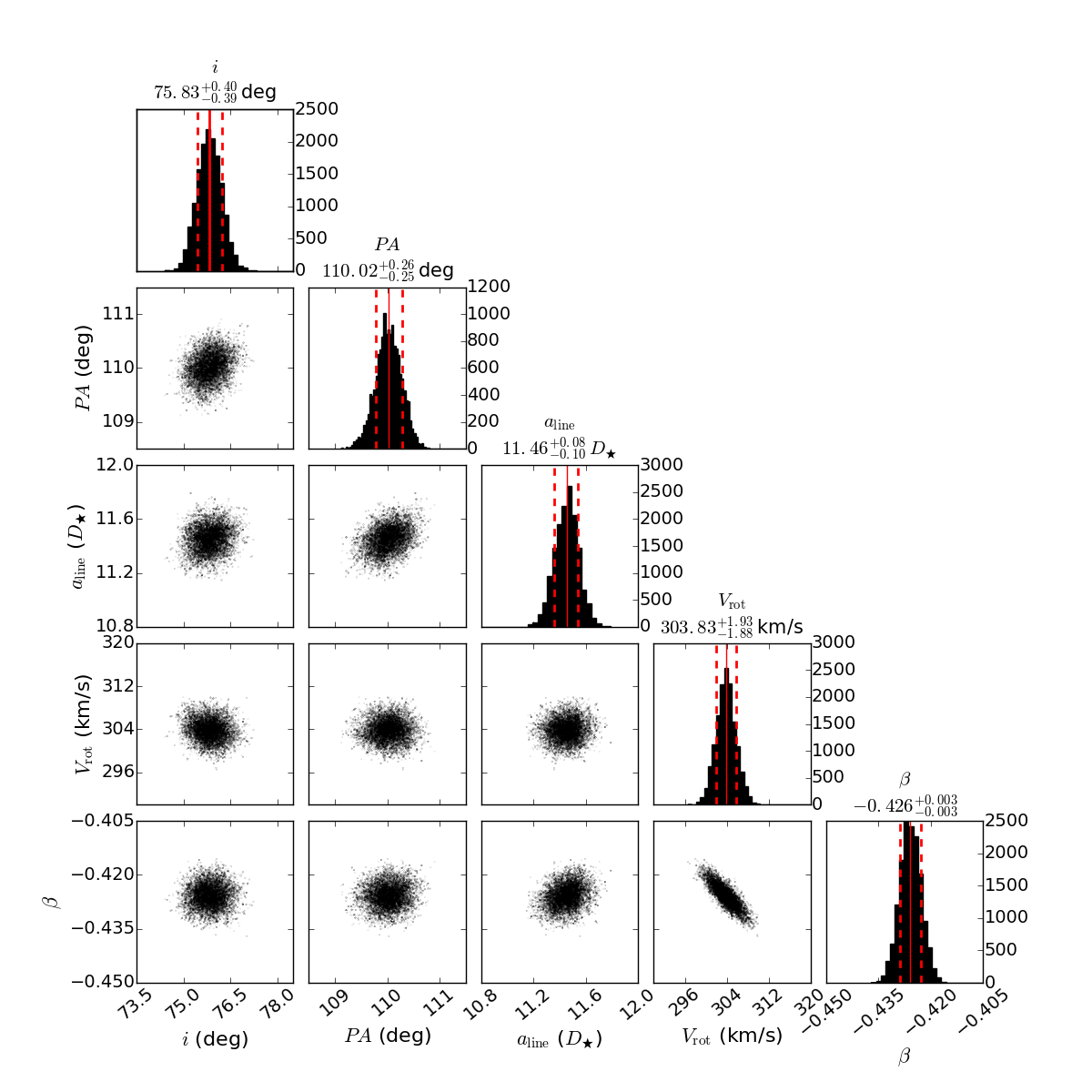}\hspace{-0.0cm}%
  \end{adjustbox}
  \vspace{-0.5cm}
  \caption{Histogram distributions and two-by-two correlations (after the burn-in phase) for the free parameters of our best-fit kinematic models using MCMC for the VEGA (left panel) and AMBER (right panel) differential data. The median values are shown in solid red lines and the first and third quartiles in dashed red lines. The median and the first and third quartiles estimated for the parameters of our best-fit models (VEGA and AMBER) are presented in Table \ref{table_mcmc_vega_amber}. In the correlation plots, darker points correspond to models with lower values of $\chi^2$. See text for discussion.}\label{mcmc_vega_amber_corner_hist}
\end{figure*}

We modeled a total of 117 (VEGA) and 24 (AMBER) measurements of differential visibility and phase. The best-fit parameters for the MCMC fit with a prior on $v \sin i$ (test iii, described above) are presented in Table \ref{table_mcmc_vega_amber}. The corresponding histograms and the two-by-two parameter correlations from this MCMC run (one for VEGA and other for AMBER) are shown in Fig. \ref{mcmc_vega_amber_corner_hist}. The corresponding histograms and correlation plots for the other two fits (tests i and ii) are shown in Figs. \ref{mcmc_other_tests_vega} and \ref{mcmc_other_tests_amber} (Appendix \ref{appendix_mcmc}). One sees that the values of $i$, $PA$, and $a_\mathrm{line}$, derived from each emission line, differ only marginally in all the fitting tests, showing the robustness of the solution for these parameters.\par

%---------------------------------------%---------------------------------------
\begin{table}
\caption{\label{table_mcmc_vega_amber} Best-fit kinematic models from test (iii) for our VEGA (H$\alpha$) and AMBER (Br$\gamma$) differential data. We show the median and the first and third quartiles for each parameter derived from the MCMC analysis. Adopted parameters stand by  ``$\equiv$''.}
\renewcommand{\arraystretch}{1.5}
\centering
\begin{tabular}{lll}
\toprule
\toprule
\multicolumn{1}{l}{Parameter} & \multicolumn{1}{c}{\textbf{VEGA diff.}} & \multicolumn{1}{c}{\textbf{AMBER diff.}}  \\
\midrule 
$i$ (deg)  &$61.2^{+1.6}_{-1.8}$ &$75.9^{+0.4}_{-0.4}$  \\
$PA$ (deg) &$108.4^{+1.9}_{-1.9}$ &$110.0^{+0.3}_{-0.3}$ \\
$a_{\mathrm{line}}$ ($\mathrm{D}_\star$) &$10.5^{+0.3}_{-0.3}$ &$11.5^{+0.1}_{-0.1}$ \\
$v_{\mathrm{rot}}$ (km s\textsuperscript{-1}) &$325^{+6}_{-6}$ &$303^{+2}_{-2}$ \\
$\beta$ &$-0.30^{+0.01}_{-0.01}$ &$-0.426^{+0.003}_{-0.003}$ \\
\midrule 
$R_{\star}$ ($\mathrm{R_\odot}$) & $\equiv$ 4.0\,\tablefootmark{a, b} & $\equiv$ 4.0\,\tablefootmark{a, b} \\
$d$ (pc) & $\equiv$ 144\,\tablefootmark{c} & $\equiv$ 144\,\tablefootmark{c} \\
$a_{c}$ ($\mathrm{D}_\star$) & $\equiv$ 0\,\tablefootmark{a} & $\equiv$ 3\,\tablefootmark{d} \\
$F_{c}$ & $\equiv$ 0\,\tablefootmark{a} & $\equiv$ 0.2\,\tablefootmark{d} \\
$EW$ (\AA) & $\equiv 19.9$\,\tablefootmark{e} & $\equiv$ 13.6\,\tablefootmark{f} \\
\midrule 
$\chi^2_\mathrm{r}$ & 4.04 & 1.57 \\
\bottomrule
\end{tabular}
\tablefoot{
\tablefoottext{a}{Based on our fit to the VEGA squared visibility.}
\tablefoottext{b}{Radius derived considering the distance adopted from \citet{gaia18}.}
\tablefoottext{c}{Distance adopted from \citet{gaia18}.}
\tablefoottext{d}{Adopted from \citet{cochetti19}.}
\tablefoottext{e}{Adopted from \citet{sigut15}.}
\tablefoottext{f}{Measured from our AMBER observations.}
}
\end{table}
%---------------------------------------%---------------------------------------

%---------------------------------------%---------------------------------------
%---------------------------------------%---------------------------------------

%---------------------------------------%---------------------------------------
\begin{figure*}
\centerline{\resizebox{1.00\textwidth}{!}{\includegraphics[angle=0]{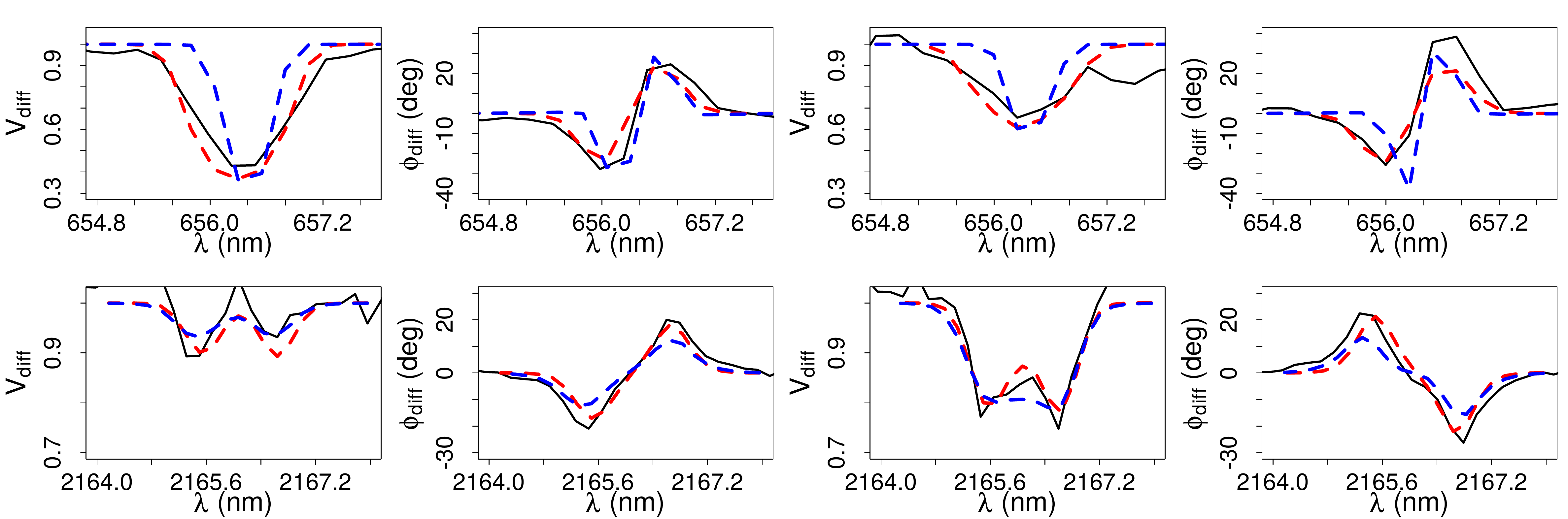}}}
\caption{Comparison between our best-fit kinematic models (dashed red; Table \ref{table_mcmc_vega_amber}) and two different VEGA (top panels) and AMBER (bottom panels) measurements (black line). Our best-fit HDUST model is also shown (dashed blue; Table \ref{table_hdust_ref}; discussion in Section \ref{sec_comparison_hdust_kinematic}). $\delta\lambda$ of the kinematic model and AMBER data is increased to 1.8 {\AA} in order to compare them to the HDUST model ($\delta\lambda$ fixed to 1.8 {\AA}).}
\label{bestfit_hdust_kinematic_model_vega_amber}
\end{figure*}
%---------------------------------------%---------------------------------------

In Fig \ref{bestfit_hdust_kinematic_model_vega_amber}, we show examples of VEGA and AMBER data in comparison to our best-fit kinematic models. For later discussion in Sect. \ref{sec_comparison_hdust_kinematic}, the visibility and phase from our best-fit HDUST model is also presented here. Our best-fit kinematic models are able to reproduce both the VEGA and AMBER differential data well. We found a reduced $\chi^2$ of $\sim$4.0 and 1.6 from fitting, in a separate way, respectively, the VEGA and AMBER datasets.\par

We derived compatible values for the disk $PA$ ($\sim$\ang{110}) from fitting the VEGA and AMBER data with an uncertainty up to $\sim$\ang{2}. This result agrees well with previous studies \citep[e.g.,][]{meilland12, touhami13, sigut15, cochetti19}. On the other hand, the inclination angle determined from the fit to the VEGA data is significantly smaller ($i$ = 61.2 $\pm$ \ang{1.8}) in comparison to the one determined from fitting AMBER ($i$ = 75.9 $\pm$ \ang{0.4}). This latter value is in good agreement with the results for $i$ found by \citet{meilland12} and \citet{cochetti19}. We also constrain the disk extension with a good precision: $a_\mathrm{line}$ = 10.5 $\pm$ 0.3 $\mathrm{D}_\star$ in the H$\alpha$ line and $a_\mathrm{line}$ = 11.5 $\pm$ 0.1 $\mathrm{D}_\star$ in the Br$\gamma$ line. These values are compatible with the ones determined by \citet{sigut15} in H$\alpha$ and \citet{meilland12} in Br$\gamma$.\par 

Another aspect concerning the disk extension in Br$\gamma$ is the significant discrepancy seen in comparison to $a_\mathrm{line}$ = 8.0 $\pm$ 0.5 $\mathrm{D}_\star$ from \citet{cochetti19}. However, these authors used a larger value for the stellar radius of 4.4 $\mathrm{R}_\sun$ and a closer distance of 134 pc \citep{vanleeuwen07}, having thus the angular size of the stellar diameter larger in $\sim$19\% than the one assumed in our kinematic analysis from our results in Sect. \ref{sec_geometric_modeling}. Considering all the other parameters fixed, this results in a smaller disk extension in $\sim$19\% than one found from our analysis. Nevertheless, the largest contribution to this discrepancy between our results and the ones from \citet{cochetti19} is due to their high value of equivalent width in the Br$\gamma$ line of 18.1 {\AA}, as discussed in Sect. \ref{sec_mcmc_method}, that also implies in a smaller disk extension in this line.\par

\begin{figure}[t]
\centerline{\resizebox{0.500\textwidth}{!}{\includegraphics{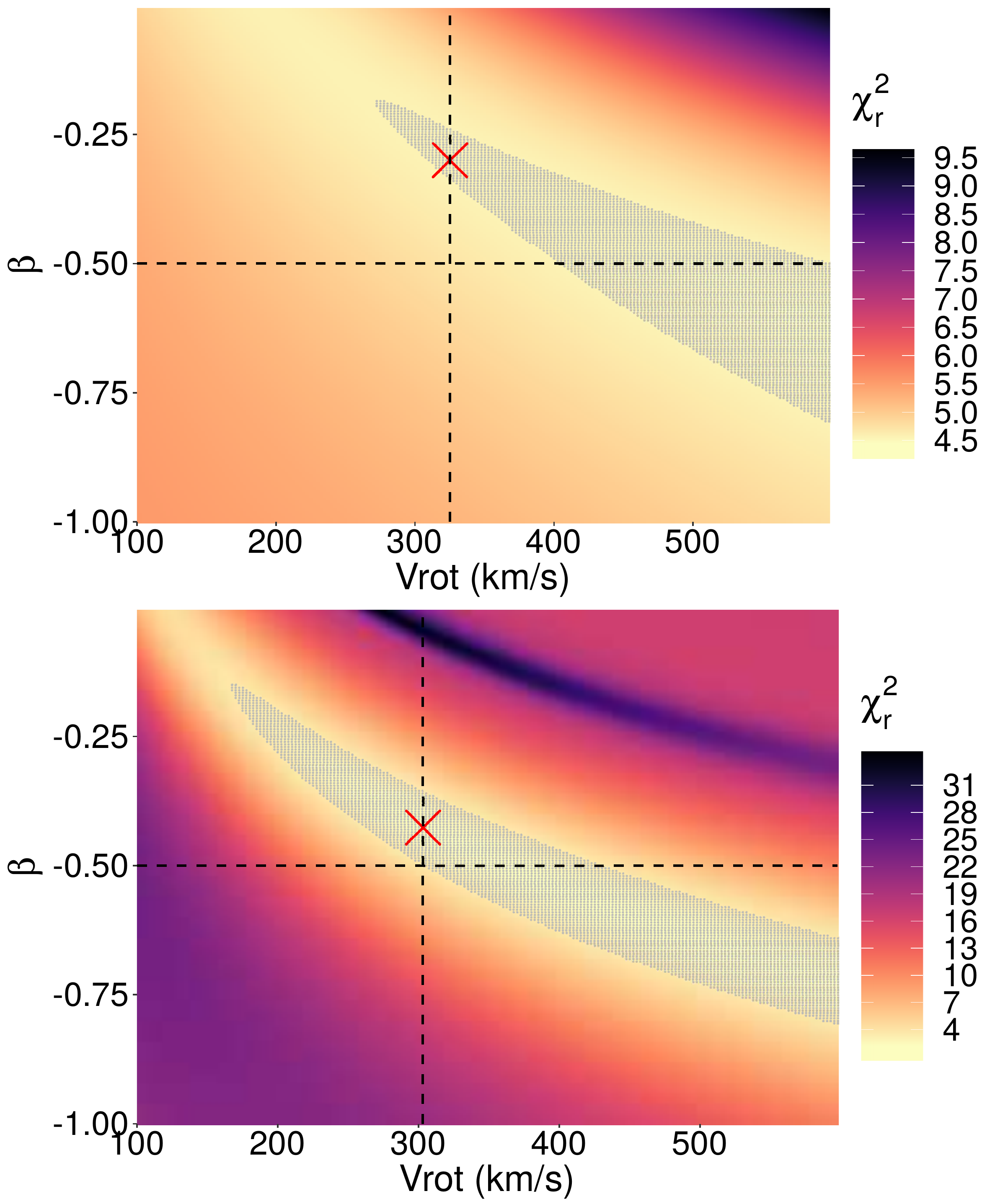}}}
\caption{ $\chi^2_\mathrm{r}$ maps of 40000 kinematic models as a function of $v_\mathrm{rot}$ and $\beta$ from the fit to VEGA (top panel) and AMBER (bottom panel) differential data. Only these two parameters were varied in a regular step in the intervals shown here. The other parameters are fixed (Table \ref{table_mcmc_vega_amber}). Our results found from the MCMC analysis for $v_\mathrm{rot}$ and $\beta$ are indicated with red crosses. In order to highlight the correlation between $\beta$ and $v_\mathrm{rot}$, the gray region corresponds to an arbitrary number of models, encompassing about the 5000 best models in both cases. The value of $\beta$ = -0.5 (Keplerian disk) and our determination for $v_\mathrm{rot}$ are marked in dashed black line. Note the strong correlation between the stellar rotational velocity and the disk velocity law exponent in both the cases. Also, note that a Keplerian disk is found from modeling the AMBER data, but not from VEGA.}
\label{vrot_beta_vega_amber}
\end{figure}

From our various tests, we showed that $\beta$ and $v_{\mathrm{rot}}$ are strongly correlated. To precisely determine their dependence, we computed a grid of kinematic models varying just these two parameters in a regular step size. The values for $i$, $PA$, $a_\mathrm{line}$ are fixed from Table \ref{table_mcmc_vega_amber}. The resulting $\chi^2_\mathrm{r}$ maps are shown in Fig. \ref{vrot_beta_vega_amber}. As expected, one sees that $v_{\mathrm{rot}}$ and $\beta$ are highly correlated for the VEGA and AMBER data. This high degeneracy can be understood since these two parameters provide the rotational velocity structure in the disk: it is hard to distinguish the effects of each one on the modeling of spectro-interferometric (and spectroscopic) data.\par

Furthermore, we see that $\beta$ = -0.5 (Keplerian disk) provides unrealistically high values for the stellar rotational velocity ($\gtrsim$ 400 km s\textsuperscript{-1}; gray region) of $\omicron$ Aquarii (VEGA analysis). For AMBER, $v_{\mathrm{rot}}$ is significantly reduced to about 300-400 km s\textsuperscript{-1}. As shown in Fig. \ref{vrot_beta_vega_amber}, our results from AMBER are consistent with a nearly Keplerian rotating disk ($\beta$ $\sim$ 0.43). However, it is conspicuous that the $\beta$ value calculated from the VEGA data ($\beta$ $\sim$ 0.30) shows such a large departure from the Keplerian case.\par

\citet{cochetti19} derived a stellar rotational velocity of 355 $\pm$ 50 km s\textsuperscript{-1} and $\beta$ = -0.45 $\pm$ 0.03. This is in fair agreement with our results for both $v_{\mathrm{rot}}$ and $\beta$. Considering our MCMC test (ii), where $v_{\mathrm{rot}}$ is fixed to the critical value and $\beta$ is a free parameter, the results for $\beta$ are shifted to higher values (more positive) with $\beta$ $\sim$ -0.42 (VEGA) and -0.54 (AMBER).\par

Therefore, regardless the MCMC fitting considered here, we verify a discrepancy of about 0.1 between the value of $\beta$ derived from the H$\alpha$ and Br$\gamma$ lines. Our results from the AMBER analysis (Br$\gamma$) seems to be consistent with a nearly Keplerian rotating disk, but we verified a larger departure from $\beta$ = -0.5 for the VEGA analysis (H$\alpha$).\par

%---------------------------------------%---------------------------------------
%---------------------------------------%---------------------------------------

%%%%%%%%%%%%%%%%%%%%%%%%%%%%%%%%%%%%%%%%%%%%%%%%%%%%%%%%%%%%%%%%%%%%%%%%%%%%%%%%%%%%%section: hdust modeling
\section{Radiative transfer modeling}\label{sec_hdust_modeling}

\subsection{The code HDUST}\label{hdust_description}

%---------------------------------------%---------------------------------------
\begin{figure*}
\centerline{\resizebox{1.00\textwidth}{!}{\includegraphics[height=1cm]{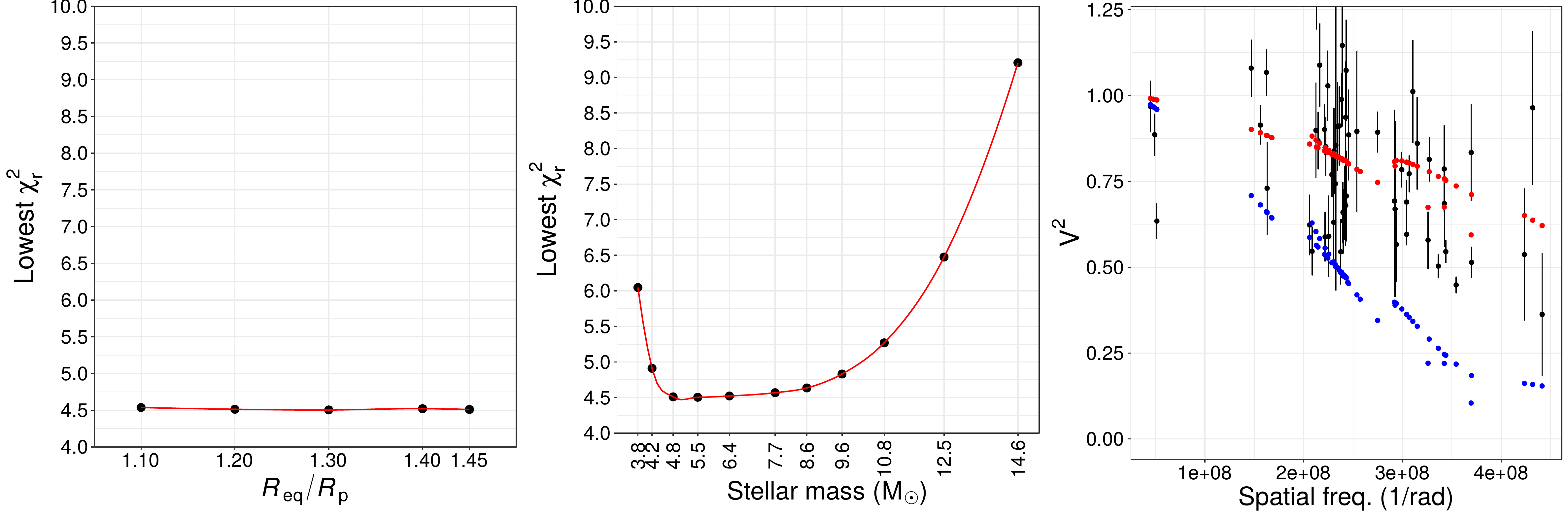}}}\caption{Analysis (i): lowest value of reduced $\chi^2$ ($\chi^2_\mathrm{r}$) for each value of $R_{\mathrm{eq}}/R_{\mathrm{p}}$ (left panel) and stellar mass (middle panel) from the HDUST fit to the VEGA $V^2$ data (642.5 nm band).  Local regression fits to $\chi^2_\mathrm{r}$, as a function of the parameter values, are shown in red line. In the right panel, the predicted visibility from our best-fit HDUST model (red points; Table \ref{table_hdust_ref}) is compared to the VEGA $V^2$ measurements in the continuum band (black points). The predicted visibility from the HDUST model with the highest mass in the BeAtlas grid (14.6 $\mathrm{M_{\odot}}$, highest $\chi^2_{\mathrm{r}}$ in the middle panel) is shown in blue points.}
\label{hdust_vega_vis2_oblat_mass}
\end{figure*}
%---------------------------------------%---------------------------------------

We used the 3-D non-LTE radiative transfer code HDUST\footnote{For access and collaborations with HDUST, please contact A. C. Carciofi.} \citep{carciofi06a, carciofi08} to perform a deeper physical analysis of $\omicron$ Aquarii. In addition to geometric and kinematic parameters, we seek to derive the density and temperature distributions in the disk, and the spectral energy distribution (SED), none of which was provided by the two simpler models considered in the two previous sections. HDUST uses a Monte Carlo method to solve the radiative transfer, statistical and radiative equilibrium equations for arbitrary density and velocity distributions in gaseous (pure hydrogen) or dusty circumstellar environments.\par

This code is well-suited to model the circumstellar environment of Be stars as it implements the VDD model. Thus, the disk velocity law is assumed to be Keplerian ($\beta$ fixed to -0.5). Many previous studies explored formal solutions of the VDD model in several limiting cases. For example, \citet{bjorkman05} investigated the isothermal, steady-state case of a disk formed by a steady mass injection rate over a long time. Effects due to non-isothermal temperature structure were studied by \citet{carciofi08}. \citet{haubois12} studied the temporal evolution of the disk structure that is subject to variable mass inject rates. Finally, the effects of a binary companion on the disk were studied by \citet{okazaki02}, \citet{oudmaijer10}, \citet{panoglou16}, and \citet{cyr17}, among others.\par

From these studies, the radial density profile in Be star disks is found to be quite complex, for example, depending on the disk age, dynamical state, or presence of a binary companion. Despite this complexity, several studies have shown that the global behavior of this density profile is successfully approximated by a simple radial power-law \citep[e.g.,][]{touhami09, vieira17}. Considering also that the vertical density structure is that of an isothermal disk (hydrostatic assumption in the $z$-axis), the disk density can be parameterized as follows:

\begin{equation}
\rho(r,z) = \rho_{0}{\left(\frac{R_{\mathrm{eq}}}{r}\right)}^{m} \exp\left( \frac{-z^{2}}{2{H(r)}^{2}}  \right),
\label{eq:rho}
\end{equation}
where $\rho_{0}$ is the disk base density, $R_\mathrm{eq}$ is the equatorial radius, and $H(r)$ is the (isothermal) disk scale height given by:

\begin{equation}
H(r) = H_{0} \left(\frac{r}{R_{\mathrm{eq}}}\right)^{3/2},
\label{eq:disk_height_scale}
\end{equation}
and $H_{0}$ is the scale height at the disk base,

\begin{equation}
H_{0} =  c_{s} R_\mathrm{eq} \, \left( \frac{G M_{\star}}{ R_\mathrm{eq}}      \right)^{-1/2}, 
\label{eq:disk_height_scale_base}
\end{equation}
where $M_{\star}$ is the stellar mass, $G$ the gravitational constant, and $c_{s}$ the sound speed velocity which depends on the local disk temperature $T$:
\begin{equation}
c_s = \sqrt{\frac{ k_{B}T  }{\mu m_{H}}},
\label{eq:sound_speek_disk}
\end{equation}
where $k_{B}$ is the Boltzmann constant, $\mu$ is the mean molecular weight of the gas, $m_{H}$ is the hydrogen mass, and $T$ is adopted as $0.72T_{\mathrm{pol}}$, where $T_{\mathrm{pol}}$ is the polar effective temperature \citep[see][]{mota19}.\par

HDUST has been used a few times to model spectro-interferometric observations \citep[e.g.,][]{carciofi09, klement15, faes15}. From the solution of the radiative transfer problem, we are able to calculate synthetic spectra and intensity maps as a function of the wavelength around specific spectral lines. We estimated the stellar and circumstellar disk parameters from the comparison of our spectro-interferometric observations (visible and near-infrared) with synthetic observables computed from the Fourier transform of HDUST monochromatic intensity maps. \par

\subsection{BeAtlas grid}\label{sec_beatlas}

%---------------------------------------%---------------------------------------
\begin{figure}
\centerline{\resizebox{0.50\textwidth}{!}{\includegraphics[angle=0]{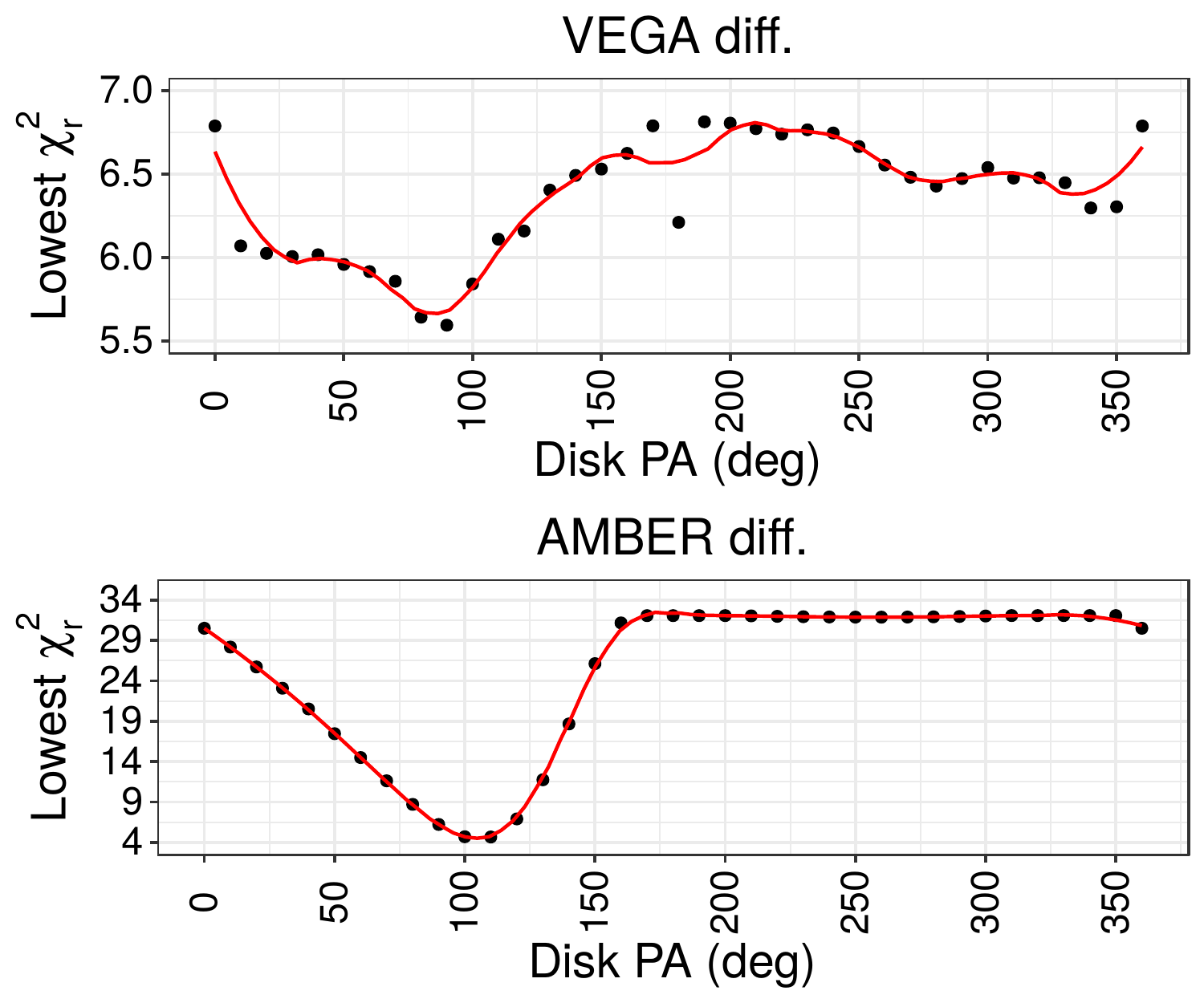}}}
\caption{Lowest value of $\chi^2_\mathrm{r}$ for each value of disk major-axis position angle from the HDUST fit to the VEGA (top panel, analysis ii) and AMBER (bottom panel, analysis iii) differential visibility and phase. Local regression fits of $\chi^2_\mathrm{r}$ as a function of the disk $PA$ are shown as a red line. }
\label{hdust_vega_amber_diskPA}
\end{figure}
%---------------------------------------%---------------------------------------

Since a few hours are needed to compute a single HDUST model, it is not possible to perform an iterative model fitting procedure similar to the one described in Sect. \ref{sec_kinematic_model_mcmc}. To overcome this issue, we used a pre-computed grid of HDUST models called BeAtlas \citep{faes15, mota19}. The BeAtlas grid is presented and described in detail by these references. It consists of $\sim$14000 models with images (specific intensity maps), SEDs, and spectra calculated in natural and polarized spectra, over several spectral regions, including the H$\alpha$ and Br$\gamma$ lines that are of interest for the analysis of our VEGA and AMBER dataset.\par

%---------------------------------------%---------------------------------------
\begin{table}
\caption{\label{table_beatlas} List of HDUST parameters in the BeAtlas grid. First row indicates the spectral type corresponding to the stellar mass \citep{townsend04}. Models are calculated with the following fixed parameters: fraction of H in the core $X_{c}$ = 0.30, metallicity $Z$ = 0.014, and disk radius = 50 $R_{\mathrm{eq}}$.}
\centering
\renewcommand{\arraystretch}{2.25}
\begin{adjustbox}{width=0.50\textwidth}
\begin{tabular}{ll}
\toprule
\toprule
Parameter &  Value \\
\midrule
Spectral type & \makecell[l]{B0.5, B1, B1.5, B2, B2.5, B3, B4, B5 \\ B6, B7, B8}\\
\midrule
$M_\star$ ($\mathrm{M_\odot}$) & \makecell[l]{14.6, 12.5, 10.8, 9.6, 8.6, 7.7, 6.4, 5.5 \\ 4.8, 4.2, 3.8}\\
$i$ (deg) & \makecell[l]{0.0, 27.3, 38.9, 48.2, 56.3, 63.6, 70.5 \\ 77.2, 83.6, 90.0}\\
Oblateness ($R_{\mathrm{eq}}/R_{\mathrm{p}}$) & 1.1, 1.2, 1.3, 1.4, 1.45\\
$\Sigma_{0}$ (g cm\textsuperscript{-2})\,\tablefootmark{a} & 0.02, 0.05, 0.12, 0.28, 0.68, 1.65, 4.00\\
$m$\,\tablefootmark{b} & 3.0, 3.5, 4.0, 4.5\\
\bottomrule

\end{tabular}
\end{adjustbox}
%}

\tablefoot{
\tablefoottext{a}{Surface density at the base of the disk.}
\tablefoottext{b}{Disk mass density law exponent.}

}
\end{table}
%---------------------------------------%---------------------------------------

In Table \ref{table_beatlas}, we show the parameter space covered by BeAtlas. Five physical parameters are varying in the grid. The stellar mass $M_\star$, the inclination angle $i$, and the stellar oblateness $R_{\mathrm{eq}}/R_{\mathrm{p}}$, fully describe the star. Other stellar parameters such as the stellar polar radius ($R_{\mathrm{p}}$), rotational velocity ($v_{\mathrm{rot}}$) and linear and angular rotational rates ($v_{\mathrm{rot}}/v_{\mathrm{crit}}$ and $\Omega/\Omega_{\mathrm{crit}}$) can be computed from $M_\star$ and $R_{\mathrm{eq}}/R_{\mathrm{p}}$ assuming rigid rotation under the Roche model \citep[see, e.g.,][]{carciofi08}. The two last parameters in Table \ref{table_beatlas} describe the circumstellar disk structure and are parameterizations of the VDD model: the base surface density ($\Sigma_{0}$) and the radial density exponent ($m$).\par

%---------------------------------------%---------------------------------------
\begin{table*}
\caption{\label{table_hdust_topmodels} First three columns: mean and standard deviation\,\tablefootmark{a} values for each HDUST parameter of the BeAtlas grid: from analysis ii (19 best-fit HDUST models), analysis iii (16 best-fit HDUST models), and analysis iv (17 best-fit HDUST models). In the bottom rows, there are shown the intervals of $\chi^2_\mathrm{r}$ between the minimum value $\chi^2_\mathrm{min, r}$ and a certain threshold $A$ ($\chi^2_\mathrm{min, r}$ + $A$\%). From modeling the AMBER data, all models have $\Sigma_{0}$ = 0.12 g cm\textsuperscript{-2} and $m$ = 3.0 up to, respectively, $\chi^2_{min, r}$ + 207\% and 240\%, thus the standard deviation shown here is null. The parameters of the HDUST models with $\chi^2_\mathrm{min, r}$ are given in the last three columns. The stellar mass is fixed to 4.2 $\mathrm{M}_{\odot}$ and disk $PA$ =  \ang{110}.}

\centering
\renewcommand{\arraystretch}{1.10}
\begin{adjustbox}{width=1.00\textwidth}
\begin{tabular}{l|ccc|ccc}
\toprule
\toprule

\multicolumn{1}{l|}{Parameter} & \multicolumn{1}{c}{ \textbf{\makecell{VEGA diff. \\ (19 best models)}} } & \multicolumn{1}{c}{\textbf{\makecell{AMBER diff. \\ (16 best models)}}}  & \multicolumn{1}{c|}{\textbf{\makecell{All interf. \\ (17 best models)}}}  & \multicolumn{1}{c}{ \textbf{\makecell{VEGA diff. \\ ($\chi^2_\mathrm{min, r}$)}} } & \multicolumn{1}{c}{ \textbf{\makecell{AMBER diff. \\ ($\chi^2_\mathrm{min, r}$)}} } & \multicolumn{1}{c}{ \textbf{\makecell{All interf. \\ ($\chi^2_\mathrm{min, r}$)}} } \\

\midrule 

$i$ (deg) &57.3 (5.3)  &71.5 (10.8) &65.3 (15.8) &56.3 &77.2 &63.6 \\
$R_{\mathrm{eq}}/R_{\mathrm{p}}$ &1.42 (0.05) &1.39 (0.07) &1.36 (0.09) &1.45 &1.45 &1.45 \\
$\Sigma_{0}$ (g cm\textsuperscript{-2}) &0.09 (0.05) &0.12 (0.00) &0.12 (0.00) &0.05 &0.12 &0.12 \\
$m$ &3.13 (0.22)\,\tablefootmark{c} &3.00 (0.00) &3.00 (0.00) &3.0 &3.0 &3.0 \\

\midrule 

$\chi^2_\mathrm{r}$ &[6.11,6.35] &[4.67,7.19] &[6.40,7.68] &6.11 &4.67 &6.40 \\

Top $A$\% best\,\tablefootmark{b} & 4\% & 54\% & 20\% &--- &--- &---  \\

\bottomrule

\end{tabular}
\end{adjustbox}

\tablefoot{
\tablefoottext{a}{These values of standard deviation are given in parenthesis since they are not error bars on the parameters.}
\tablefoottext{b}{``Top $A$\% best'' stands by the HDUST models with $\chi^2_\mathrm{min, r} \leq \chi^2_\mathrm{r} \leq \chi^2_\mathrm{min, r}$ + $A$\%, where $\chi^2_\mathrm{min, r}$ is the minimum $\chi^2_\mathrm{r}$. These thresholds are chosen to encompass about the same number of HDUST models ($\sim$15-20 models).}
\tablefoottext{c}{Mean and standard deviation calculated from 16 models since three out of 19 models, in this $\chi^2_\mathrm{r}$ threshold, are non-parametric models of the BeAtlas grid.}
}

\end{table*}
%---------------------------------------%---------------------------------------

The previously described volume mass density (Eq. \ref{eq:rho}) and the surface mass density are related as follows:

\begin{equation}
\Sigma(r) \equiv \int_{-\infty}^{+\infty} \rho(r,z) dz,
\end{equation}

\begin{equation}
\rho(r,z) = \frac{\Sigma(r)}{H(r) \sqrt{2\pi}} \exp\left( \frac{-z^{2}}{2{H(r)}^{2}}  \right).
\label{eq:rho_sigma}
\end{equation}

From that, to facilitate the comparison to other disk models, we note that the relation between the volume and surface mass densities at the base of the disk is given by:

\begin{equation}
\rho_{0} = \Sigma_{0} \sqrt{\frac{GM_{\star}}{ 2\pi {c_s}^{2} {R_{\mathrm{eq}}}^{3}     }   }.
\end{equation}

The range of values for $\Sigma_{0}$ and $m$ in the grid encompasses somewhat extreme cases in the literature for the circumstellar disk of Be stars. For example, see Fig. 7 of \citet{vieira17}. The listed values of $\Sigma_{0}$ correspond to $\rho_{0}$ from $\sim 10^{-12}$ g cm\textsuperscript{-3} to $\sim 10^{-10}$  g cm\textsuperscript{-3}. Parametric models with $m$ = 3.5 are equivalent to the steady-state solution of the viscous diffusion equation considering an isothermal disk scale height. Thus, concerning the mass density law exponent $m$, models with $m > 3.5$ would represent a disk in an accretion phase, while the ones with $m < 3.5$ a disk in an ongoing process of dissipation \citep[see, e.g.,][]{haubois12, vieira17}.\par

\subsection{Results}\label{sec_hdust_analysis}

%---------------------------------------%---------------------------------------
\begin{table*}[b]
\caption{\label{table_hdust_ref} Parameters of our best-fit HDUST model in the BeAtlas grid to explain the joint analysis of our interferometric data: VEGA calibrated and differential data and AMBER differential data. A part of these parameter values are presented in the last column of Table \ref{table_hdust_topmodels}. The polar radius and the stellar rotational velocity are obtained from $M_{\star}$ and $R_{\mathrm{eq}}/R_{\mathrm{p}}$. The linear rotational rate is also shown here \citep[$v_{\mathrm{crit}}$ from ][]{fremat05}.}
\centering
\renewcommand{\arraystretch}{1.00}
\begin{adjustbox}{width=0.80\textwidth}
\begin{tabular}{lccccc|ccc}
\toprule
\toprule
$M_{\star}$ ($\mathrm{M}_{\odot}$) & $R_{\mathrm{eq}}/R_{\mathrm{p}}$ & $i$ (deg) & $PA$ (deg) & $\Sigma_{0}$ (g cm\textsuperscript{-2}) & $m$  & $R_{\mathrm{p}}$ ($\mathrm{R_{\odot}}$) & $v_{\mathrm{rot}}$ (km s\textsuperscript{-1}) & $v_{\mathrm{rot}}$/$v_{\mathrm{crit}}$ \\
\midrule 
4.2 &1.45 &63.6 &110 &0.12 &3.0 &3.7 &368 &0.96\\
\bottomrule
\end{tabular}
\end{adjustbox}

\end{table*}
%---------------------------------------%---------------------------------------

We performed four different analyses of our data using different subsets. For that, the reduced $\chi^2$ between the predicted interferometric observables from each HDUST model and the data was calculated as follows:

\begin{enumerate}[label=(\roman*)]
\setlength\itemsep{1em}

\item  calibrated VEGA $V^2$ in the 642.5 nm band (close-by continuum to H$\alpha$).

\item VEGA differential visibility and phase (H$\alpha$ line).

\item AMBER differential visibility and phase (Br$\gamma$ line).

\item All the quantities above analyzed together. 
\end{enumerate}

Analysis (i) was performed to evaluate the constraint on the stellar mass $M_\star$ and oblateness $R_{\mathrm{eq}}/R_{\mathrm{p}}$. In Fig. \ref{hdust_vega_vis2_oblat_mass}, we show the lowest value of  $\chi^2_\mathrm{r}$ for each value of stellar oblateness and mass from fitting the VEGA $V^2$ data in the continuum band. The predicted $V^2$ from our best-fit BeAtlas model (with $M_\star$ = 4.2 $M_{\odot}$; Table \ref{table_hdust_ref}) is overplotted to the VEGA measurements. For comparison, the predicted visibility curve from the BeAtlas model with the highest stellar mass, $M_{\star}$ = 14.6 $\mathrm{M_{\odot}}$, is also overplotted to the data. These two models have the same values of $i$, $R_{\mathrm{eq}}/R_{\mathrm{p}}$, $\Sigma_{0}$, and $m$. In Sect. \ref{sec_geometric_modeling}, we presented a similar analysis, but in terms of simple geometric models. For better visualisation, we show in Fig. \ref{hdust_vega_vis2_oblat_mass} the local regression fits of $\chi^2_\mathrm{r}$ as a function of $R_{\mathrm{eq}}/R_{\mathrm{p}}$ and $M_\star$. Like all such calculations in this paper, all these regression fits of $\chi^2_\mathrm{r}$ are performed with the LOESS method\footnote{As implemented in R: \url{https://stat.ethz.ch/R-manual/R-devel/library/stats/html/loess.html}.}.\par

%---------------------------------------%---------------------------------------
\begin{figure*}
\centerline{\resizebox{0.95\textwidth}{!}{\includegraphics{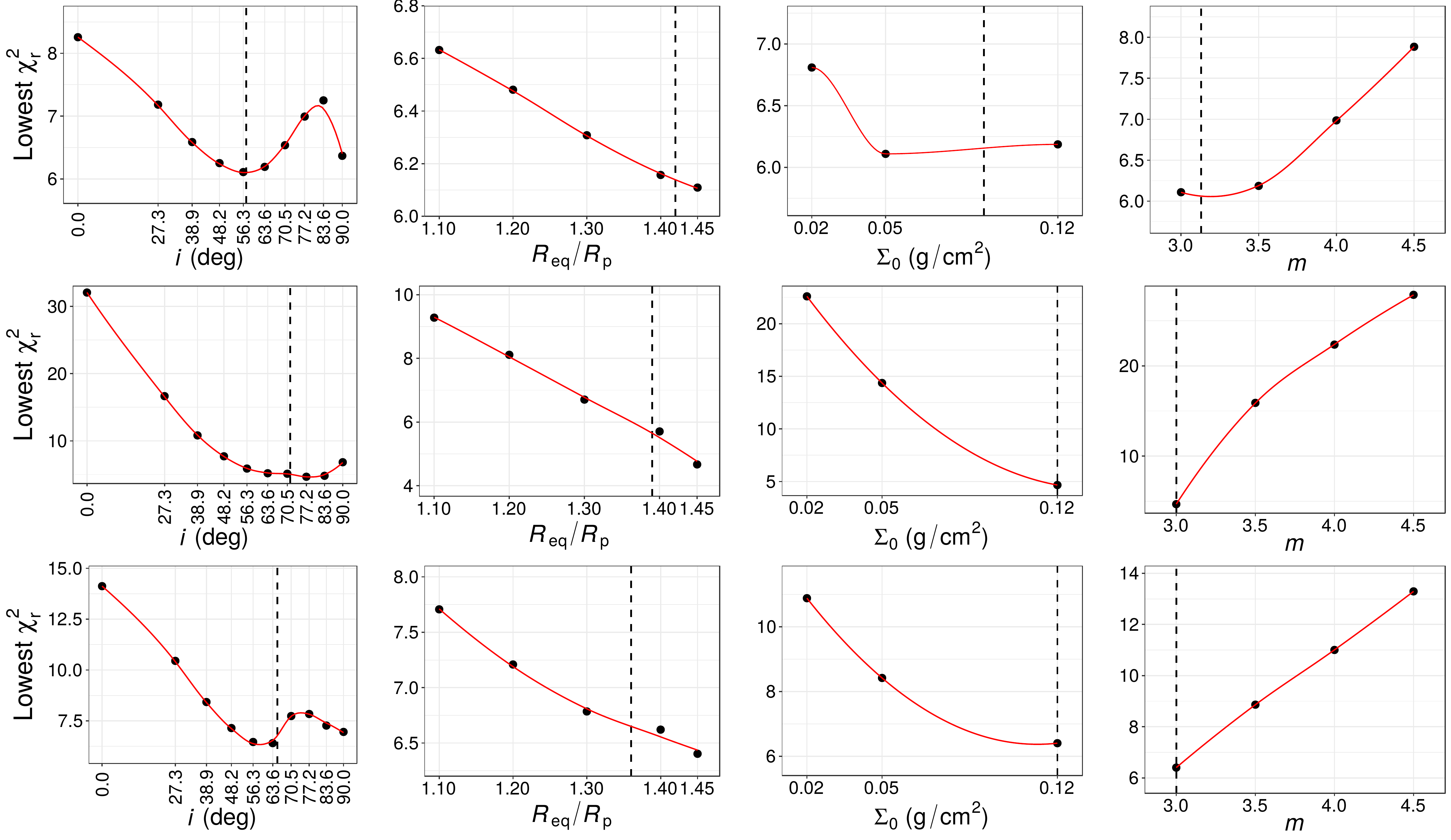}}}
\caption{Lowest value of $\chi^2_\mathrm{r}$ for each value of stellar inclination angle, oblateness, base disk surface density, and disk density law exponent from the HDUST fit to the: VEGA differential data (top, analysis ii), AMBER differential data (middle, analysis iii), and all the interferometric data considered in this section (bottom panel, analysis iv). The stellar mass is fixed to 4.2 $M_{\odot}$ and disk $PA$ to \ang{110}. Local regression fits to $\chi^2_\mathrm{r}$, as a function of the parameter values, are shown as a red line. The mean parameter values for the sets of best models (Table \ref{table_hdust_topmodels}) are marked in dashed black line. Our best-fit BeAtlas model to fit all the interferometric data is shown in Table \ref{table_hdust_ref}. See text for discussion.}
\label{hdust_params_full}
\end{figure*}
%---------------------------------------%---------------------------------------

As in the analysis with geometric models, we cannot constrain the stellar oblateness using VEGA $V^2$ data. On the other hand, the mass is better constrained with $M_\star$ $\sim$ 4.8 $\mathrm{M}_{\odot}$ (B6 dwarf). From Fig. \ref{hdust_vega_vis2_oblat_mass}, one sees how the measured $V^2$ are mismatched by the HDUST model with $M_{\star}$ = 14.6 $\mathrm{M_{\odot}}$ (unrealistic mass value for $\omicron$ Aquarii) due to the larger polar radius of $\sim$7.4 $\mathrm{R_{\odot}}$ in this model. Among all the values for $M_\star$ in the grid, $M_\star$ = 4.2 $\mathrm{M}_{\odot}$ corresponds to a B7 dwarf star \citep{townsend04}. Since $\omicron$ Aquarii shows luminosity class III-IV, it could be expected to have a mass somewhat higher than a dwarf of same spectral type, which is compatible with our results.\par

In Fig. \ref{hdust_vega_amber_diskPA}, we show the lowest $\chi^2_\mathrm{r}$ for each value of disk major-axis position angle $PA$ from the fit to the VEGA and AMBER differential visibilities and phases: analyses (ii) and (iii). Here, the stellar mass is fixed to $M_\star$ = 4.2 $\mathrm{M}_{\odot}$ from analysis (i), which also allows a better comparison to other studies of $\omicron$ Aquarii \citep[e.g.][]{sigut15}. In both cases, $\chi^2_\mathrm{r}$ of the models is minimized around $PA$ = \ang{110}, a value that we adopt in the remaining of this section. This is in good agreement to our results found with the kinematic model in Sect. \ref{sec_kinematic_model_mcmc}.\par

In Fig. \ref{hdust_params_full}, we present our results from modeling the VEGA and AMBER differential visibility and phase in a separate way -- analyses (ii) and (iii) -- as well as from the simultaneous fit to all the interferometric data (analysis iv). The lowest $\chi^2_\mathrm{r}$ is shown as a function of the following HDUST parameters: the inclination angle, stellar oblateness, base disk surface density, and the radial disk density law exponent. In Table \ref{table_hdust_topmodels}, we show the statistics from these parameters calculated from the HDUST models within a certain threshold of $\chi^2_\mathrm{r}$, which, in each case, is chosen to match a similar number of models ($\sim$15-20 best-models). In Table \ref{table_hdust_topmodels}, the parameters for the models with the lowest value of $\chi^2_\mathrm{r}$ are also shown. In Table \ref{table_hdust_ref}, we show the parameters for the best BeAtlas model to explain simultaneously all our different interferometric datasets.\par

Since our HDUST analysis is limited to the pre-computed BeAtlas grid (limited parameter space and selected parameter values), we stress that the results presented here do not correspond to the real $\chi^2$ minimum to explain our datasets in the framework of HDUST. Furthermore, the values for the standard deviation are shown in parenthesis in Table \ref{table_hdust_topmodels} since these are not determinations for the error bars on the parameters. They are just an evaluation for the dispersion on the parameters values of the BeAtlas best-models (within in a certain threshold of $\chi^2_\mathrm{r}$). For example, from fitting AMBER, we found that all the BeAtlas models have $\Sigma_{0}$ = 0.12 g cm\textsuperscript{-2}, and $m= 3.0$, up to, respectively, the top 207\% and top 240\% best-models. For this reason, it is shown, in this case, null standard deviation in Table \ref{table_hdust_topmodels} for these parameters (top 54\% best-models).\par

From the separate analysis of the VEGA and AMBER differential datasets, we are able to describe the stellar and disk parameters, in H$\alpha$ and Br$\gamma$, by the same HDUST model with: $R_{\mathrm{eq}}/R_{\mathrm{p}}$ = 1.45, $\Sigma_{0}$ = 0.12 g cm\textsuperscript{-2}, and $m$ = 3.0. One clear exception is found for the inclination angle. From the H$\alpha$ analysis, $\chi^2_\mathrm{r}$ is minimized for $i$ = \ang{56.3}. On the other hand, this is achieved with $i$ = \ang{77.2} in the Br$\gamma$ line. Such discrepancy of $\sim\ang{20}$ is in agreement with the one found from our kinematic modeling. As expected, the joint analysis to all the data provides an intermediate mean value of $\sim$\ang{65} for the inclination angle, showing a larger dispersion (higher standard deviation) in comparison to the results found from the separate analysis for VEGA and AMBER. One sees that the mean value for stellar oblateness is somewhat decreased, when considering all the datasets. However, in this case, the dispersion is significantly increased ($\pm$0.09) when compared to the separate VEGA and AMBER differential fits ($\pm$0.05-0.07). This happens due to the inclusion of the calibrated VEGA data in the joint analysis that do not allow us to properly infer this parameter (see, again, Fig. \ref{hdust_vega_vis2_oblat_mass}).\par

%---------------------------------------%---------------------------------------
\begin{figure*}
\begin{adjustbox}{minipage=\textwidth,scale=1.00}
\centerline{\resizebox{1.15\textwidth}{!}{\includegraphics{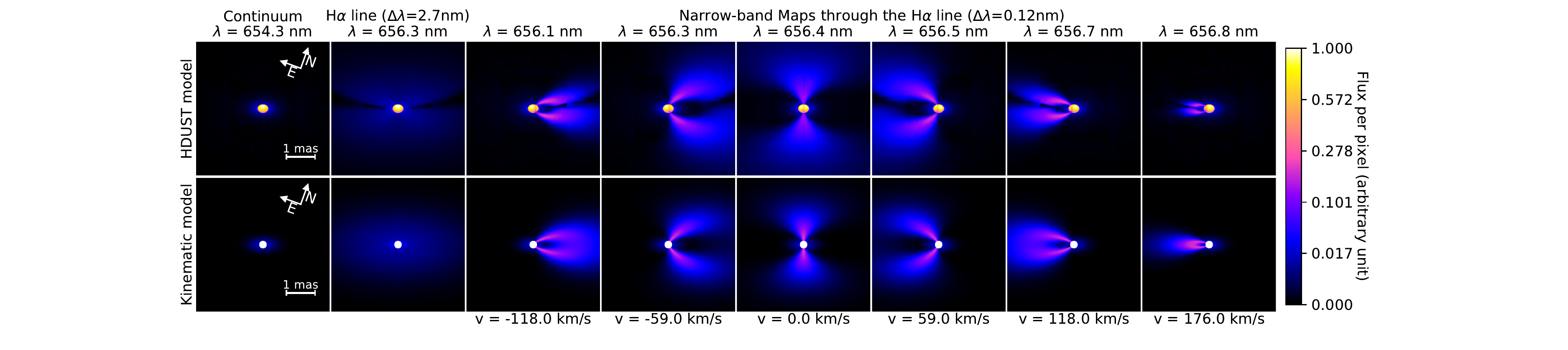}}}
\centerline{\resizebox{1.15\textwidth}{!}{\includegraphics{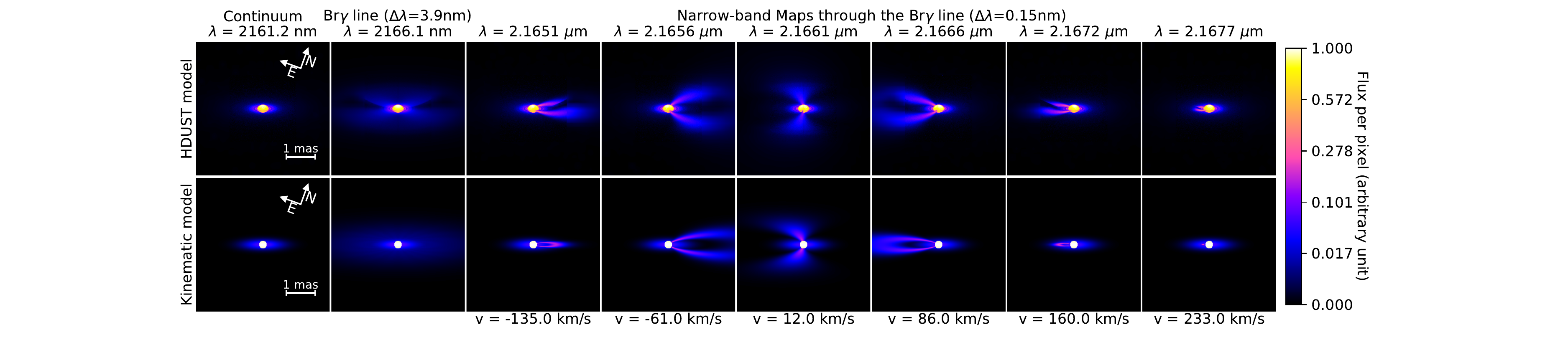}}}
\caption{Intensity maps of our best-fit HDUST and kinematic models at different wavelengths around the H$\alpha$ line (first two rows) and the Br$\gamma$ line (last two rows). Flux/pixel is in arbitrary units with the same scale in H$\alpha$ and Br$\gamma$. The image integrated in wavelength around each of these lines ($\Delta \lambda$ = 2.7 nm around H$\alpha$ and 3.9 nm around Br$\gamma$) are shown in the second column.}
\label{halpha_brgamma_maps}
\end{adjustbox}
\end{figure*}
%---------------------------------------%---------------------------------------

\section{Comparison between kinematic and HDUST best-fit models}\label{sec_comparison_hdust_kinematic}

In Fig. \ref{bestfit_hdust_kinematic_model_vega_amber}, we compare the synthetic differential visibility and phase from our best-fit kinematic and HDUST models to the actual VEGA and AMBER data for a few baselines.
Comparisons to non-interferometric observables (spectral energy distribution and line profiles) are presented in Sect.~\ref{sec_spectrum_sed}. Our best-fit models are compared to all the AMBER data in Fig. \ref{models_all_amber} (Appendix \ref{appendix_bestfits_all_data}). One sees that our best-fit kinematic models do a better job of reproducing both the VEGA and AMBER data. From the separate kinematic modeling of the VEGA and AMBER differential data, the $\chi^2_\mathrm{r}$ of the model is lower than with HDUST (BeAtlas grid). Fixing the stellar mass to a reliable value for $\omicron$ Aquarii (4.2 $\mathrm{M}_{\odot}$), our best-fit HDUST model has $\chi^2_\mathrm{r}$ $\sim$ 6.1 and 4.7 for VEGA and AMBER, respectively. From the kinematic modeling, we found $\chi^2_\mathrm{r}$ $\sim$ 4.0 and 1.6 to explain these same datasets.\par

For VEGA, in particular, our best-fit HDUST model adjustment for the measured visibility width is worse than with the kinematic model. This particular issue in modeling the VEGA data can be explained; in HDUST, the disk velocity law exponent is fixed by $\beta$ = -0.5 (Keplerian disk rotation), while in the kinematic model it is a free parameter. As shown in Sect. \ref{sec_kinematic_models_results}, we find values for $\beta$ that are higher than -0.5, and this is accentuated from the analysis of the VEGA data ($\beta$ $\sim$ -0.3).\par

Apart from this issue regarding the analysis in H$\alpha$, we are able to describe well the disk density with the same physical parameters in both the H$\alpha$ and Br$\gamma$ lines: $\Sigma_0$ = 0.12 g cm\textsuperscript{-2} and $m$ = 3.0. As will be later discussed, this result found using HDUST is consistent with the ones presented in Sect. \ref{sec_kinematic_models_results}, showing a similar disk extension in these lines.\par

In Fig. \ref{halpha_brgamma_maps}, the intensity maps for each model are shown at the close-by continuum region and at different wavelength values in both the H$\alpha$ and Br$\gamma$ emission lines. The integrated intensity map (around each of these lines) is also presented. For a more realistic comparison, here we consider our best-fit kinematic model with a small flux contribution of 5\% from the disk in the continuum nearby to H$\alpha$ and $a_{c}$ = 2 $\mathrm{D_{\star}}$. As shown in Table \ref{sec_kinematic_models_results}, these parameters were adopted as null in the kinematic analysis for the VEGA data, since we were not able to resolve the disk from our analysis of VEGA $V^2$ measurements in the continuum band (Sect. \ref{sec_geometric_modeling}). Regarding the continuum region close to Br$\gamma$, the disk extension and flux contribution are given in Table \ref{table_mcmc_vega_amber} for the AMBER analysis.\par

The major difference between the intensity maps in H$\alpha$ and Br$\gamma$ is the disk flattening which is due to the different inclination angle derived from these two regions, $i$ $\sim$ \ang{57} (H$\alpha$) and $\sim$\ang{72} (Br$\gamma$), from the best models provided in Table \ref{table_hdust_topmodels}. Moreover, as seen in the images, the stellar flattening is taken into account in the HDUST modeling, but not in the kinematic model (the star is modeled as a uniform disk). Apart from these departures, we see that our best-fit HDUST model presents a fairly similar distribution to the one computed with the kinematic code: a Gaussian distribution represents the circumstellar disk. This can be better noted considering the full integrated images around the emission lines.\par

%%%%%%%%%%%%%%%%%%%%%%%%%%%%%%%%%%%%%%%%%%%%%%%%%%%%%%%%%%%%%%%%%%%%%%%%%%%%%%%%%%%%%section: sed and spectra
\section{Comparison to non-interferometric observables}\label{sec_spectrum_sed}
%---------------------------------------%---------------------------------------
\begin{figure}
\centerline{\resizebox{0.45\textwidth}{!}{\includegraphics[angle=0]{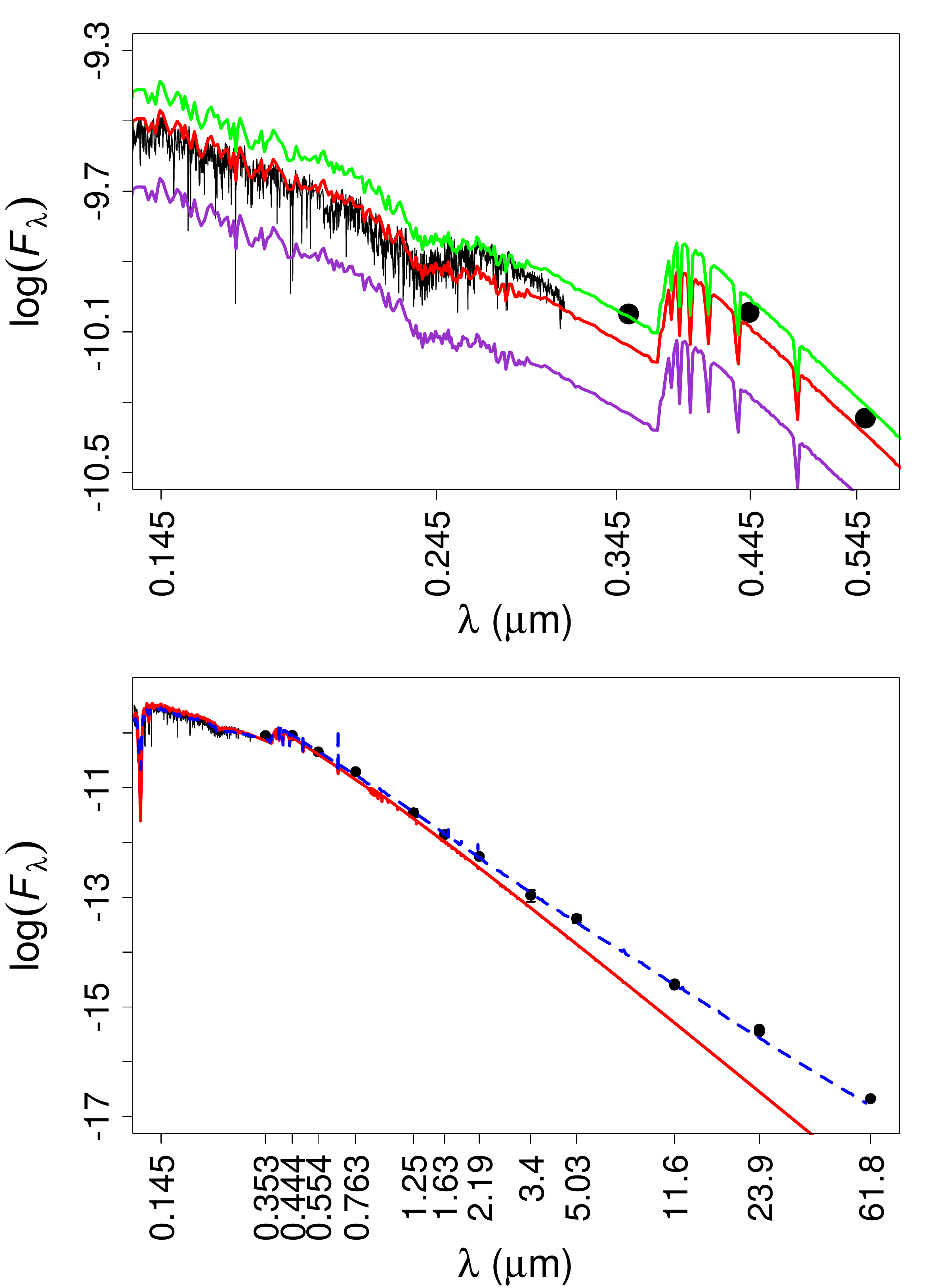}}}
\caption{Comparison between the observed $\omicron$ Aquarii and model SEDs from the ultraviolet to the far-infrared region. Flux unit is in erg cm\textsuperscript{-2} s\textsuperscript{-1} {\AA}\textsuperscript{-1} and wavelength is shown in logarithmic scale. IUE/SWP and IUE/LWP spectra are shown in black line and photometric data in black points. Top panel: purely photospheric models (color lines) with variation in the stellar radius (no inclusion of geometrical oblateness): $R_{\star}$ = 3.2 $\mathrm{R_{\odot}}$ (orchid), 4.0 $\mathrm{R_{\odot}}$ (red), and 4.4 $\mathrm{R_{\odot}}$ (green). Bottom panel: photospheric model with 4.0 $\mathrm{R_{\odot}}$ (red) and our best-fit HDUST model from fitting all the interferometric data (dashed blue line; Table \ref{table_hdust_ref}). Note that the UBV-bands are better reproduced with $R_{\star}$ = 4.0-4.4 $\mathrm{R_{\odot}}$. Our best HDUST model reproduces the observed IR excess due to the circumstellar disk well.}
\label{sed_kurucz_hdust}
\end{figure}
%---------------------------------------%---------------------------------------

In this section, we compare our best-fit models, found from the analysis of interferometric observables, to the observed spectral energy distribution (SED) and line profiles (H$\alpha$ and Br$\gamma$) of $\omicron$ Aquarii. With respect to polarimetric data, it is discussed in Sect. \ref{sec_variability_polarimetry} when addressing the disk stability.\par

\subsection{Spectral Energy Distribution}\label{sec_sed}

In Fig. \ref{sed_kurucz_hdust}, we present the spectral energy distribution (SED) of $\omicron$ Aquarii from the ultraviolet (IUE/SWP and IUE/LWP spectra\footnote{Public data available in the Barbara A. Mikulski Archive for Space Telescopes (MAST): \url{https://archive.stsci.edu/iue/}.}) to the far-infrared region. References for the photometric data are given as follows: UBVJHK-bands \citep{anderson12}, i-band \citep{henden16}, LM-bands \citep{bourges17}, and IRAS 12, 25, and 60 $\mu$m bands \citep{abrahamyan15}.\par

%---------------------------------------%---------------------------------------
\begin{figure}[t]
\centerline{\resizebox{0.50\textwidth}{!}{\includegraphics{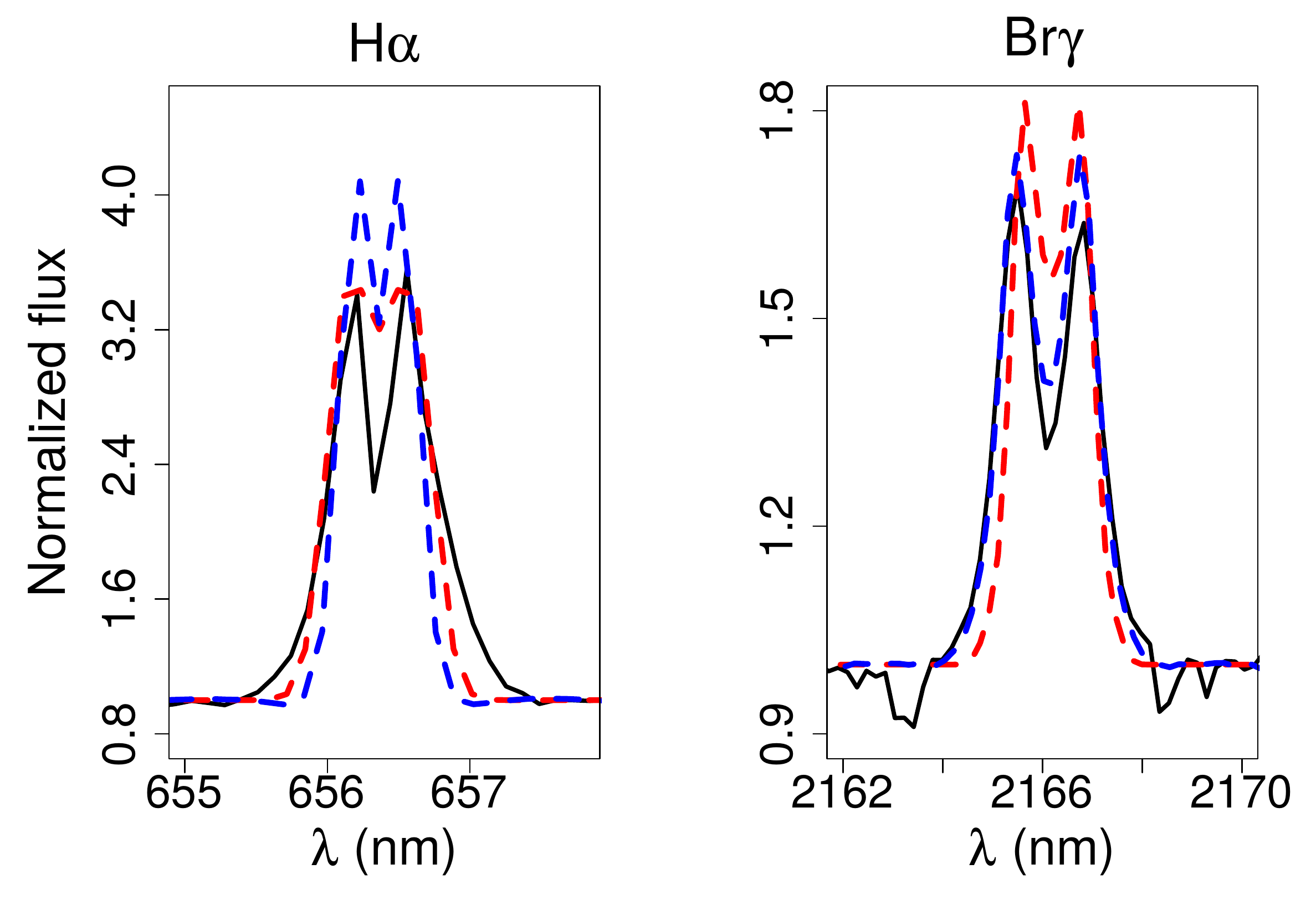}}}
\caption{Comparison between our best-fit kinematic models (dashed red; Table \ref{table_mcmc_vega_amber}) and HDUST model (dashed blue, Table \ref{table_hdust_ref}) in the H$\alpha$ and Br$\gamma$ line profiles. Mean observed line profiles of H$\alpha$ (BeSOS) and Br$\gamma$ (AMBER) are shown in black line. Our best-fit kinematic and HDUST models provide reasonable synthetic profiles to the observed ones in both H$\alpha$ and B$\gamma$ }
\label{hdust_kinematic_model_profiles}
\end{figure}
%---------------------------------------%---------------------------------------

For the spectral region up to the V-band, we compare the data to the SEDs of purely photospheric atmosphere models with solar metallicity \citep{castelli04}. In this region, the circumstellar disk flux level is much lower than the photospheric flux, thus allowing a proper probe of the stellar radius \citep[e.g.,][]{meilland09}. The surface gravity was fixed at $\log g$ = 4.0, this being the closest value in \citet{castelli04} to $\log g$ = 3.9 that is given by our results of $M_{\star}$ = 4.2 $\mathrm{M_{\odot}}$ and $R_{\star}$ = 4.0 $\mathrm{R_{\odot}}$. The effective temperature was fixed at 13000 K, following \citet{cochetti19}. As in the previous sections, we consider the distance to be 144 pc, from the Gaia DR2 parallax.

These synthetic SEDs were calculated for three different stellar radius values, $R_{\star}$: 3.2 $\mathrm{R_{\odot}}$ \citep{sigut15}, 4.0 $\mathrm{R_{\odot}}$, and 4.4 $\mathrm{R_{\odot}}$ \citep{cochetti19}. The value of 4.0 $\mathrm{R_{\odot}}$ corresponds to the stellar radius determined from the fit to the VEGA $V^2$ data using a two-component model: 4.0 $\pm$ 0.3 $\mathrm{R_{\odot}}$. The effect of interstellar medium extinction is not included in these models since it is negligible for $\omicron$ Aquarii. Assuming a total to selective extinction ratio of $R_{V}$ = 3.1, \citet{touhami13} derived a color excess of $E(B-V)$ = 0.015 $\pm$ 0.008 for this star from their fit to the SED. This means the observed flux is $\sim$96\% of the intrinsic one in the V-band (lower by $\sim$0.02 dex). It is beyond the scope of this paper to estimate the extinction due to the circumstellar disk, however, from the comparison to purely photospheric models, we see in Fig. \ref{sed_kurucz_hdust} that the effect of extinction (due to the interstellar and circumstellar matter) is conspicuously weak on the 0.220 $\mu$m bump.\par

From Fig. \ref{sed_kurucz_hdust}, we see that the UV and visible regions are better reproduced for a stellar radius of about 4.0-4.4 $\mathrm{R_{\odot}}$, when compared to 3.2 $\mathrm{R_{\odot}}$, adopted in \citet{sigut15}, which corresponds to the expected polar radius for a B7 dwarf. We stress that the radius derived by \citet{cochetti19} is closer to our results from the fit to the VEGA $V^2$ data (Sect. \ref{sec_geometric_modeling}). Their result of $R_{\star}$ = 4.4 $\mathrm{R_{\odot}}$ corresponds to a uniform disk diameter of $\theta$ $\sim$ 0.28 mas ($d$ = 144 pc). A better comparison to \citet{cochetti19} is hard since they do not provide error bars on $R_{\star}$ from fitting the SED. Furthermore, they derived $R_{\star}$ = 4.4 $\mathrm{R_{\odot}}$ for $\omicron$ Aquarii using a distance of 134 pc from \citet{vanleeuwen07}, rather than the value of 144 pc adopted here. From Fig. \ref{sed_kurucz_hdust}, this implies a larger discrepancy between the observed and synthetic SED for $R_{\star}$ = 4.4 $\mathrm{R_{\odot}}$, overestimating the observed flux.\par

We also compare the predicted SED of our best-fit HDUST model (Table \ref{table_hdust_ref}) to the SED of the purely photospheric model with 4.0 $\mathrm{R_{\odot}}$. Despite being able to reproduce the UBV-bands well, one sees that a purely photospheric model clearly underestimates the observed flux beyond the near-infrared due to the flux contribution from the circumstellar disk \citep[e.g.,][]{poeckert78, waters86}. From Fig. \ref{sed_kurucz_hdust}, it is evident that the SED is much better reproduced up to the far-infrared region when taking into account the IR excess from the gaseous circumstellar disk present in our best-fit HDUST model.\par

%%%%%%%%%%%%%%%%%%%%%%%%%%%%%%%%%%%%%%%%%

\subsection{H$\alpha$ and Br$\gamma$ profiles}\label{sec_spectrum}

Our H$\alpha$ spectra taken with the VEGA instrument (20 spectra, period from 2012 to 2016) are not analyzed in this work since they are saturated. This is a known effect seen in previous works on Be stars and correlated to the magnitude of the object. We stress that this instrumental saturation effect does not impact the visibilities and phases extracted from the fringes measured with VEGA  \citep[see, e.g.,][]{delaa11}. To overcome this problem we used H$\alpha$ line profiles from the BeSOS\footnote{Be Stars Observation Survey.} catalog \citep[][]{arcos18, vanzi12}, obtained between 2012 and 2015, and thus covering a similar period to our VEGA observations. The typical spectral resolution of the BeSOS spectra is $\sim$0.1 {\AA}. \par

In Fig. \ref{hdust_kinematic_model_profiles}, we compare the H$\alpha$ and Br$\gamma$ profiles from our best-fit models to observed profiles, namely, the mean H$\alpha$ line profiles from BeSOS (7 profiles\footnote{Public data available at: http://besos.ifa.uv.cl}) and the mean Br$\gamma$ line profiles from our AMBER observations (8 profiles). The observed profiles in Fig. \ref{hdust_kinematic_model_profiles} were binned in wavelength in order to have a spectral resolution equal to one of the synthetic profiles from the kinematic and HDUST models: 1.3 {\AA} (H$\alpha$) and 1.8 {\AA} (Br$\gamma$). The mean $EW$ in H$\alpha$ from the BeSOS data is 19.1 {\AA}. This is in agreement with the mean value of 19.9 {\AA} found in \citet{sigut15}, based on contemporaneous spectra, and adopted in our analysis with the kinematic code (Sect. \ref{sec_kinematic_models_results}).\par

First, we note that our best-fit kinematic and HDUST models provide a fairly reasonable match to the observed H$\alpha$ and Br$\gamma$ line profiles. The kinematic models correspond to our best-fits obtained from modeling the VEGA and AMBER differential data separately (Sect. \ref{sec_kinematic_model_mcmc}). On the other hand, our best-fit HDUST model shown in H$\alpha$ and Br$\gamma$ is derived from the simultaneous fit to all our interferometric data (Table \ref{table_hdust_ref}). Moreover, we stress the difficulty found by \citet{sigut15}, using the radiative transfer code BEDISK, to reproduce the line wings and central absorption in the H$\alpha$ profile of $\omicron$ Aquarii (see their Fig. 5).\par

However, it can be seen in Fig. \ref{hdust_kinematic_model_profiles} that both our best-fit kinematic and HDUST model are not able to properly reproduce, in particular, the wings of the H$\alpha$ profile. On the other hand, the wings of the Br$\gamma$ profile are fairly well reproduced by both of them, especially with HDUST.\par

Therefore, this inability to reproduce the wings of the H$\alpha$ profile well is likely due to physical processes in the disk that are not taken into account in our models. It is known that the H$\alpha$ profile wings of Be stars can be highly affected by non-coherent scattering, thus resulting in non-kinematic line-broadening in this transition \citep[see, e.g.,][]{hummel92, delaa11}. It is beyond the scope of this paper to quantify this possible effect in the H$\alpha$ line of $\omicron$ Aquarii.\par

%%%%%%%%%%%%%%%%%%%%%%%%%%%%%%%%%%%%%%%%%%%%%%%%%%%%%%%%%%%%%%%%
%%%%%%%%%%%%%%%%%%%%%%%%%%%%%%%%%%%%%%%%%%%%%%%%%%%%%%%%%%%%%%%%
%%%%%%%%%%%%%%%%%%%%%%%%%%%%%%%%%%%%%%%%%%%%%%%%%%%%%%%%%%%%%%%%section: discussion
\section{Discussion}\label{sec_discussion}

\subsection{Disk extension in H$\alpha$ and Br$\gamma$}\label{sec_disk_extension}

In Sect. \ref{sec_kinematic_models_results}, we showed that the disk extension is similar in the H$\alpha$ and Br$\gamma$ lines. Interestingly, from previous studies, we could expect to find a larger disk extension in H$\alpha$ than Br$\gamma$. For example, \citet{meilland11} found that $\delta$ Scorpii (B0.3IV), which was also observed with the VEGA and AMBER instruments, shows a circumstellar disk 1.65 times larger in H$\alpha$ than in Br$\gamma$. Furthermore, \citet{gies07} derived the angular sizes of four Be stars ($\gamma$ Cassiopeiae, $\phi$ Persei, $\zeta$ Tauri, and $\kappa$ Draconis) in the K-band region using interferometric data from the CHARA/CLASSIC instrument. They showed that the disk of these stars was significantly larger (up to $\sim$1.5-2.0 times) in the H$\alpha$ line than in the K-band. However, \citet{carciofi11} investigated theoretically, using the code HDUST, the formation loci of H$\alpha$ and Br$\gamma$, and found them to be quite similar at least in the parameter space explored by the authors (see their Fig. 1). Moreover, \citet{stee01}, using the code SIMECA, found that Be star disks can be larger (up to two times) in Br$\gamma$ than in H$\alpha$.\par

%---------------------------------------%---------------------------------------
\begin{figure}[t]
\centerline{\resizebox{0.475\textwidth}{!}{\includegraphics{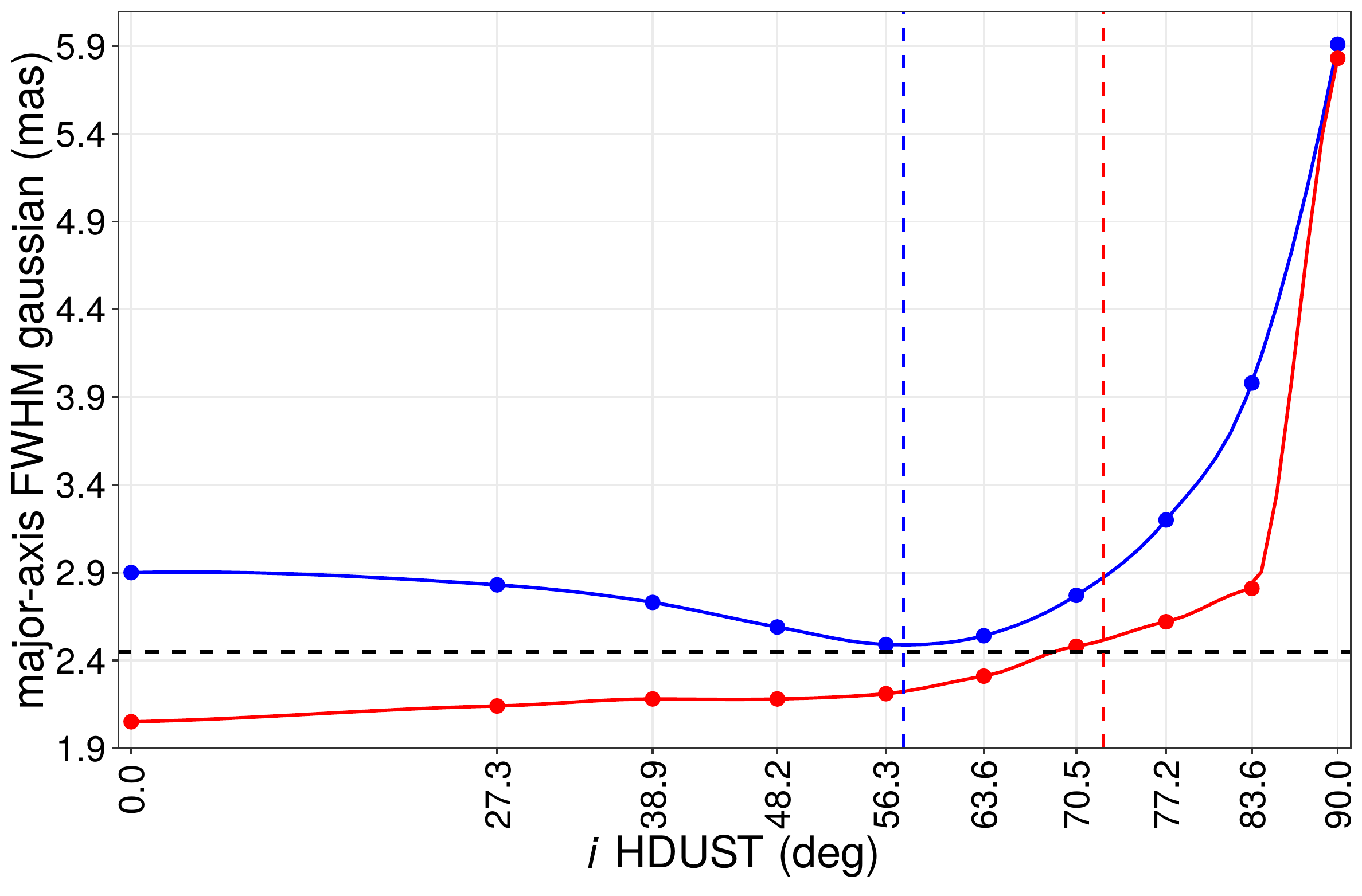}}}
\caption{Major-axis FWHM of Gaussian distribution (fitted from our best-fit HDUST model) as a function of the HDUST inclination angle. All the other HDUST parameters are fixed. Blue points correspond to the fit in H$\alpha$ and red points in Br$\gamma$. The vertical dashed lines mark our values for inclination angle derived from the HDUST analysis, fitting the data in H$\alpha$ (blue) and Br$\gamma$ (red). Note that the equivalent Gaussian fits show a similar extension (2.45 mas, marked in horizontal dashed line) for these values of $i$.}
\label{hdust_gaussian_fwhm}
\end{figure}
%---------------------------------------%---------------------------------------

For a quantitative comparison of the disk extension in H$\alpha$ and Br$\gamma$, we fitted simple Gaussian distributions to the intensity map of our best-fit HDUST model for all the values of inclination angle in BeAtlas. In order to remove the contribution from the star and disk continuum, we removed the image from the continuum before performing the fit and we hide the central part of the image which is affected by the stellar contribution.\par

In Fig. \ref{hdust_gaussian_fwhm}, we show the major-axis FWHM from our fit as a function of the inclination angle for the H$\alpha$ and Br$\gamma$ lines. First, one sees that the disk size-extension (major-axis FWHM) varies differently in the H$\alpha$ and Br$\gamma$ lines as a function of the inclination angle. The disk extension increases in Br$\gamma$ with the inclination angle. On the other hand, it decreases significantly in H$\alpha$ up to $i \sim$ \ang{56} and increases after this value. One sees that the ratio between the extension in these lines decreases from about 1.50 at zero inclination to about 1.05 at \ang{63.5}. Furthermore, we note that the disk extensions in these lines are very close to each other for $i \sim$ \ang{56} (H$\alpha$) and $i \sim$ \ang{72} (Br$\gamma$): major-axis FWHM $\sim$ 2.45 mas. Considering $d$ = 144 pc, the disk size is $\sim$10 $\mathrm{D_\star}$ (close to our findings from the kinematic modeling).\par

Therefore, from this simple analysis using HDUST models, we verify our findings using the kinematic model: a similar circumstellar disk extension in H$\alpha$ and Br$\gamma$. This arises since the (equivalent) Gaussian disk to our best-fit HDUST model presents quite different changes on its extension in these lines as a function of the inclination angle. Based on that, we can also explain the difference between $\delta$ Scorpii and $\omicron$ Aquarii. The former is seen under a low inclination angle ($\sim$\ang{30}) and exhibits a high ratio between the H$\alpha$ and Br$\gamma$ disk sizes. The latter is seen under a higher inclination angle and shows similar disk sizes in both lines. On the other hand, as discussed above, $\phi$ Persei and $\zeta$ Tau show larger disks in H$\alpha$ than in the K-band and these stars are seen close to edge-on with $i$ = \ang{78} \citep{mourard15} and \ang{85} \citep{carciofi09}, respectively. Thus, this similarity in the disk extensions, found for quite different values of inclination angle, could indicate a more complex physical structure of the circumstellar disk than the one assumed by our best-fit HDUST model (based on a vertically isothermal disk).\par

%%%%%%%%%%%%%%%%%%%%%%%%%%%%%%%%%%%%%%%%%%%%%%%%%%%%%%%%%%%%%%%%%%
\subsection{Inclination angle and vertical disk structure}\label{sec_incl_non_isothermal}

From our H$\alpha$ and Br$\gamma$ differential data analysis, using the kinematic model, we achieved good precision in the determination of the stellar inclination angle: $i$ $\sim$ 61.2 $\pm$ \ang{1.8} (VEGA) and $i$ = 75.9 $\pm$ \ang{0.4} (AMBER). Nevertheless, there is a clear discrepancy between the inclination angle found from fitting the VEGA and AMBER datasets. The value determined from VEGA is about \ang{15} lower than the one found in the analysis of the AMBER data. We can show that this issue does not stem from an intrinsic limitation of the kinematic code (2-D model) for Be stars seen under high inclination angle ($i$ $\gtrsim$ \ang{60}). Indeed, by using a sophisticated 3-D radiative transfer model (HDUST), not subjected to such a limitation, we verified the same discrepancy on $i$ from the fit to these datasets separately (see, again, in Fig. \ref{hdust_params_full}, the trend of $\chi^2_{\mathrm{r}}$, as a function of $i$).\par

It may be argued that the difference found in inclination angle is hiding a difference in the disk thickness in these lines. Assuming a non-geometrically thin disk, for an ellipse with major and minor axes denoted, respectively, by $a$ and $b$, the ratio between $a$ and $b$, the circumstellar disk flattening, is given by \citep[see, e.g.,][]{meilland07a}:

\begin{equation}
\frac{a}{b} = \frac{1}{\cos i + 2\sin \frac{\Theta}{4} \sin \left(i - \frac{\Theta}{4}\right) }\,,
\label{eq:incl_opening_angle}
\end{equation}
where $i$ is the stellar inclination angle and $\Theta$ the disk opening angle. Since the $i$ derived from H$\alpha$ using our physical models is much lower than from Br$\gamma$ (a reliable value when compared to other results in the literature), this would imply a disk thicker (higher opening angle $\Theta$) in H$\alpha$ than in Br$\gamma$. Considering the values described above, the disk opening angle in H$\alpha$ would be $\Theta$ $\sim$ \ang{37} larger in H$\alpha$ than in Br$\gamma$ (assuming a geometrically thin disk in Br$\gamma$). Such a high value of opening angle is far beyond what is measured and expected by the VDD model, typically less than $\sim$\ang{10} \citep[cf.][]{rivinius13}. This might indicate the necessity of more complex physical assumptions in the physical properties of our disk model. \par

Since the code HDUST provides a pure hydrogen modeling for the photosphere plus disk regions, this disagreement between the VEGA (H$\alpha$) and AMBER (Br$\gamma$) analyses in the determination of $i$ could be due to an opacity effect. It is well-known that the inclusion of heavy elements can impact the density and temperature stratifications in the circumstellar disk of Be stars by shielding emission from the star. \citep[see, e.g.,][]{sigut07}. Furthermore, we stress that our best-fit HDUST model is a parametric model (based on a vertically isothermal structure). Departures from vertically isothermal disks are well-known in the literature. For example, using the radiative transfer code BEDISK, \citet{sigut09} verified that isothermal and self-consistent hydrostatic models can present large differences regarding the temperature stratification in the disk of Be stars. Using HDUST, \citet{carciofi08} also found that non-isothermal effects can be significant for denser Be star disks. Thus, further investigation is needed concerning this effect on the determination of $i$ for $\omicron$ Aquarii, but that is beyond the scope of this paper.\par

Finally, another possibility to explain the difference in apparent inclination angle found in our modeling could be a non-negligible contribution of a polar wind. Clues of the presence of polar wind, or at least of circumstellar material in the polar regions, have been found by \citet{domiciano06} and \citet{meilland07b}. In our models, we assume that all the circumstellar material is in the thin equatorial disk. If a non negligible fraction of the material is located near the poles, although we would expect it to be quite diluted and optically thin (at least in the continuum), it might affect the line emission with a different magnitude in H$\alpha$ and in Br$\gamma$. If one assumes that the hydrogen level populations favor H$\alpha$ emission over Br$\gamma$, the polar contribution of H$\alpha$ would be higher, and the environment might look less flattened in this line than in Br$\gamma$.

%%%%%%%%%%%%%%%%%%%%%%%%%%%%%%%%%%%%%%%%%%%%%%%%%%%%%%%%%%%%%%%%%%
\subsection{Stellar and disk rotation}\label{sec_rotation}

In Sect. \ref{sec_hdust_modeling}, our results are presented in terms of the stellar oblateness $R_{\mathrm{eq}}$/$R_{\mathrm{p}}$ (denoted by $f$ in Eq. \ref{omega_omegacrit_oblateness}). First, we give the relation between the oblateness and the angular $\Omega/\Omega_{\mathrm{crit}}$ and linear $v_\mathrm{{rot}}/v_{\mathrm{crit}}$ rotational rates as follows:

\begin{equation}
\frac{\Omega}{\Omega_{\mathrm{crit}}} = \frac{v_{\mathrm{rot}}}{v_{\mathrm{crit}}} \, \frac{R_{\mathrm{eq,crit}}}{R_\mathrm{eq}} = \left(\frac{3}{2}\right)^{3/2} \, \left[\frac{2(f-1)}{f^3}\right]^{1/2},
\label{omega_omegacrit_oblateness}
\end{equation}
where $R_{\mathrm{eq,crit}}$ and $R_\mathrm{eq}$ (in units of polar radius) are, respectively, the stellar equatorial radius in the case of critical velocity and the actual one \citep[see, e.g.,][]{fremat05, ekstrom08}.\par

%---------------------------------------%---------------------------------------
\begin{figure}
\centerline{\resizebox{0.50\textwidth}{!}{\includegraphics{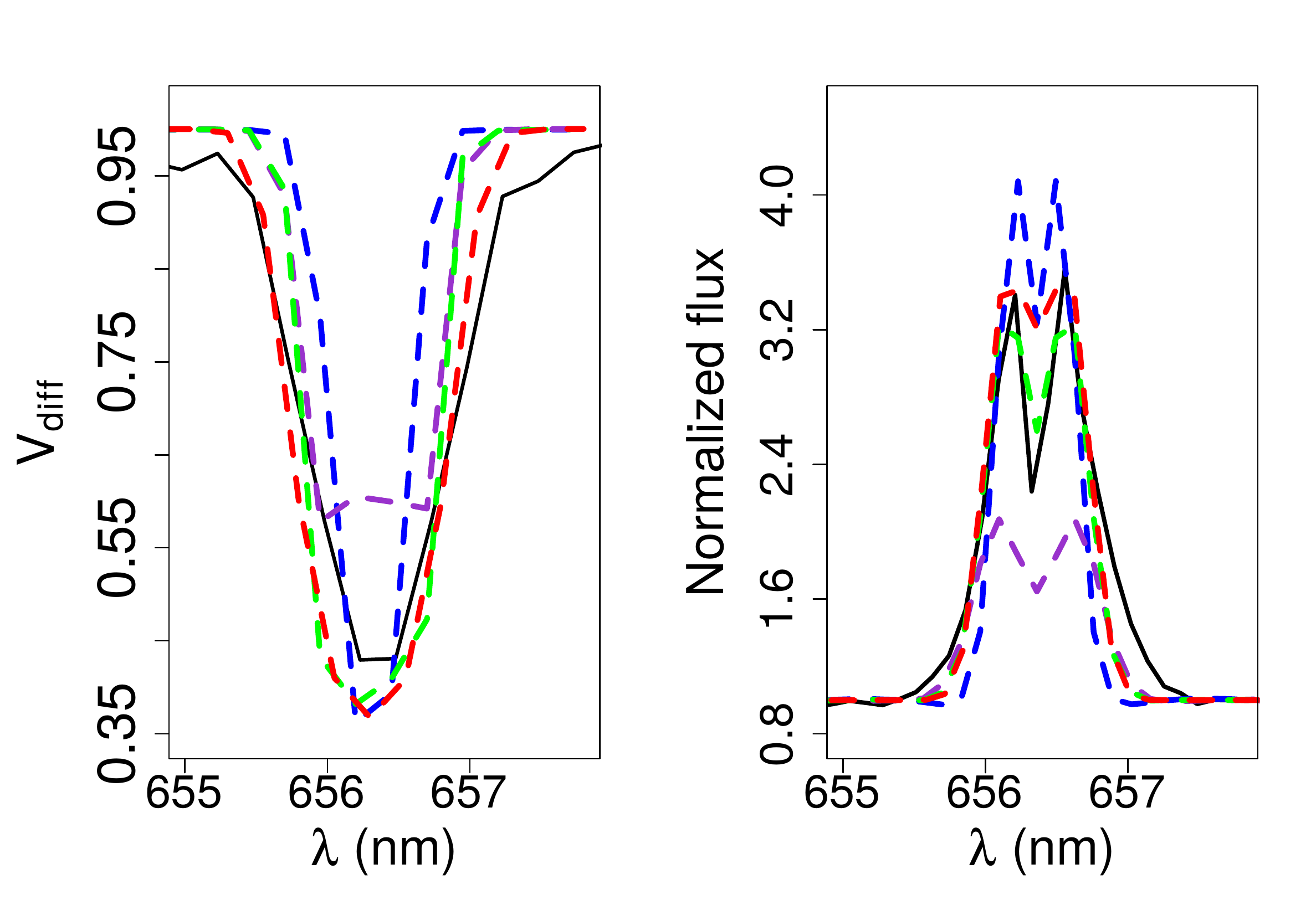}}}
\caption{Bias effect of the disk velocity on H$\alpha$ modeling. One VEGA measurement and observed H$\alpha$ profile (BeSOS, as in Fig. \ref{hdust_kinematic_model_profiles}) are shown in black lines. Our best-fit kinematic (dashed red) and HDUST (dashed blue) models are shown in H$\alpha$ visibility and line profile. They are compared to HDUST models with a higher mass of 10.8 $\mathrm{M}_{\odot}$ with: $\Sigma_{0}$ = 0.12 g cm\textsuperscript{-2} (dashed orchid) and $\Sigma_{0}$ = 0.28 g cm\textsuperscript{-2} (dashed green). See text for discussion.}
\label{hdust_kinematic_vis_profiles}
\end{figure}
%---------------------------------------%---------------------------------------

Considering only the uncertainties on $v_{\mathrm{rot}}$ (Table \ref{table_mcmc_vega_amber}), with the critical velocity $v_{\mathrm{crit}}$ fixed to 391 km s\textsuperscript{-1} \citep{fremat05}, we obtain a linear rotational rate of $v_{\mathrm{rot}}$/$v_{\mathrm{crit}}$ = 0.83 $\pm$ 0.02 ($v_{\mathrm{rot}}$ = 325 $\pm$ 6 km s\textsuperscript{-1}, VEGA) and 0.775 $\pm$ 0.005 ($v_{\mathrm{rot}}$ = 303 $\pm$ 2 km s\textsuperscript{-1}, AMBER). From the HDUST analysis, we find $v_{\mathrm{rot}}$/$v_{\mathrm{crit}}$ = 0.96 (VEGA and AMBER) from our best-fit model (no error bars). This difference between the kinematic and HDUST analysis can be explained since the $\beta$ exponent (velocity law in the disk) is fixed in the HDUST analysis (Keplerian disk, $\beta$ = -0.5), while it is a free parameter in the kinematic model. We derived values for $\beta$ from the kinematic analysis that are significantly higher (more positive) than -0.5 (see Table \ref{table_mcmc_vega_amber}).\par 

Apart from these differences, our analysis is consistent with a high rotational rate for $\omicron$ Aquarii, showing $v_{\mathrm{rot}}$/$v_{\mathrm{crit}}$ from $\sim$0.8 up to 1.0, depending on the particular analysis considered. The BeAtlas fits to the VEGA and AMBER differential data are significantly worsened (Fig. \ref{hdust_params_full}), when considering $R_{\mathrm{eq}}$/$R_{\mathrm{p}}$ = 1.20-1.30 ($\Omega/\Omega_{\mathrm{crit}}$ = 0.88-0.96). Thus, our HDUST analysis indicates that $\omicron$ Aquarii rotates faster than $\Omega/\Omega_{\mathrm{crit}}$ = 0.96, disfavouring the lower range of $\Omega/\Omega_{\mathrm{crit}}$ between 0.86 and 0.93 that is derived by \citet{cochetti19}.\par

In Sect. \ref{sec_kinematic_models_results}, we found a strong correlation between the velocity at the base of the disk and the $\beta$ exponent of the rotation law $\beta$. The inferred degeneracy, stronger in the case of the VEGA data, which have a lower spectral-resolution with higher uncertainties, prevents us from independently constraining these two parameters with our kinematic model when fitting only our spectro-interferometric data. However, the addition of an external constraint, the measured $v \sin i$, removed this degeneracy, allowing us to derive a more accurate value of $\beta$ in comparison to the other MCMC fitting tests (Appendix \ref{appendix_mcmc}). 

From our MCMC fit to the AMBER dataset, with a preset $v \sin i$, we derived a $\beta$ of -0.426 $\pm$ 0.003. Thus, the disk appears to be rotating in a nearly Keplerian fashion. Despite this very low error on $\beta$, note that the error bars on $\beta$ change with respect to the presented MCMC tests, up to $\pm$ 0.008 (see Fig. \ref{mcmc_other_tests_amber}). On the other hand, the value derived from the fit of the VEGA data is about 0.1 higher than from AMBER. We stress that this discrepancy cannot be explained by a radial dependent rotational law because both lines roughly stem from the same region in the disk (similar disk extensions in these lines).\par

Moreover, this apparent higher value of $\beta$ in H$\alpha$ was also the origin of some biases that we found when modeling the VEGA differential data alone using HDUST. Without fixing $M_\star$, the VEGA analysis with HDUST favours unrealistically high values of stellar mass up to $\sim$11 $\mathrm{M_{\odot}}$. This happened due to the fact the higher mass models also correspond to higher rotational velocity at the base of the disk. As the value of $\beta$ is fixed to -0.5 in the BeAtlas grid of models, this was the only way to increase the rotational velocity in the disk. In Fig. \ref{hdust_kinematic_vis_profiles}, our best-fit kinematic and HDUST models are compared to a higher mass HDUST model for one VEGA differential measurement and the H$\alpha$ profile. When compared to our best HDUST model (mass fixed to 4.2 $\mathrm{M_{\odot}}$), the visibility drop and the H$\alpha$ profile are better reproduced with HDUST models with higher value of mass, but also considering a larger value of $\Sigma_{0}$ (0.28 g cm\textsuperscript{-2}). This happens since the stellar radius is also increased for a higher mass model and the flux contribution from the star is larger in the line. In this case, our BeAtlas model is able to produce more similar synthetic H$\alpha$ visibility and profile to the ones from our best-fit kinematic model.\par

One possible explanation for the discrepancy between the value of $\beta$ determined from H$\alpha$ and Br$\gamma$ could be the higher effects of non-kinematic broadening on H$\alpha$. This is already evidenced by the larger wings, in terms of Doppler shift, for this emission line . Such effects are known to be due to non-coherent scattering in the circumstellar environment, as explained, for example, in \citet{Auer68}. Global effects on interferometric data were discussed by \citet{stee12b} in the case of the Be star $\gamma$ Cassiopeiae observed with VEGA. These authors used a similar kinematic model, but with two additional parameters to quantify the non-coherent scattering and found that about half the flux in the line was affected by such an effect. Nevertheless, the possible bias on the measurement of $\beta$ in a line strongly affected by such non-kinematic broadening should be investigated further. \par

We also note that a possible close companion could influence $v_{\mathrm{rot}}$, as well as the disk structure, as previously mentioned. However, the presence of a close companion with a detectable influence on the measured parameters seems excluded from the observed calibrated $V^2$, and in particular from the spectro-interferometric differential observables, which both show signatures well reproduced by a symmetric rotating disk.\par

%%%%%%%%%%%%%%%%%%%%%%%%%%%%%%%%%%%%%%%%%%%%%%%%%%%%%%%%%%%%%%%%%%
\subsection{Disk variability: a multi-technique analysis}\label{sec_variability}

\subsubsection{Spectroscopy}\label{sec_variability_spectroscopy}

The Be star $\omicron$ Aquarii is known to possess a stable H$\alpha$ line profile for up to several years. For example, \citet{sigut15} verified that the $EW$ in the H$\alpha$ line is stable (within about 5\%) up to about nine years (from 2005 to 2014). \par 

\begin{figure}
\centerline{\resizebox{0.44\textwidth}{!}{\includegraphics{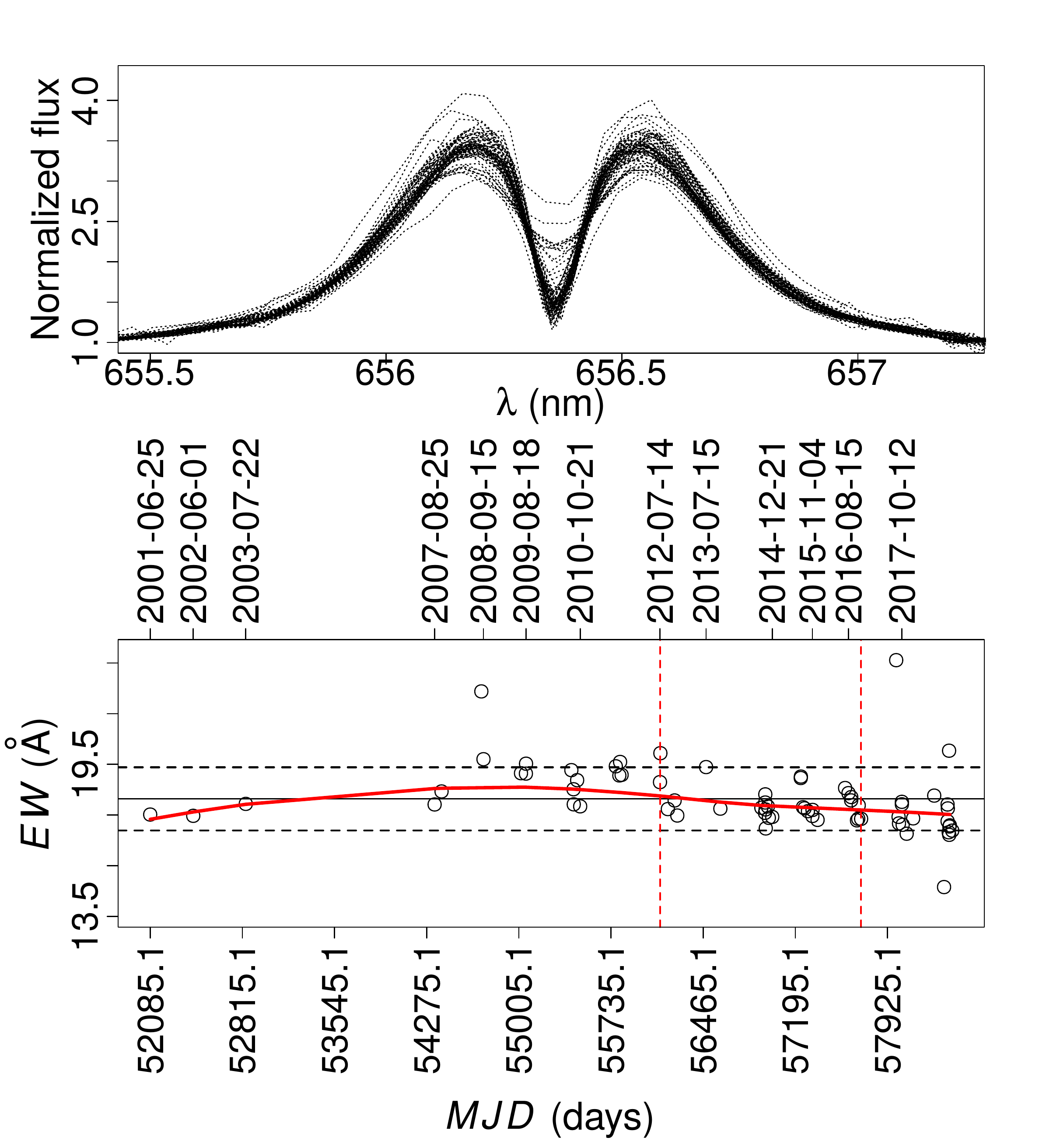}}}
\caption{Top panel: 70 observed H$\alpha$ profiles of $\omicron$ Aquarii from the BeSS database, covering about 17 years of observations (2001-2018). Bottom panel: H$\alpha$ equivalent width as a function of the observation time (modified Julian date). Civil dates are indicated for a part of the measured $EW$. The mean $EW$ (solid) within the standard deviation (dashed lines) is marked in black. Local regression fit of $EW$, as a function of time, is shown as a solid red line. The time interval covered by our interferometric observations (VEGA) is indicated with dashed red lines. See text for discussion.}
\label{bess_halpha_calcul_quantities}
\end{figure}

To go further in the analysis of the disk stability, we analyzed 70 H$\alpha$ line profiles, spanning from 2001 to 2018, from the BeSS database\footnote{Public data available at: \url{http://basebe.obspm.fr}.} \citep{neiner11}. Since these observations are performed with several instruments, the line profiles shown here are interpolated to have spectral resolution of 0.5 {\AA} (lowest resolution in the dataset). From these observations, we calculated the equivalent width ($EW$) in the H$\alpha$ line. In Fig. \ref{bess_halpha_calcul_quantities}, we show the analyzed H$\alpha$ spectra together to the temporal evolution of the H$\alpha$ $EW$.\par

%---------------------------------------%---------------------------------------
%---------------------------------------%---------------------------------------

%---------------------------------------%---------------------------------------
\begin{figure*}
\centerline{\resizebox{1.00\textwidth}{!}{\includegraphics{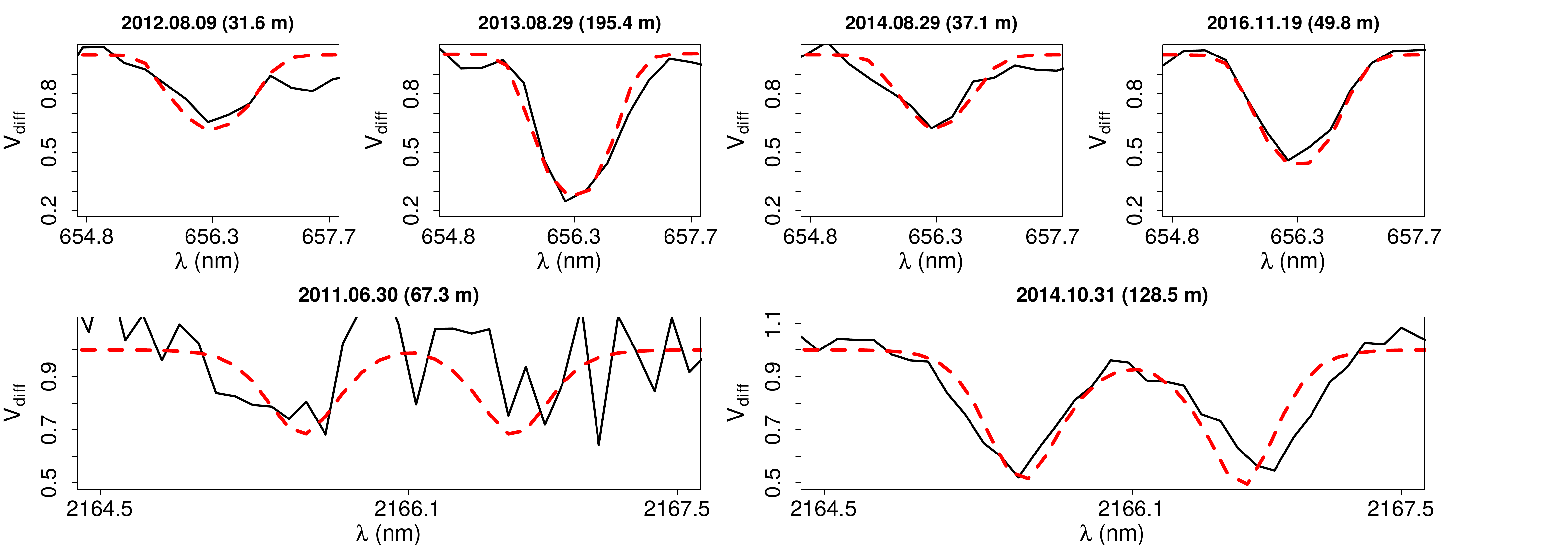}}}
\caption{VEGA (top panels) and AMBER (bottom panels) differential visibilities extracted from observations at different epochs (black line). VEGA measurements span four years and the AMBER ones span three years. The observation date and the baseline length (projected onto the sky) are indicated in the top of each panel. Our best-fit kinematic models derived from the fit, in a separate way, to each dataset (VEGA and AMBER) are shown in dashed red line. Note that our best-fit kinematic models match well to the differential visibilities obtained at different epochs.}
\label{vis_vega_amber_varying_time}
\end{figure*}
%---------------------------------------%---------------------------------------

We found that the disk is fairly stable over this 17-year time span with a mean value of $\overline{EW}$ = 18.1 $\pm$ 1.2 {\AA}. This value agrees well to older results in the literature. \citet{slettebak78} measured H$\alpha$ $EW$ = 18.80 $\pm$ 0.11 {\AA} in 1975 and 18.58 $\pm$ 0.21 {\AA} in 1976. From the H$\alpha$ profile observed in 1981, \citet{andrillat83} measured $EW$ = 17.2 {\AA}. Thus, this supports an even longer global disk stability up to at least 40 years. However, a slight increasing trend in $EW$ is seen between 2001 and 2012. This could suggest an augmentation in the disk density of $\omicron$ Aquarii in this period. Considering the period of our interferometric observations (from 2012 to 2016), it is hard to observe any trend of H$\alpha$ $EW$ as a function of time.\par

\subsubsection{Interferometry}\label{sec_variability_interferometry}

These results are consistent with our ability to model, with the same model parameters, simultaneously all our VEGA and AMBER data regardless of the epoch. In Fig. \ref{vis_vega_amber_varying_time}, we present a temporal evaluation of our spectro-interferometric data in the H$\alpha$ and Br$\gamma$ lines. Since the drop in visibility is expected to change due to possible variations in the disk extension, we only show here the differential visibilities from the VEGA and AMBER observations. These measurements are chosen to cover the whole period of our observations from 2011 to 2016. For a more robust comparison, we chose measurements obtained with different baseline lengths (projected onto the sky), and, thus, covering different levels of spatial resolution.\par

One sees that, regardless of the period of time of the observations, our final kinematic models provide a very reasonable match to both the VEGA and AMBER data. Thus, considering our interferometric data, we are not able to detect any conspicuous variation of the circumstellar disk extension within a period up to five years (from 2011 to 2016). This is in agreement with previous interferometric studies of $\omicron$ Aquarii by \citet{sigut15}. Besides that, this analysis supports our approach of fitting each one of the interferometric datasets (VEGA and AMBER) without imposing any discrimination based on the observation time.\par

\subsubsection{Polarimetry}
\label{sec_variability_polarimetry}

\begin{figure}
\centerline{\resizebox{0.50\textwidth}{!}{\includegraphics{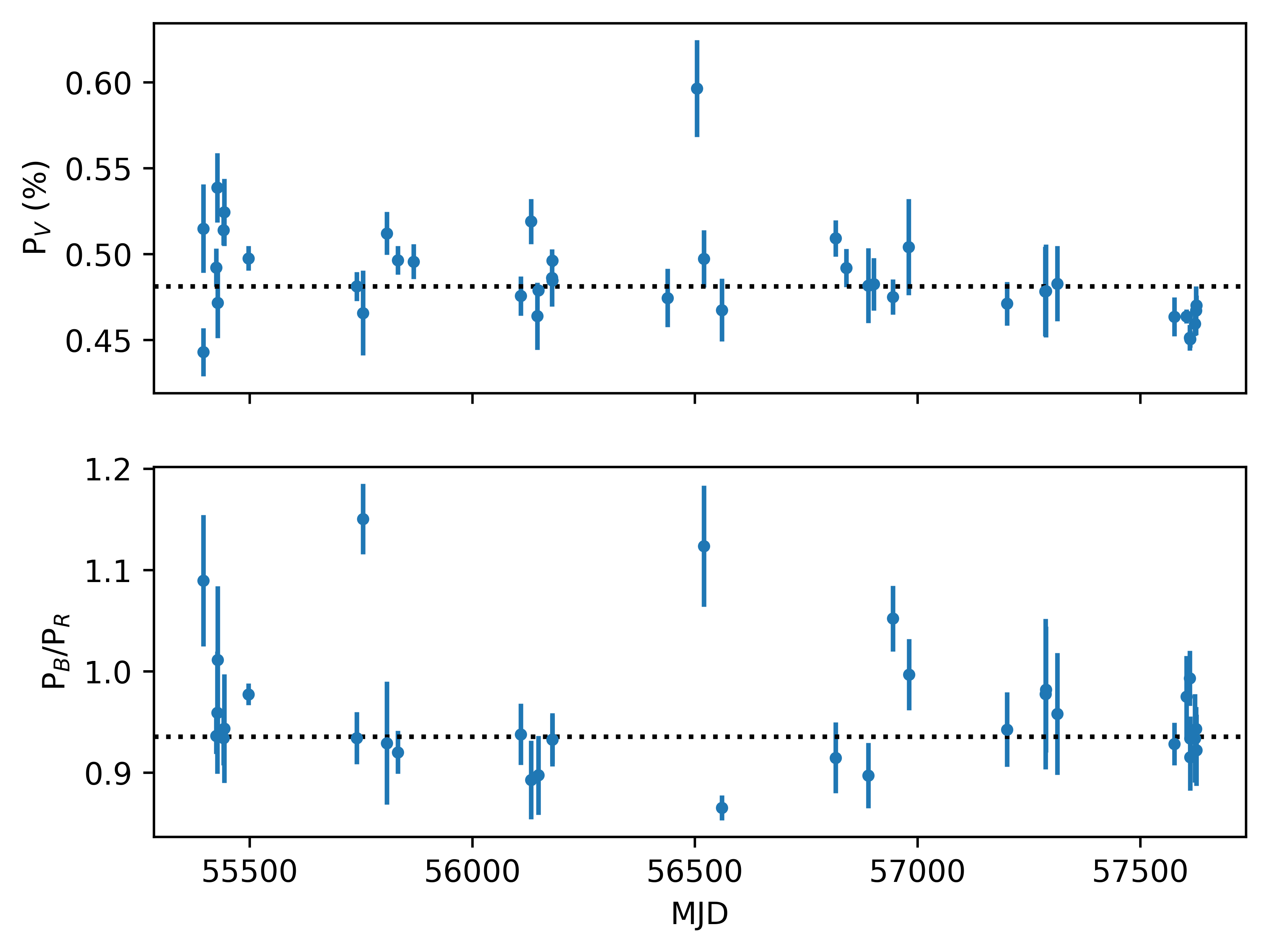}}}
\caption{Polarimetric quantities of $\omicron$ Aquarii, as a function of the observation time, spanning about six years. Top panel: observed V-band polarization (44 measurements). Bottom panel: ratio between the observed B- and R-bands polarization (40 measurements). The mean values of these quantities are shown in dashed line. See text for discussion.}
\label{plot_vs_mjd}
\end{figure}

Additional multi-epoch polarimetric data also support our findings of a stable disk for $\omicron$ Aquarii, close to the steady-state regime. In Fig \ref{plot_vs_mjd}, we show the temporal evolution of broad-band linear polarimetry in the V-band ($P_{V}$) of $\omicron$ Aquarii, as well as the ratio between the B- and R-bands polarization ($P_{B}/P_{R}$).\par 

These data were obtained over 43 nights, from June 2010 to August 2016, with the IAGPOL polarimeter \citep{magalhaes96}, mounted on the 0.6 m Boller \& Chivens telescope at Observatório do Pico dos Dias (OPD/LNA). This polarimeter is composed by a rotating half-wave retarder and a Savart Plate used as analyser to provide the modulation of the light polarization, and then the polarimetric quantities. Details of data reduction are found in \citet{magalhaes84} and \citet{bednarski16}.\par

From Fig. \ref{plot_vs_mjd}, the mean value of the observed V-band polarization is $\overline{P_{V}}$ = 0.48 $\pm$ 0.03\%. This value, derived from the mean and standard deviation of the Stokes Q and U parameters, is compatible to the one determined by \citet{yudin01}, namely, 0.52 $\pm$ 0.05\%. Since the observations from \citet{yudin01} predate our OPD/LNA observations by more than a decade, we conclude that the polarization values of $\omicron$ Aquarii remained very constant for over 20 years.

%%%----------------------------------------------------
\begin{table}[t]
\caption{\label{table_interstellar_polarization_omicron_aquarii} Interstellar parameters derived for $\omicron$ Aquarii: the Serkowski parameters, $P_{\mathrm{max}}$ and $\lambda_{\mathrm{max}}$, with the polarization angle $PA_{\mathrm{IS}}$.}
\centering
\renewcommand{\arraystretch}{1.00}
\begin{adjustbox}{width=0.300\textwidth}
\begin{tabular}{lcc}
\toprule
\toprule
$P_{\mathrm{max}}$ (\%) & $\lambda_{\mathrm{max}}$ ($\mu$m) & $PA_{\mathrm{IS}}$ (deg) \\
\midrule
0.11 $\pm$ 0.01 & 0.49 $\pm$ 0.18 & 132 $\pm$ 4 \\
\bottomrule
\end{tabular}
\end{adjustbox}
\end{table}
%%%----------------------------------------------------

In order to determine the intrinsic value of polarization, the interstellar contribution to the observed values quoted above must be removed. For that, we observed four main sequence stars, in the BVRI-bands, which are angularly close to $\omicron$ Aquarii. A MCMC method was implemented to process the four BVRI data of each field star, generating a sample of the likelihood function in terms of the interstellar Serkowski parameters $P_{\mathrm{max}}$ and $\lambda_{\mathrm{max}}$ \citep{serkowski75, wilking82}. The best estimates for these parameters are shown in Table \ref{table_interstellar_polarization_field_stars} (Appendix \ref{appendix_interstellar_polarization}).\par

There is a good agreement among the $PA$ values of the field stars. Moreover, by using Gaia DR2 distances, we found that $P_{\mathrm{max}}$ increases linearly along the line of sight of $\omicron$ Aquarii (see Fig. \ref{field_stars_om_aqr_distance_gaiadr2} in Appendix \ref{appendix_interstellar_polarization}). In this case, it suggests that the alignment of the grains at the interstellar medium is nearly homogeneous \citep[e.g.,][]{mclean79}. Thus, from a simple linear fit to $P_{\mathrm{max}}$ vs distance for the field fields, we determined $P_{\mathrm{max}}$ for $\omicron$ Aquarii. The derived interstellar polarization parameters for $\omicron$ Aquarii are shown in Table \ref{table_interstellar_polarization_omicron_aquarii}, which are in reasonable agreement with the ones reported in \citet{yudin01} of $P_{\mathrm{max}}$ = 0.20\% and $PA_{\mathrm{IS}}$ = \ang{125} (no error bars).\par

Taking into account our results for the interstellar polarization components, we found for the intrinsic V-band polarization and position angle $P^{\mathrm{int}}_{V}$ = 0.49 $\pm$ 0.03\% and $PA^{\mathrm{int}}$ = 2.5 $\pm$ \ang{2.7}, respectively. \citet{yudin01} determined $P^{\mathrm{int}}_{V}$ = 0.60\% with $PA^{\mathrm{int}}$ = \ang{6.0}, which is close to our $PA^{\mathrm{int}}$ value. Moreover, both estimates for $PA^{\mathrm{int}}$ are consistent with our determination for the disk major-axis position angle ($\sim$\ang{110}), being almost perpendicular to the polarization vector, as expected.\par

Furthermore, our best-fit HDUST model (Table \ref{table_hdust_ref}) predicts a polarization degree of 0.41\% in the V-band. This agrees well with our measurement for the average intrinsic polarization of the OPD/LNA data. Therefore, besides the independent checks provided by the SED and spectroscopic data (Sect. \ref{sec_spectrum_sed}), our polarimetric data also support our physical model for $\omicron$ Aquarii, which was derived purely from the fit to interferometric data (as discussed in Sect. \ref{sec_hdust_analysis}).\par

%---------------------------------------%---------------------------------------
%---------------------------------------%---------------------------------------

Lastly, Fig. \ref{plot_vs_mjd} shows that both the polarization degree in the V-band and the ratio between the B- and R-bands are almost constant in time, showing a small scatter around the mean value. In particular, this latter quantity is related to the density scale at the inner portion of the disk \citep[][]{haubois14}. From the theoretical investigation of \citet{panoglou19}, the variation on the polarization degree in the V-band ($\Delta P_{V}$) can reach up to about 0.1\% due to asymmetries in the disk density structure, caused by a binary companion. Moreover, \citet{haubois14} predicted $\Delta P_{V}$ of
up to 2\% due to temporal changes in the mass decretion rate. The standard deviation of our $P_{V}$ distribution (approximately Gaussian), namely, $\sim$0.03\%, is quite a bit lower than the above values. It is well explained in terms of the precision of our polarimetric data, as the typical error bar on $P_{V}$ is $\sim$0.01-0.02\% (Fig. \ref{plot_vs_mjd}).\par

\subsubsection{A stable disk}\label{sec_variability_final_remarks}

Besides the analysis of the H$\alpha$ $EW$ and broad-band polarimetric quantities, our modeling with the code HDUST indicates that the disk must be close to the steady-state regime: having a radial density law exponent of 3.0 \citep[e.g.][]{haubois12, vieira17}. Other studies of $\omicron$ Aquarii are in fair agreement to our findings from HDUST. Using the radiative transfer code BEDISK, \citet{silaj10} derived $m$ = 3.5 from the fit to the H$\alpha$ profile, while \citet{sigut15} found $m$ = 2.7 as a representative value from the analysis of all the different observables.\par 

Previous and ongoing studies of Be stars with stable disks found similar results to ours. For example, \citet{klement15} found $m$ = 2.9 for the late-type Be star $\beta$ Canis Minoris (B8Ve). \citet{mota19} derived $m$ = $2.44^{+0.27}_{-0.16}$ for $\alpha$ Arae (B2Vne). The B9Ve star $\alpha$ Columbae shows $m$ = $2.54^{+0.06}_{-0.13}$ (A. Rubio, priv. comm). Thus, the radial density exponent is consistently equal or somewhat less than 3.0 for these Be stars with stable disks. Also, from analysing the temporal variation of the disk density, \citet{vieira17} identified a slightly extended range of $m$ (between $\sim$3.0 and $\sim$3.5) for the steady-state regime, in comparison to the canonical value of 3.5. As pointed out by these authors, this canonical value is based on simplifications of the standard theory, which assumes, for example, vertically isothermal disks and isolated systems (single stars). One possibility to explain the measured $m$ lower than 3.5 could thus rely on non-isothermal effects in the disk structure \citep[see, e.g.,][]{carciofi08}.\par

Finally, we note that such long-term stability of $\omicron$ Aquarii's disk is consistent with other results in the literature: late-type Be stars are more likely to have more stable disks than earlier Be stars \citep[e.g.,][]{vieira17, labadie18, rimulo18}. As discussed in Sect. \ref{sec_rotation}, the stellar rotation seems to be very close ($\sim$96\%) to the critical value \citep[391 $\pm$ 27 km s\textsuperscript{-1} from][]{fremat05}, in particular regarding the HDUST analysis: $v_{\mathrm{rot}}$ = 368 km s\textsuperscript{-1} (Table \ref{table_hdust_ref}). This is consistent with the results from \citet{cranmer05}: Be stars with lower effective temperature $T_{\mathrm{eff}} \lesssim 21000$ K -- that is, later spectral types such as our target -- are more likely to have a rotation rate close to one than the earlier Be stars. Thus, one possibility to explain such a long-term stability of the disk of $\omicron$ Aquarii could rely on its fast rotation, ensuring in this case a nearly constant mass-injection rate into the disk.\par

%%%%%%%%%%%%%%%%%%%%%%%%%%%%%%%%%%%%%%%%%%%%%%%%%%%%%%%%%%%%%%%%section: conclusions

\section{Conclusions}\label{sec_conclusions}

We analyzed VEGA $V^2$, as well as VEGA and AMBER differential visibility and phase of the Be-shell star $\omicron$ Aquarii. To date, the spectro-interferometric dataset analyzed in this paper is the largest for a Be star, considering quasi-contemporaneous observations in both the H$\alpha$ (VEGA) and Br$\gamma$ (AMBER) lines.\par

For the first time, we measured $\omicron$ Aquarii's stellar radius ($R_{\star}$ = 4.0 $\pm$ 0.3 $\mathrm{R_{\odot}}$) and determined the disk extension in the H$\alpha$ and Br$\gamma$ lines as, respectively, 10.5 $\pm$ 0.3 $\mathrm{D_{\star}}$ and 11.5 $\pm$ 0.1 $\mathrm{D_{\star}}$. Using radiative transfer models computed with the code HDUST, we explained the quasi-identical extension of the emission in these lines by an opacity effect found for disks seen under a high inclination angle.\par

We showed that the inclination angle derived from H$\alpha$ is about \ang{15} lower than the one determined in Br$\gamma$, when analysing each line separately with HDUST. More complex physical models, for example, with non-isothermal vertical scaling of the disk or the addition of heavier elements, could resolve this issue and should be investigated in the future.\par

Our simple kinematic model highlighted the high correlation between the rotational velocity at the base of the disk and the rotational law exponent $\beta$. Assuming external constraints, such as $v \sin i$, we managed to constrain this parameter and showed that the disk rotation is nearly Keplerian ($\beta$ $\sim$ 0.43) from the analysis in the Br$\gamma$ emission line. As for the inclination angle, the determination of $\beta$, using the H$\alpha$ line ($\beta$ $\sim$ 0.30), seems to be significantly biased. Other studies also verified such a large deviation from the Keplerian rotation for Be stars when analysing interferometric quantities measured in H$\alpha$ \citep[see, e.g.,][]{delaa11}. One possible explanation would be the higher effect of non-coherent scattering on the H$\alpha$ line formation than on Br$\gamma$.\par

Despite being derived purely from the fit to interferometric data, our best-fit HDUST model provides a very reasonable match to non-interferometric observables of $\omicron$ Aquarii: the observed SED, H$\alpha$ and Br$\gamma$ line profiles, and polarimetric quantities. Thus, this cross-check provides an independent validation of our best-fit physical model. We found using HDUST a satisfying common physical description for the circumstellar disk in both H$\alpha$ and Br$\gamma$: a base disk surface density $\Sigma_{0}$ = 0.12 g cm\textsuperscript{-2} ($\rho_{0}$ = $5.0\e{-12}$ g cm\textsuperscript{-3}) and a radial density law exponent $m$ = 3.0, that is, close to the steady-state regime according to the VDD model ($m$ = 3.5). This result agrees with recent studies of other Be stars with stable disks, and may indicate the necessity to revise $m$ = 3.5 (steady-state standing for single stars with vertically isothermal disks) that is predicted by the VDD theory. Otherwise, this could indicate non-isothermal effects on the disk vertical structure of $\omicron$ Aquarii. The long-term stability of the $\omicron$ Aquarii's disk is verified by our analysis of a large sample of H$\alpha$ profiles and polarimetric data, spanning about 20 and six years, respectively. Combined with older results in the literature, a longer global disk stability is suggested for up to at least 40 years.\par

The stellar rotation seems to be very close ($\sim$96\%) to the critical value (391 km s\textsuperscript{-1}), in particular accordingly to our HDUST analysis: $v_{\mathrm{rot}}$ = 368 km s\textsuperscript{-1} from the best-fit HDUST model with fixed $M_{\star}$ = 4.2 $\mathrm{M_{\odot}}$ (cf., Sects. \ref{sec_beatlas} and \ref{sec_hdust_analysis}). One possibility to explain such a long-term stability in the disk of $\omicron$ Aquarii could rely on its own high stellar rotation, being, in this case, a main source for the mass injection from the stellar surface to the disk. Thus, apart from the mass decretion due to other possible mechanisms in Be stars, this would provide a constant rate of mass injection. In short, our results on the stellar rotation and on the disk stability are consistent with the literature results showing that late-type Be stars are more likely to be fast rotators and have stable disks (see Sect. \ref{sec_variability_final_remarks}).\par

Finally, to further investigate these issues, our multi-wavelength and multi-emission line modeling approach must be performed on a larger sample of Be stars with disks of different densities and seen under different inclination angles. The implementation of a MCMC model fitting procedure with the kinematic model, and the use of our grid of HDUST models (BeAtlas), are very promising for the spectro-interferometric analysis of a large survey of Be stars, providing robust model parameters and associated uncertainties. A future project will attempt this task on a few dozen objects observed with VEGA and AMBER. 

\bibliographystyle{aa} % style aa.bst

\begin{acknowledgements}

We thank the anonymous referee for helping to improve this paper. E. S. G. de Almeida thanks OCA and the ``Ville de Nice'' (Nice, France) for the financial support to this work through the ``Bourse Doctorale Olivier Chesneau'' during the period of 2016-2019. E. S. G. de Almeida acknowledges A. Rubio for relevant information about her work on $\alpha$ Columbae. R. Ligi has received funding from the European Union's Horizon 2020 research and innovation program under the Marie Sk\l odowska-Curie grant agreement n. 664931. D. M. Faes acknowledges FAPESP (grant 2016/16844-1). A. C. Carciofi acknowledges support from CNPq (grant 307594/2015-7). This work was supported by the "Programme National de Physique Stellaire" (PNPS) of CNRS/INSU co-funded by CEA and CNES". This work made use of the computing facilities of the Laboratory of Astroinformatics (IAG/USP, NAT/Unicsul), whose purchase was made possible by the Brazilian agency FAPESP (grant 2009/54006-4) and the \mbox{INCT-A}. This work is based upon observations obtained with the Georgia State University Center for High Angular Resolution Astronomy Array at Mount Wilson Observatory. The CHARA Array is supported by the National Science Foundation under Grants No. AST-1211929 and AST-1411654. This work used BeSOS Catalogue, operated by the Instituto de Física y Astronomía, Universidad de Valparaíso, Chile : http://besos.ifa.uv.cl and funded by Fondecyt iniciación N 11130702. Based on observations collected at the European Southern Observatory under ESO programmes 087-D.0311 and 094.D-0140. This work has made use of the BeSS database, operated at LESIA, Observatoire de Meudon, France: http://basebe.obspm.fr. Some of the data presented in this paper were obtained from the Mikulski Archive for Space Telescopes (MAST). STScI is operated by the Association of Universities for Research in Astronomy, Inc., under NASA contract NAS5-26555. Support for MAST for non-HST data is provided by the NASA Office of Space Science via grant NNX13AC07G and by other grants and contracts. This research has made use of the Jean-Marie Mariotti Center (JMMC) services \texttt{LITpro}, \texttt{SearchCal}, and \texttt{AMHRA}, co-developped by CRAL, IPAG and Lagrange. This work has made use of the SIMBAD and VizieR databases, operated at CDS, Strasbourg, France.

\end{acknowledgements}

\bibliography{references.bib}

%%%%%%%%%%%%%%%%%%%%%%%%%%%%%%%%%%%%%%%%%%%%%%%%%%%%%%%%%%%%%%%%section: appendices
\begin{appendix}
\onecolumn

\section{Observational logs}
\label{appendix_observational_logs}

%---------------------------------------%---------------------------------------
\begin{table}[!h]
\caption{\label{calib_vega_amber} List of stellar calibrators used for the VEGA observations.}
\centering
\renewcommand{\arraystretch}{1.1}
\begin{adjustbox}{width=0.50\textwidth}
\begin{tabular}{ccccc}
\toprule
\toprule

\makecell{Star \\ (HD)} &Spec. type  &\makecell{R \\ (mag)} &\makecell{K \\ (mag)} & \makecell{Diameter \\ (mas)}\\
%\hline
%\multicolumn{5}{c}{VEGA}\\
\midrule 
194244 	& B9V	& 6.1	& 6.1 &	0.161 $\pm$ 0.011\\
210424 	& B5III	& 5.5	& 5.7 & 0.177 $\pm$ 0.012\\
211924 	& B5IV	& 5.4	& 5.5 & 0.219 $\pm$ 0.015\\
224926 	& B7III-IV & 5.2 & 5.4	& 0.197 $\pm$ 0.014\\
%\hline
\bottomrule
%\multicolumn{5}{c}{AMBER}\\
%\hline
%210434 	& G8/K0III & 6.0 & 4.1	& 0.713$\pm$0.049\\
%\hline
\end{tabular}
\end{adjustbox}
\end{table}
%---------------------------------------%---------------------------------------

%---------------------------------------%---------------------------------------
\begin{table}[!h]
\caption{\label{log_vega_amber} List of the VEGA and AMBER observations. In the third column, the number of measurements are shown accordingly to the presented UT interval (second column). CHARA (VEGA) and VLTI (AMBER) telescope configurations are shown in the fourth column.}
\centering
\renewcommand{\arraystretch}{1.1}
\begin{adjustbox}{width=0.50\textwidth}
\begin{tabular}{cccc}
\toprule
\toprule

UTC (date)  & UTC (hh:mm)   &\textnumero\ of measur.& Tel. conf.  \\
%            &               &                       &              \\
\midrule 
\multicolumn{4}{c}{\textbf{VEGA}}  \\
\midrule 
2012-08-28 &09:40           & 1 &	S1-S2 	\\
2013-08-28 &09:07 - 09:43   & 2 &	S2-S1-E2\\
2013-08-28 &07:05           & 1 &	W2-W1-E1\\
2013-08-30 &10:35	        & 1 &   S2-S1-W2\\
2013-10-31 &05:45           & 1 &	W2-W1	\\
%%%2013-12-13 &03:09           & 1 &	W2-S1-E2\\ %%%obs in 720nm (PI: Stee)
2013-08-29 &07:55	        & 1 &   S2-S1-E2\\
2013-08-29 &06:35           & 1 &   W2-W1-E1\\
2014-07-03 &10:38 - 11:27   & 2 &	E2-E1	\\
2014-07-04 &08:24 - 11:32   & 3 &	E2-S2-W2\\
2014-07-06 &08:06           & 1 &	E2-E1	\\
2014-07-08 &10:57 - 11:49   & 4 &	E2-S2-W2\\
2014-07-10 &08:07 - 11:57   & 9 &	E2-S2-W2	\\
2014-08-22 &07:12 - 07:41   & 2 &	E2-S2-W2	\\
2014-08-23 &06:19 - 06:49   & 2 &	E2-S2-W2\\
2014-08-25 &06:40           & 1 &	E2-E1	\\
2014-08-28 &07:19 - 09:39   & 3 &	W2-W1	\\
2014-08-29 &06:41 - 09:51   & 3 &	E2-E1	\\
2014-10-17 &03:52           & 1 &	S1-E1-W1	\\
2014-10-19 &03:40 - 05:32   & 4	&   W2-S2-W1	\\
2014-10-20 &02:38 	        & 1 &   W2-S2-W1  	\\
2014-10-23 &03:13 - 03:43   & 2 &	S1-E1-W1\\
2016-11-19 &01:44 - 04:53   & 3 &	E2-E1	\\
\midrule 
\multicolumn{4}{c}{\textbf{AMBER}}  \\
\midrule 
2011-06-20 & 10:08 & 1 & D0-I1-H0\\ 
2014-10-29 & 00:48 - 01:27 & 2 & A1-G1-I1\\
2014-10-30 & 01:06 - 03:31 & 2 & A1-G1-J3\\
2014-10-31 & 00:25 - 03:39 & 3 & A1-K0-J3\\
\bottomrule
\end{tabular}
\end{adjustbox}
\end{table}
%---------------------------------------%---------------------------------------

\clearpage
\section{MCMC fitting tests: fits to the VEGA and AMBER data with the kinematic model}\label{appendix_mcmc}

%---------------------------------------%---------------------------------------
\begin{figure}[!h]
  \begin{adjustbox}{minipage=\linewidth,scale=1.25}
  \centering
  \hspace{-4.3cm}
  \includegraphics[width=0.45\linewidth]{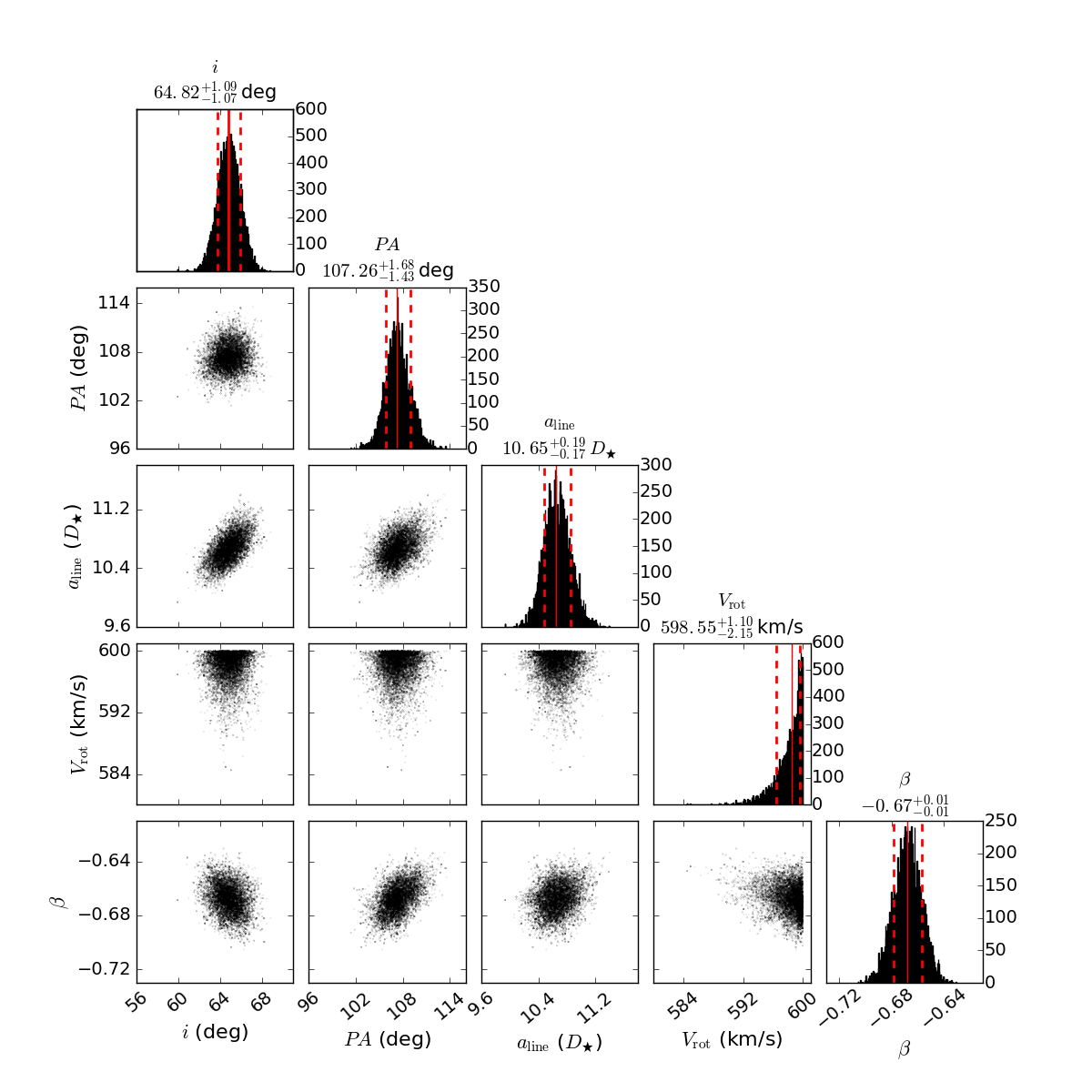}%\hspace{-0.50cm}%
  \medskip
  \hspace{-0.5cm}%
  \includegraphics[width=0.45\linewidth]{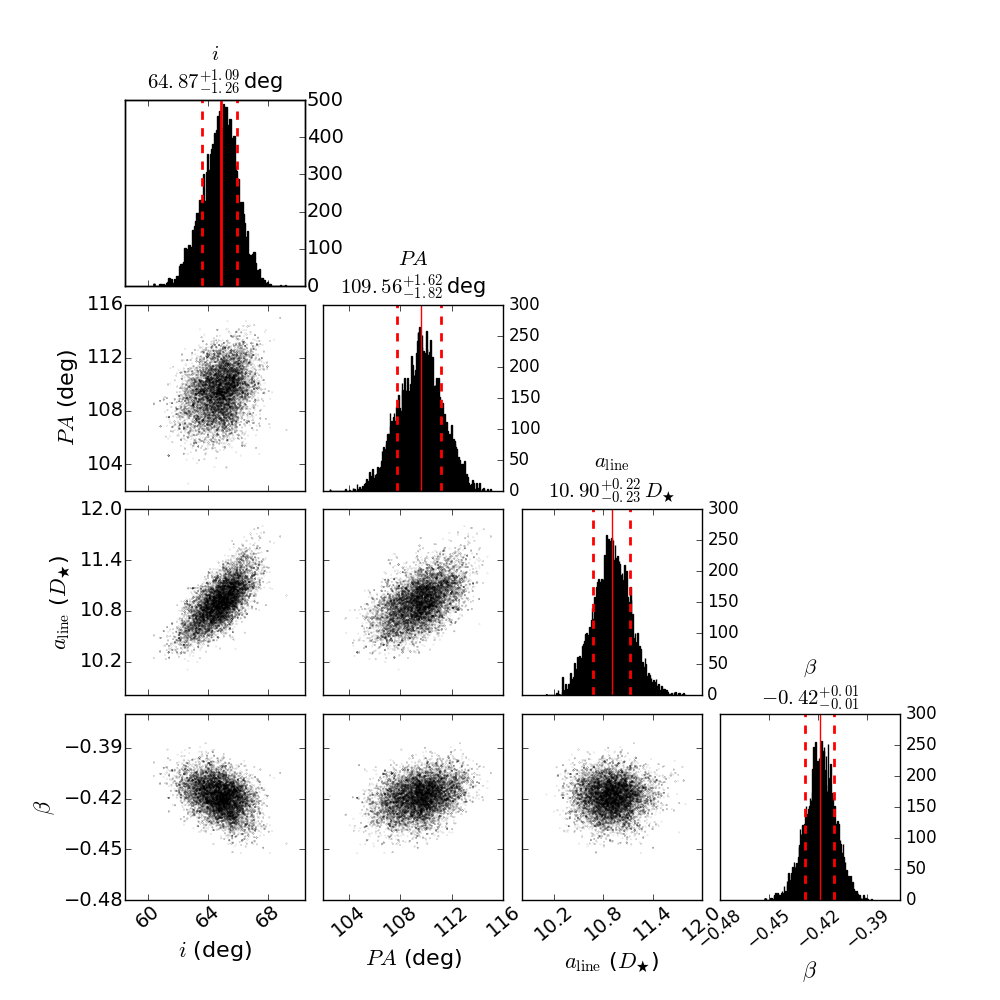}
  \end{adjustbox}
  \vspace{-0.5cm}
  \caption{As in Fig. \ref{mcmc_vega_amber_corner_hist}, but for the other MCMC fitting tests (test i in the left and test ii in the right) to fit the VEGA differential data.}\label{mcmc_other_tests_vega}
\end{figure}
%---------------------------------------%---------------------------------------

\vspace{-0.75cm}

%---------------------------------------%---------------------------------------
\begin{figure}[!h]
  \begin{adjustbox}{minipage=\linewidth,scale=1.25}
  \centering
  \hspace{-4.3cm}
  \includegraphics[width=0.45\linewidth]{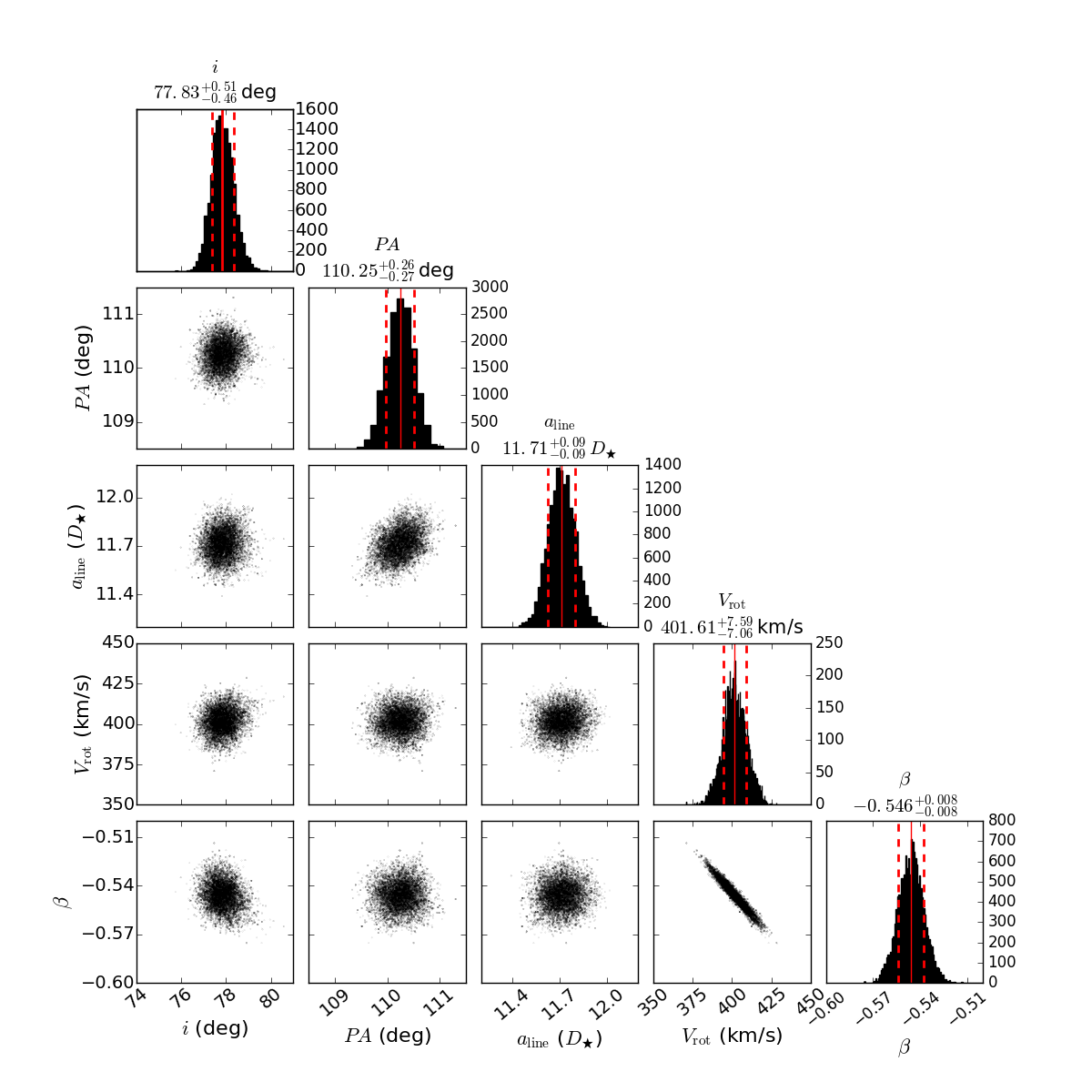}%\hspace{-0.50cm}%
  \medskip
  \hspace{-0.5cm}%
  \includegraphics[width=0.45\linewidth]{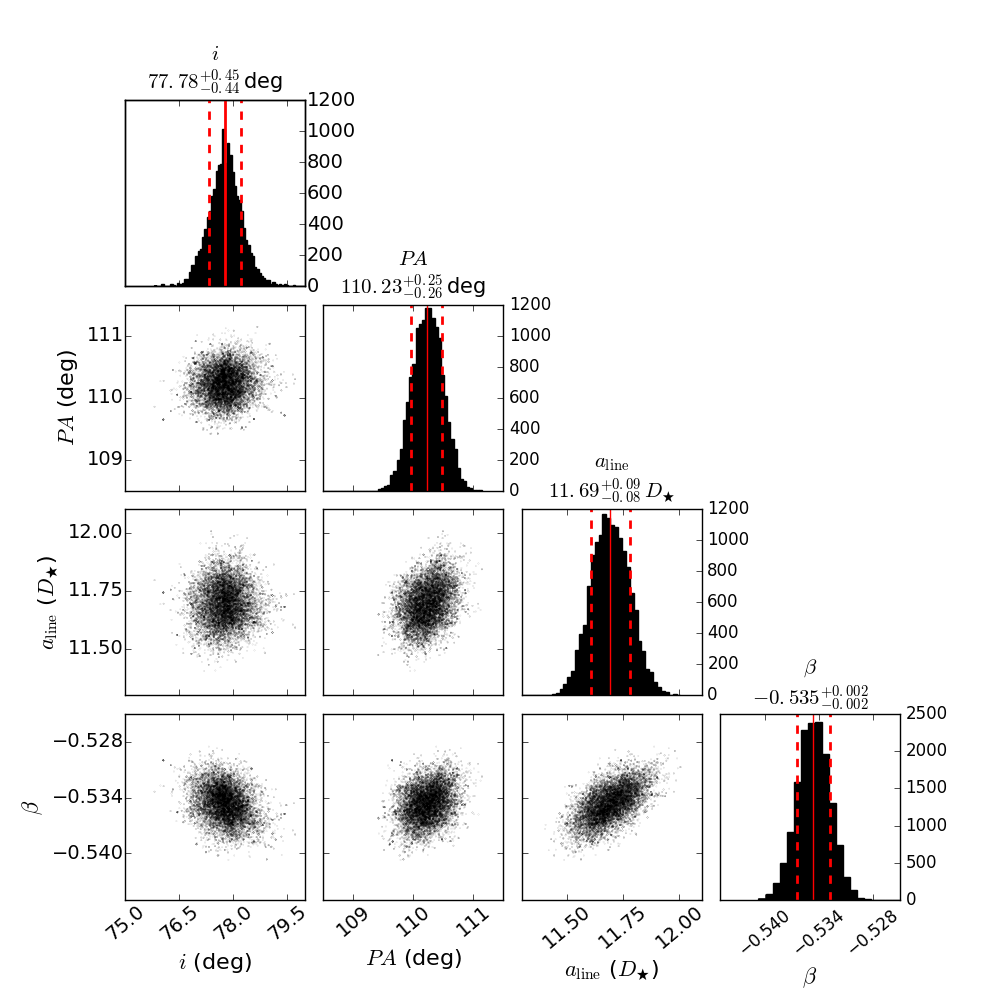}
  \end{adjustbox}
  \vspace{-0.5cm}
  \caption{As in Fig. \ref{mcmc_vega_amber_corner_hist}, but for the other MCMC fitting tests (test i in the left and test ii in the right) to fit the AMBER differential data.}\label{mcmc_other_tests_amber}
\end{figure}
%---------------------------------------%---------------------------------------

%%%\clearpage
\section{Best-fit kinematic and HDUST models: AMBER}
\label{appendix_bestfits_all_data}

%---------------------------------------%---------------------------------------
%\begin{sidewaysfigure*}
%\centerline{\resizebox{1.04\textwidth}{!}{\includegraphics{./kinematic_hdust_all_amber.eps}}}
%\caption{Comparison between our best-fit kinematic (dashed red; Table \ref{table_mcmc_vega_amber}) and HDUST (dashed blue; Table \ref{table_hdust_ref}) models to all the AMBER measurements (black).}
%\label{models_all_amber}
%\end{sidewaysfigure*}
%---------------------------------------%---------------------------------------

%---------------------------------------%---------------------------------------
\begin{sidewaysfigure*}
\centerline{\resizebox{1.04\textwidth}{!}{\includegraphics{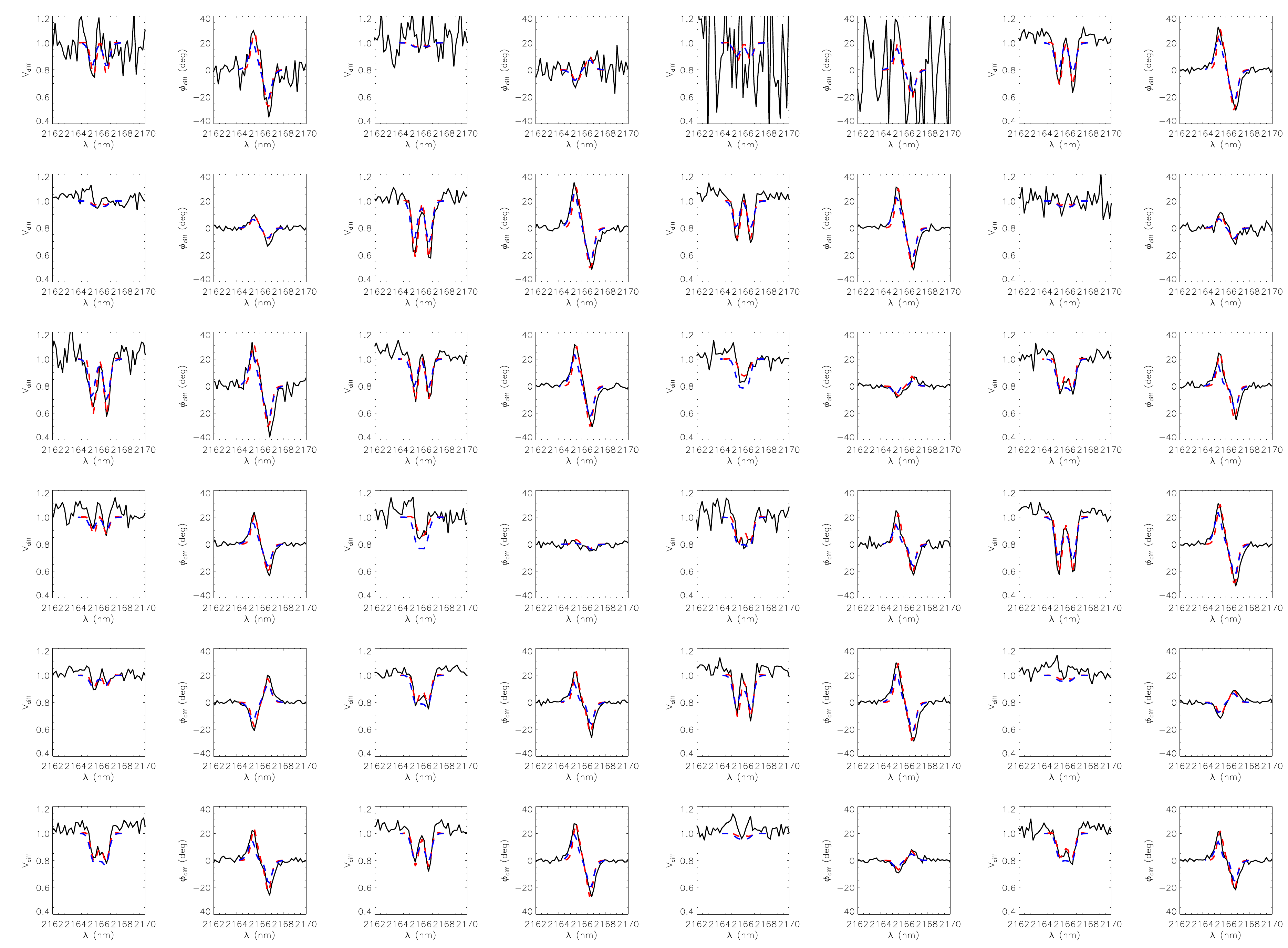}}}
\caption{Comparison between our best-fit kinematic (dashed red; Table \ref{table_mcmc_vega_amber}) and HDUST (dashed blue; Table \ref{table_hdust_ref}) models to all the AMBER measurements (black).}
\label{models_all_amber}
\end{sidewaysfigure*}
%---------------------------------------%---------------------------------------

\clearpage
\section{Interstellar polarization}
\label{appendix_interstellar_polarization}

%---------------------------------------%---------------------------------------

\begin{table*}[!h]
\caption{\label{table_interstellar_polarization_field_stars} Fitted Serkowski parameters, with the polarization angle, for the field stars used to derive the interstellar polarization of $\omicron$ Aquarii. We show the median and the 15.87th and 84.13th percentiles for $P_{\mathrm{max}}$ and $\lambda_{\mathrm{max}}$ from the MCMC analysis \citep[more details in][]{bednarski16}. The interstellar polarization angle estimated for each field star is the mean value among the observations in the BVRI-bands ($\langle PA_{\mathrm{IS}} \rangle$).}

\centering
\renewcommand{\arraystretch}{1.75}
\begin{adjustbox}{width=0.75\textwidth}
\begin{tabular}{lccccc}
\toprule
\toprule
Star & \makecell{RA (J2000) \\ (hh:mm:ss)} & \makecell{DEC (J2000) \\ (deg:arcmin:arcsec)} & $P_{\mathrm{max}}$ (\%) & $\lambda_{\mathrm{max}}$ ($\mu$m) & $\langle PA_{\mathrm{IS}} \rangle$ (deg) \\
\midrule 

HD 208719  & 21 58 20.0 & -01 49 46.7 & $0.095^{+0.006}_{-0.005}$ & $0.75^{+0.05}_{-0.06}$ & 130.7 $\pm$ 2.9 \\ %%%OK
HD 209348  & 22 02 48.4 & -02 28 44.4 & $0.012^{+0.005}_{-0.005}$ & $0.40^{+0.34}_{-0.24}$ & 128 $\pm$ 44 \\ %%%OK
2MASS J22025363-0229207 & 22 02 53.6 & -02 29 20.7 & $0.88^{+0.17}_{-0.18}$ & $0.34^{+0.21}_{-0.17}$ & 135.4 $\pm$ 9.1 \\ %%%OK
2MASS J22025544-0230058 & 22 02 55.4 & -02 30 05.8 & $0.975^{+0.017}_{-0.014}$ & $0.49^{+0.02}_{-0.02}$ & 136.8 $\pm$ 0.6\\ %%%OK

\bottomrule
\end{tabular}
\end{adjustbox}

\end{table*}
%---------------------------------------%---------------------------------------

%---------------------------------------%---------------------------------------
\begin{figure*}[!h]
\centerline{\resizebox{0.60\textwidth}{!}{\includegraphics{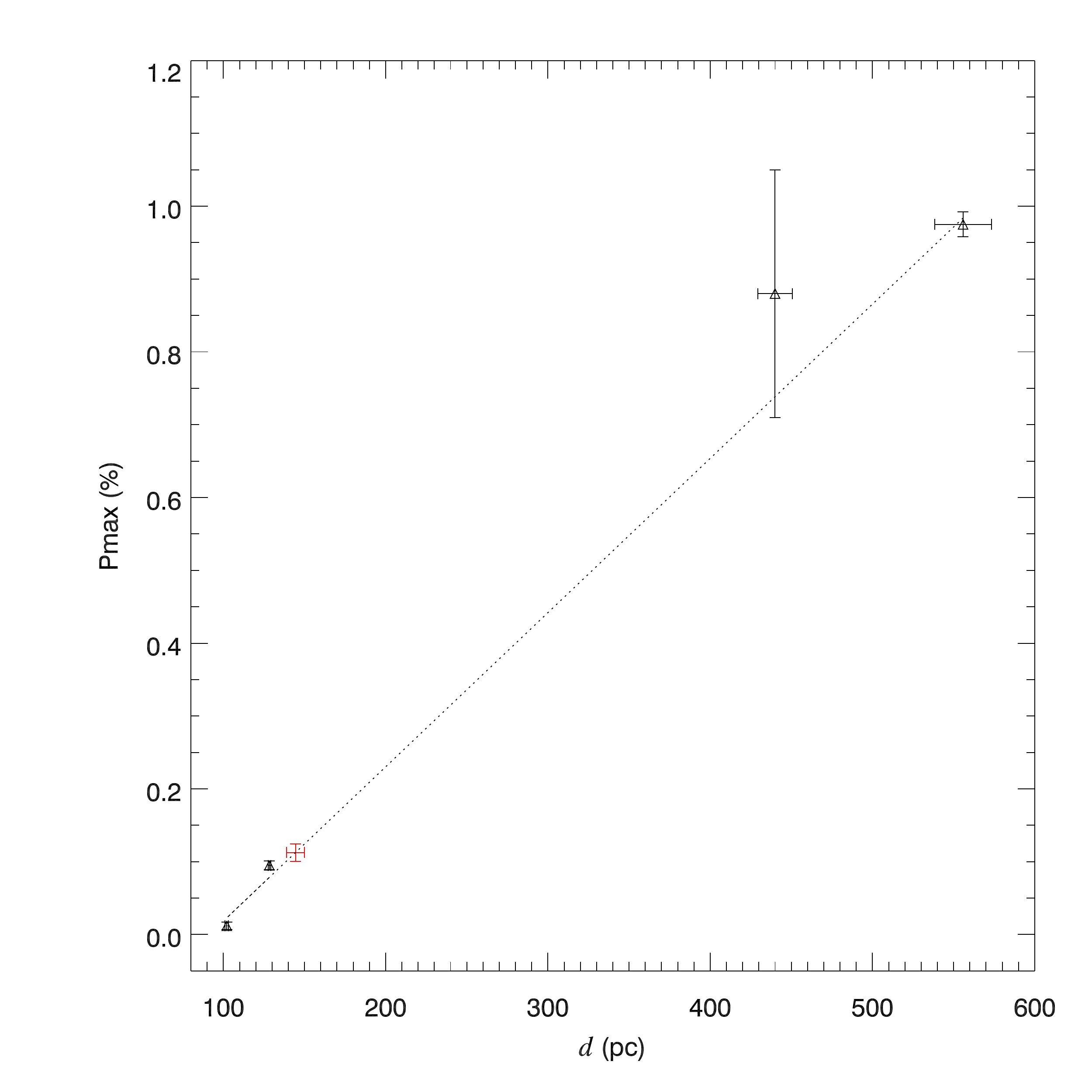}}}
\caption{Fitted $P_\mathrm{max}$ for the field stars (Table \ref{table_interstellar_polarization_field_stars}, open triangles) as a function of the Gaia DR2 distance. From the linear fit to $P_\mathrm{max}$ vs $d$ for the field stars (dotted line), we determined $P_\mathrm{max}$ for $\omicron$ Aquarii (red cross).}
\label{field_stars_om_aqr_distance_gaiadr2}
\end{figure*}
%---------------------------------------%---------------------------------------

\end{appendix}

\end{document}